\documentclass[a4paper,11pt]{article}
\usepackage{graphicx,amssymb,bm,latexsym,color,epsf,cite}
\makeatletter
\@addtoreset{equation}{section}

\makeatother
\newcommand{\bea}{\begin{eqnarray}}
\newcommand{\eea}{\end{eqnarray}}
\newcommand{\vs}[1]{\vspace{#1 mm}}
\newcommand{\hs}[1]{\hspace{#1 mm}}
\renewcommand{\a}{\alpha}
\renewcommand{\b}{\beta}
\renewcommand{\c}{\gamma}
\newcommand{\G}{\Gamma}
\renewcommand{\d}{\delta}

\newcommand{\s}{\sigma}
\renewcommand{\t}{\theta}
\newcommand{\vp}{\varphi}
\newcommand{\la}{\lambda}
\newcommand{\pa}{\partial}

\newcommand{\nn}{\nonumber\\}
\newcommand{\p}[1]{(\ref{#1})}

\newcommand{\lra}{\leftrightarrow}
\newcommand{\bF}{\bar F}
\newcommand{\br}{\bar R}
\newcommand{\bg}{\bar g}
\newcommand{\bpsi}{\bar \psi}
\newcommand{\bp}{\bar \phi}
\newcommand{\ba}{\bar A}
\newcommand{\bd}{\bar D}

\newcommand{\bnabla}{\bar\nabla}
\newcommand{\tr}{{\rm tr\,}}
\newcommand{\Tr}{{\rm Tr}}

\newcommand{\Dsl}{D \kern-0.65em /\,}

\newcommand{\tz}{\tilde z}

\begin{document}

\begin{flushright}
KU-TP 074 \\
\today
\end{flushright}

\begin{center}
{\Large\bf Effective Action from the Functional Renormalization Group}
\vs{10}

{\large
Nobuyoshi Ohta$^{a,}$\footnote{e-mail address: ohtan@phys.kindai.ac.jp}
and Les\l{}aw Rachwa\l{}$^{b,}$\footnote{e-mail address: grzerach@gmail.com}
} \\
\vs{10}
$^a${\em Department of Physics, Kindai University,
Higashi-Osaka, Osaka 577-8502, Japan}

$^b${\em Department of Nuclear Physics and Physical Engineering, Czech Technical University,
B\v{r}ehova 7, Prague 11519, Czech Republic
}

\vs{10}
%%%%%%%%%%%%%%%%%%%%%%%%%%%%%%%%
{\bf Abstract}
\end{center}

We study the quantum gravitational system coupled to a charged scalar, Dirac fermions, and electromagnetic fields.
We use the ``exact'' or ``functional'' renormalization group equation to derive the effective action $\G_0$
by integrating the flow equation from the ultraviolet scale down to $k=0$. The resulting effective action consists
of local terms and nonlocal terms with unique coefficients.
%, which could be tested by comparing it with observation.

%%%%%%%%%%%%%%%%%%%%%%%%%%%%
\section{Introduction}
%%%%%%%%%%%%%%%%%%%%%%%%%%%%
\label{sec1}

It is one of the most urgent problems in theoretical physics to understand quantum property of the gravitational
system interacting with various matter fields. The Einstein gravity is non-renormalizable in perturbation theory.
If one believes that gravity must be described by some quantum theory at some level,
then we are led to expect that the theory will contain terms quadratic in curvatures which are necessarily
generated by quantum effects, with coefficients of order unity.
In fact, we expect to find in the action all possible diffeomorphism-invariant terms constructed with the metric
and its derivatives.
It had been shown long ago that if one includes such higher derivative terms, then the theory is perturbatively
renormalizable~\cite{Stelle1}. The price one must pay is that perturbative unitarity is lost.
The superstring theory is supposed to circumvent the problem, but it is still difficult to understand
quantum geometric aspects of the spacetime physics in superstrings.

Another approach to get insight into the quantum effects of gravity is to use the functional renormalization group (FRG).
It enables us to study the RG flow of  infinitely many couplings as functions of a cutoff $k$.
It has been used to study the ultraviolet (UV) behavior of gravity and establish the existence of a nontrivial
fixed point (FP) which may be used to define a continuum limit~\cite{Percacci:2017,RS,Eichhorn:2018}.
To formulate this, one defines the effective average action (EAA) $\G_k$ by performing the path integral
over field modes with the momentum scales equal or bigger than a scale given by a cutoff $k$~\cite{wetterich,Morris1994},
and following the usual procedure of defining effective action. This EAA itself is not the usual effective action
and it is still divergent in general in the UV limit.
The $k$-dependence of the EAA is described by the FRG equation (FRGE)
\begin{equation}
\pa_t \G_k=\frac12 {\rm STr}\left[\left(\frac{\d^2 \G_k}{\d\phi\d\phi} +R_k \right)^{-1} \pa_t R_k \right],
\label{eq1}
%\label{frge1}
\end{equation}
where $t=\ln k$, and $R_k$ is a cutoff kernel which goes to zero when its argument $z$ is greater than
the cutoff scale $k^2$. Because the FRGE only sees the variation of the EAA, it is free from the UV divergence and
well defined in contrast to EAA. Technically this comes from the fact that $\pa_t R_k$, which is inside the trace of FRGE,
falls off fast in the UV. This FRGE describes the flow of the theory when the energy scale is changed.
In this sense, it is also called flow equation.

We can fix initial point for the EAA at some UV scale $\Lambda$ and identify it as the ``bare'' action of the theory.
One can investigate how the EAA behaves in the limit $\Lambda\to\infty$ by integrating
the FRGE in the direction of larger $k$.
If the flow reaches a FP, we may expect that all physical observables to be well defined,
and the theory is UV-complete. Together with the assumption of finite dimensionality of the critical surface
on which the putative FP lies, this is what is called asymptotic safety. There has been considerable evidence
that the nontrivial FP exists in pure gravity with higher order terms and also gravity theory coupled to
matter~\cite{Reuter,DouP,RS2,PP,BS1,Litim,CP,CPR,BMS2009,Nied,NP,DonaP,DM,OP2013,DEP,FLNR2014,DSZ,OP2015,OPV2016,
GKLS,DPR,ELPRS,KRS}. 

Most of the work so far focused on the existence of the UV FP, leading to the scenario of asymptotically safe gravity.
There were also some works about the situation in the infrared (IR) regime~\cite{Christiansen:2012rx,Wetterich:2018poo,
Jizba:2019oaf} pertaining to solve the problems with infrared divergences. We also note that many of
the interesting results amenable to observations happen at finite energy for which the problems of UV-completion
are irrelevant provided some form of IR-UV decoupling holds (e.g. using Appelquist-Carazzone theorem for massive modes).

It is often considered that the EAA, which is obtained by integrating over the modes with scale larger
than $k$, describes physical quantities around the energy scale $k$.
However, there is a debate if the $k$-dependent EAA itself has physical meaning since
the cutoff $k$ is just a figurative energy scale used to cut off the path integral~\cite{Donoghue2019}.
How do we then study physical aspects of the asymptotically UV-safe gravity?
We note that it is the effective average action at $k=0$, which is nothing but the quantum effective
action as obtained in the standard quantum field theory, that contains information on the quantum effects on the physical
quantities such as scattering amplitudes and transition probabilities at all energy scales.
% There is no doubt in that effective action
%is physical, fully describes quantum physics and it is valid at all energy scales.

It is thus significant to compute the effective action. The FRGE can be used as a tool to calculate the effective action
nonperturbatively. For this purpose, we integrate the FRGE in the direction of small $k$ and the infrared endpoint of
the flow of EAA gives the quantum effective action $\Gamma_0$.
We should mention that this method of obtaining quantum effective action is not restricted only to
asymptotic safe theories, but is valid for any quantum field theories.
The advantage of this method, compared to calculating directly a functional integral, is that we never encounter divergences
since the FRGE does not suffer from divergences and is well defined.
Some examples have been discussed in \cite{SCM:2010,CPRT2015}, in which gravity theory coupled to a scalar is considered
among others. Here we would like to extend this calculation to the more  general case of gravity coupled to
gauge, Dirac and charged scalar fields.
Integrating the flow, we can derive the nonlocal, finite parts of the effective action up to two powers of
the gauge field strengths and curvatures of the metric.
The usual method to calculate the rhs of the FRGE is to use the asymptotic expansion for the trace of the heat kernel
coefficients, which gives only local expressions. To calculate nonlocal, finite parts of the effective action, we need
a more sophisticated one which includes an infinite number of heat kernel coefficients.
This expansion has been developed in \cite{BV,Avramidi}.
An alternative derivation has been given in \cite{CZ2012}.
We use this nonlocal heat kernel expansion to calculate not only the local but also nonlocal parts of the effective action
and to check them against standard computation using local heat kernel methods.
We will see that these nonlocal terms have unique coefficients, but local terms are subject to renormalization
and have arbitrary coefficients depending on the renormalization conditions.
These unique nonlocal terms can be used to describe physical phenomena and obtain scattering amplitudes.
They constitute a piece of generically genuine and universal information that we can extract
in the effective field theory of gravitational interactions at low energies.

Nonlocal effective actions and their applications have been discussed in several works in different approaches.
The nonlocal form of the anomaly-induced gravitational action was first indicated in \cite{DDI}.
The effective action was calculated perturbatively in the weak field limit
using the covariant nonlocal expansion of the heat kernel in \cite{BV3,BV4}.
Nonlocal action from long distance fluctuations of massless particles in QED
and generation of magnetic fields during inflation were also considered in~\cite{El-Menoufi}.
In \cite{DE:2015}, it was used to derive trace anomaly in QED and violation of equivalence principle.
Refs.~\cite{BBD2017,DK2018} derived it using the Weyl anomaly, and discussed the generation of cosmological
magnetic fields in the universe. Nonlocal effective gravitational actions have very interesting cosmological
phenomenology such as explaining problems of dark matter~\cite{Barvinsky:2011hd},
dark energy~\cite{Barvinsky:2011hd,Maggiore:2014sia} and inflation \cite{Koshelev:2016xqb}.
It has quite unusual classical features~\cite{Woodard:2018gfj,Li:2015bqa}.
A comprehensive review of its quantum properties can be found in \cite{Modesto:2017sdr}.

The low-energy effective action has been studied in a number of papers.
A nice review of this approach is given in \cite{Burgess}. In this approach, all independent terms compatible with
the symmetries of the theory are written down in the order of increasing dimension~\cite{RSW}. It is important to
sort out which interactions are important in the low energy. The coefficients of these terms may be calculated
by perturbations on flat background. On the other hand, our effective action is equivalent to that obtained by integrating
out the quantum fluctuations in the path integral on general backgrounds and
is not just valid for low energy but for all energy scale.
The one-loop effective action for pure gravity and gravity coupled to a scalar field is directly calculated
using local and nonlocal heat kernel expansion~\cite{CJ}. Attempt to give the approximate form of the effective action
as the sum of Feynman diagrams is made in \cite{DeWitt}.

In this paper, we compute the effective action for the gravitational system coupled to some matter.
In the gravitational sector, we restrict to a theory described by the Einstein action (Ricci scalar only) for simplicity,
while a gravitational theory with higher derivatives is one of our future goals of an extension of this project.
In the matter sector, we consider the action consisting of one abelian massless $U(1)$ gauge field, one charged massive
scalar field and a general number $N_F$ of massive Dirac fermions coupled both to electromagnetism as well as to gravity.
This is the simplest but quite general theory allowing us to see how the final result is affected by various fields.
The reason for inclusion of only one abelian gauge field is because this is the only vector field that survives in
the IR macroscopic limit (classically observable). On the other side, we consider an arbitrary number of fermions
since there are plenty of them in Standard Model of particle physics and extensions thereof.
Further generalization to include gauge fields in non-abelian gauge group, arbitrary numbers of fermions
and scalars in various representations, and Yukawa interactions is a future goal.

This paper is organized as follows. In the next section, we first summarize the methodology of nonlocal heat kernel
expansion and explain how to incorporate it in our calculation of the FRGE.
The result is given in the form of a master formula.
This formula requires the Hessian of the action for quantum fields (this is the operator of the second variational
derivative of the action with respect to fluctuating fields, which we denoted by $\frac{\d^2 \G_k}{\d\phi\d\phi}$
in (\ref{eq1})).
In section \ref{sec3}, we calculate the Hessian in a mutually coupled system of Einstein gravity, electromagnetic (EM)
gauge field, Dirac and charged scalar fields.
In section \ref{sec4}, we plug these results into the master formula to derive the FRGE, and
integrate it (over $k$) from the UV scale $k=\Lambda$ down to $k=0$ to obtain the nonlocal terms with definite coefficients
in the effective action $\Gamma_0$. Section \ref{sec5} is devoted to discussions.
Some technical details are relegated to three appendices.

%%%%%%%%%%%%%%%%%%%%%%%%%%%%%%%%%%%%%%%%%%%%%%%%%%%
\section{Nonlocal heat kernel expansion}
%%%%%%%%%%%%%%%%%%%%%%%%%%%%%%%%%%%%%%%%%%%%%%%%%%%
\label{sec2}

Suppose that our Laplacian (or Hessian) takes the form
\bea
\bm \Delta = -\nabla^2 {\bf 1} + {\bf U},
\label{deltaform}
\eea
where $\nabla$ is a covariant derivative with respect to all background fields and
{\bf U} is a non-derivative term. The main computation will be performed in a spacetime with Euclidean signature.
We will need the trace of a following expression
\bea
h_k(\bm\Delta,\omega) = \frac{\pa_t R_k(\bm\Delta)}{\bm\Delta +\omega + R_k(\bm\Delta)}.
\label{s}
\eea
In the above formula, $\omega$ plays the role of the mass of the mode and $R_k(z)$ is the cutoff function.
The FRGE then has the form
\bea
\pa_t \G_k=\frac12\Tr\, h_k(\bm\Delta,\omega)\,,
\label{frge2}
\eea
where the functional trace (denoted by ${\rm Tr}$) is both with respect to internal and external indices,
so the spacetime dependence requires here also to take the volume integral.
Introducing the Laplace transform
\bea
h_k(\bm\Delta,\omega) =\int_0^\infty ds\, \tilde h_k(s,\omega) e^{-s\bm\Delta},
\eea
\p{frge2} reduces to
\bea
\pa_t\G_k = \frac12 \int_0^\infty ds\, \tilde h_k(s,\omega)\, \Tr\, e^{-s\bm\Delta}.
\label{frge3}
\eea

The commutator of the covariant derivatives $\nabla$ gives the curvature tensor $\bm{\Omega}_{\mu\nu}$
(depending on a specific representation on which these derivatives act):
\bea
[\nabla_\mu,\nabla_\nu]=\bm{\Omega}_{\mu\nu}\,.
\eea
We note that the tensor $\bm{\Omega}_{\mu\nu}$ is antisymmetric and may have indices in particular matrix representation
of internal indices. From standard (metric) Riemann tensor $R_{\mu\nu\rho\sigma}$ using raising of indices with
the inverse metric tensor $g^{\mu\nu}$, we construct other curvature tensors, such as Ricci tensor $R_{\mu\nu}$
and Ricci scalar (curvature scalar) $R$. Together with ${\bf U}$ and $\bm{\Omega}_{\mu\nu}$,
these last three tensors constitute a set of generalized curvatures.

The nonlocal heat kernel expansion up to the quadratic order in generalized curvatures
${\cal R}=({\bf U},\bm{\Omega}_{\mu\nu},R_{\mu\nu\rho\sigma},R_{\mu\nu},R)$ is given by~\cite{CZ2012}
\bea
\Tr (e^{-s\bm\Delta}) \hs{-2}&=&\hs{-2} \frac{1}{(4\pi s)^{d/2}} \int d^d x \sqrt{g}\;
\tr \Bigg\{ {\bf 1} + s \Big({\bf 1}\frac{R}{6}-{\bf U} \Big)
+s^2 \Big[ {\bf 1} R_{\mu\nu} f_{Ric}(-s\nabla^2) R^{\mu\nu} \nn
&& \hs{-5} + {\bf 1} R f_{R}(-s\nabla^2) R + R f_{RU}(-s\nabla^2) {\bf U}
+ {\bf U} f_{U}(-s\nabla^2) {\bf U} +\bm{\Omega}_{\mu\nu}f_\Omega (-s\nabla^2)\bm{\Omega}^{\mu\nu} \Big] \Bigg\},~~~~
\label{nonlocal}
\eea
where the structure functions $f$'s are given by
\bea
f_{Ric}(x) \hs{-2}&=&\hs{-2} \frac{1}{6x}+\frac{1}{x^2}[f(x)-1] \sim \frac{1}{60} - \frac{x}{840} + \frac{x^2}{15120}
+\ldots, \nn
f_{R}(x) \hs{-2}&=&\hs{-2} \frac{1}{32}f(x)+\frac{1}{8x}f(x)-\frac{7}{48x}-\frac{1}{8x^2}[f(x)-1]
\sim \frac{1}{120} - \frac{x}{336} + \frac{11}{30240} x^2 +\ldots, \nn
f_{RU}(x) \hs{-2}&=&\hs{-2} -\frac{1}{4}f(x)-\frac{1}{2x}[f(x)-1]
\sim -\frac{1}{6} + \frac{x}{30} - \frac{x^2}{280} + \ldots, \nn
f_{U}(x) \hs{-2}&=&\hs{-2} \frac{1}{2}f(x)
\sim \frac{1}{2} - \frac{x}{12} + \frac{1}{120}x^2 +\ldots, \nn
f_\Omega (x) \hs{-2}&=&\hs{-2} -\frac{1}{2x}[f(x)-1]
\sim \frac{1}{12} - \frac{x}{120} + \frac{x^2}{1680} + \ldots.
\label{nlff}
\eea
Here the basic structure function (form-factor) $f(x)$ reads
\bea
f(x)= \int_0^1 d\xi e^{-x\xi(1-\xi)} \sim 1 - \frac{x}{6} + \frac{1}{60}x^2 - \frac{1}{840}x^3 +\ldots,
\label{basicff}
\eea
where we have also displayed the Maclaurin expansions for small $x$.
The first constant terms in~Eqs.~(\ref{nlff})-(\ref{basicff}) represent the local heat kernel expansion coefficients.
The full nonlocal heat kernel has the infinite number of heat kernel coefficients, which can be collected
in the form of nonlocal form-factors $f(x)$, $f_{Ric}(x)$, $f_R(x)$, $f_{RU}(x)$, $f_U(x)$, $f_\Omega(x)$.
Substituting \p{nonlocal} into Eq.~\p{frge3}, we obtain
\bea
\pa_t\G_k \hs{-2}&=&\hs{-2} \frac12 \frac{1}{(4\pi)^{d/2}} \int d^d x \sqrt{g}\, \tr \Bigg\{
{\bf 1} \left[ \int_0^\infty ds \tilde h_k(s,\omega) s^{-d/2} \right]
+\left({\bf 1}\frac{R}{6}-{\bf U}\right)\left[ \int_0^\infty ds \tilde h_k(s,\omega) s^{-d/2+1} \right] \nn
&&\hs{-5}
 +\ {\bf 1} R_{\mu\nu} \left[ \int_0^\infty ds \tilde h_k(s,\omega) s^{-d/2+2} f_{Ric}(sz)\right]R^{\mu\nu}
+ {\bf 1} R \left[ \int_0^\infty ds \tilde h_k(s,\omega) s^{-d/2+2} f_{R} (sz) \right]R \nn
&& \hs{-5} +\, R \left[ \int_0^\infty ds \tilde h_k(s,\omega) s^{-d/2+2} f_{RU} (sz) \right] {\bf U}
+ {\bf U} \left[ \int_0^\infty ds \tilde h_k(s,\omega) s^{-d/2+2} f_{U} (sz) \right] {\bf U} \nn
&& \hs{-5} +\, \bm{\Omega}_{\mu\nu} \left[ \int_0^\infty ds \tilde h_k(s,\omega) s^{-d/2+2} f_\Omega (sz) \right]
\bm{\Omega}^{\mu\nu} +O({\cal R}^3) \Bigg\} \\
\hs{-2}&=&\hs{-2} \frac12 \frac{1}{(4\pi)^{d/2}} \int d^d x \sqrt{g}\, \tr \Big\{
{\bf 1} Q_{\frac{d}{2}}[h_k] + \Big({\bf 1}\frac{R}{6}-{\bf U}\Big) Q_{\frac{d}{2}-1}[h_k]
+{\bf 1} R_{\mu\nu}\, g_{Ric}\, R^{\mu\nu}+{\bf 1}R\, g_{R} R  \nn
&& +\, R\, g_{_{RU}} {\bf U}+{\bf U} g_{_{U}} {\bf U}
+\bm{\Omega}_{\mu\nu} g_\Omega \bm{\Omega}^{\mu\nu} + \ldots \Big\},
\label{master}
\eea
which is our master formula to derive the FRGE. In the above, we have used ``small'' traces ${\rm tr}(\ldots)$
to denote traces only in the internal space of indices.
Here and in what follows, $z=-\nabla^2$ and we have defined
\bea
&& g_A = g_A(z,\omega,k)=\int_0^\infty ds\, \tilde h_k (s,\omega)f_A(sz) s^{-d/2+2},~~(A=Ric, R, RU, U,\Omega), \\
&& Q_n[h_k] = \frac{1}{\G(n)}\int_0^\infty dx\, x^{n-1}h_k(x).
\eea
More explicitly, we have
\bea
g_{Ric} \hs{-2}&=&\hs{-2} \int_0^\infty ds \tilde h(s,\omega) f_{Ric}(sz) s^{-d/2+2} \nn
&=&\hs{-2} \frac{1}{6z} Q_{\frac{d}{2}-1}[h_k] -\frac{1}{z^2} Q_{\frac{d}{2}}[h_k]
+\frac{1}{z^2} \int_0^1 d\xi Q_{\frac{d}{2}}[h_k^{z\xi(1-\xi)}], \nn
g_{R} \hs{-2}&=&\hs{-2} -\frac{7}{48z} Q_{\frac{d}{2}-1}[h_k] +\frac{1}{8z^2} Q_{\frac{d}{2}}[h_k]
+\frac{1}{2} \int_0^1 d\xi Q_{\frac{d}{2}-2}[h_k^{z\xi(1-\xi)}] \nn
&& +\frac{1}{8z} \int_0^1 d\xi Q_{\frac{d}{2}-1}[h_k^{z\xi(1-\xi)}]
-\frac{1}{8z^2} \int_0^1 d\xi Q_{\frac{d}{2}}[h_k^{z\xi(1-\xi)}], \nn
g_{RU} \hs{-2}&=&\hs{-2} \frac{1}{2z} Q_{\frac{d}{2}-1}[h_k]
-\frac{1}{4} \int_0^1 d\xi Q_{\frac{d}{2}-2}[h_k^{z\xi(1-\xi)}]
-\frac{1}{2z} \int_0^1 d\xi Q_{\frac{d}{2}-1}[h_k^{z\xi(1-\xi)}],\nn
g_{U} \hs{-2}&=&\hs{-2} \frac{1}{2} \int_0^1 d\xi Q_{\frac{d}{2}-2}[h_k^{z\xi(1-\xi)}], \nn
g_\Omega \hs{-2}&=&\hs{-2} \frac{1}{2z} Q_{\frac{d}{2}-1}[h_k]
-\frac{1}{2z} \int_0^1 d\xi Q_{\frac{d}{2}-1}[h_k^{z\xi(1-\xi)}],
\label{g's}
\eea
where we have defined
\bea
Q_{n}[h_k^{z\xi(1-\xi)}] = \int_0^\infty ds \tilde h(s,\omega)s^{-n} e^{-sz\xi(1-\xi)}.
\eea
To evaluate this, we first recall that
\bea
h_k(x,\omega)=\int_0^\infty ds \tilde h(s,\omega) e^{-sx}.
\eea
From this, we find
\bea
\int_0^\infty dx x^{n-1} h_k(x+z \xi(1-\xi),\omega)
& =& \int_0^\infty dx x^{n-1} \int_0^\infty ds \tilde h(s,\omega) e^{-sx-sz\xi(1-\xi)} \nn
&=& \G(n) Q_{n}[h_k^{z\xi(1-\xi)}] .
\label{qn}
\eea
Using the definition~\p{s} and the optimized cutoff~\cite{Litim2}
\bea
R_k(z)=(k^2-z) \t(k^2-z),
\label{litimcf}
\eea
in the first expression of \p{qn}, we find
\bea
Q_n[h_k] &=& \frac{2 k^{2n}}{\G(n+1)}\frac{1}{1+\tilde\omega}, \\
Q_n[h_k^{z\xi(1-\xi)}] &=& \frac{1}{\G(n+1)} \frac{2}{1+\tilde\omega} [k^2-z\xi(1-\xi)]^n \t[k^2-z\xi(1-\xi)],
\eea
where $\tilde\omega=\omega/k^2$.
We then obtain
\bea
\int_0^1 d\xi Q_0[h_k^{z\xi(1-\xi)}] &=& \frac{2}{1+\tilde\omega} \left[1-\sqrt{1-\frac{4}\tz}
\t(\tilde z-4)\right], \nn
\int_0^1 d\xi Q_1[h_k^{z\xi(1-\xi)}] &=& \frac{2k^2}{1+\tilde\omega} \left[1-\frac\tz{6}
+\frac{\tz}{6}\sqrt{1-\frac{4}{\tz}} \t(\tz-4)\right], \nn
\int_0^1 d\xi Q_2[h_k^{z\xi(1-\xi)}] &=& \frac{k^4}{1+\tilde\omega} \left[1-\frac{\tz}{3}
+\frac{\tz^2}{30}-\frac{\tz^2}{30}\left(1-\frac{4}{\tz}\right)^{\frac{5}{2}} \t(\tz-4) \right],
\eea
with $\tz=z/k^2$.
Substituting these into \p{g's}, we obtain
\bea
g_{Ric} \hs{-2}&=&\hs{-2} \frac{1}{30} \frac{1}{1+\tilde\omega}
\left[ 1-\left(1-\frac{4k^2}{z}\right)^{\frac52} \t(z-4k^2)\right], \nn
g_{R} \hs{-2}&=&\hs{-2}  \frac{1}{1+\tilde\omega}
\left[ \frac{1}{60}-\frac{1}{16}\left(1-\frac{4k^2}{z}\right)^{\frac12}\t(z-4k^2)
+\frac{1}{24}\left(1-\frac{4k^2}{z}\right)^{\frac32}\t(z-4k^2) \right. \nn
&& \hs{20}\left. +\, \frac{1}{240}\left(1-\frac{4k^2}{z}\right)^{\frac52}\t(z-4k^2)\right], \nn
g_{RU} \hs{-2}&=&\hs{-2}  \frac{1}{1+\tilde\omega}
\left[ -\frac13+\frac12\left(1-\frac{4k^2}{z}\right)^{\frac12}\t(z-4k^2)
-\frac16\left(1-\frac{4k^2}{z}\right)^{\frac32}\t(z-4k^2)\right], \nn
g_{U} \hs{-2}&=&\hs{-2}  \frac{1}{1+\tilde\omega}
\left[ 1-\left(1-\frac{4k^2}{z}\right)^{\frac12}\t(z-4k^2)\right], \nn
g_\Omega \hs{-2}&=&\hs{-2}  \frac{1}{6(1+\tilde\omega)}
\left[ 1-\left(1-\frac{4k^2}{z}\right)^{\frac32}\t(z-4k^2)\right].
\label{gs}
\eea
We will use these results to derive the flow equation.

We note that the intermediate results of the RG flow depend slightly on the choice of the cutoff kernel function $R_k(z)$.
However, as discussed in \cite{Percacci:2017, RS, CPR, Nagy:2017zvc}, the evidences for existence of FPs
(realizing the idea of asymptotic safety in the UV) are independent of this choice and the scheme of regularization
and renormalization. The same is also true in the $k\to 0$ limit of the EAA, while the interpolating effective action
$\Gamma_k$ at finite $k>0$ shows a bit of dependence. On the other hand, the nonlocal terms in the effective action
$\Gamma_0$ are universal and unambiguous and they do not rely on such details as the choice of the IR cutoff function.
Here we chose the optimized cutoff function in (\ref{litimcf}) in order to have an analytic control over all functions
and integrals involved in the process of taking functional traces in (\ref{frge2}) and to have compact form
for all intermediate expressions of the RG flow also in the case of finite $k$.

%%%%%%%%%%%%%%%%%%%%%%%%%%%%%%%%%%%%%%%%%%%%%%%%%%%%%%%%%%%%%%%%%%%%%%%%%%%%%%%%%%%%%
\section{Hessian for gravity coupled to EM, Dirac and charged scalar fields}
%%%%%%%%%%%%%%%%%%%%%%%%%%%%%%%%%%%%%%%%%%%%%%%%%%%%%%%%%%%%%%%%%%%%%%%%%%%%%%%%%%%%%
\label{sec3}

We consider the total action $S_T$ of the system consisting of gravity coupled to EM (gauge), Dirac and charged scalar
fields:
\bea
S_T %[h,\bar C,C;g]
= S_H[g] %+ S_{gf}[h,\bg]+S_{gh}[h,\bar C,C,\bg]
+S_V[g,V_\mu]+S_f[g,\psi] +S_S[g,\phi],
\eea
where $S_H, S_V, S_f$ and $S_S$ are the actions for the (Einstein) gravity, EM field $V_\mu$,
Dirac fermion $\psi$ and a charged scalar field $\phi$. Their explicit forms are given below.

To calculate the flow equation, we have to derive the Hessian in each sector, to which we now turn.

\subsection{Hessian for graviton}
\label{sec3.1}

Let us first consider the Einstein action (Ricci scalar of the metric $g$) together with the gauge fixing:
\bea
S_H[g] = -\frac{1}{\kappa^2}\int d^d x\sqrt{g}\, R(g)
+ \frac{1}{2} \int d^d x\sqrt{g}f_\mu f^\mu,
\label{actionsh}
\eea
with $g=\det(g_{\mu\nu})$ and where $f_\mu$ is the gauge-fixing function
\bea
f_\mu=\bnabla^\nu h_{\mu\nu}-\frac{1}{2} \bnabla_\mu h.
\label{gfdif}
\eea

The quantum fluctuations are defined by
\bea
g_{\mu\nu} &=& \bg_{\mu\nu} + h_{\mu\nu}\quad{\rm and}\quad h\,\,\,=\,\,\,\bg^{\mu\nu}h_{\mu\nu}\,.
\label{qf}
\eea
For simplicity, we set $\kappa=1$ and consider $d=4$ spacetime dimensions in what follows.
Background quantities are denoted by bars over them.
Separating the background and quantum fields~\p{qf}, we find the Hessian ${\bf H}_G$ in the following expression
from the expansion of the action (\ref{actionsh}) to the second order in fluctuations $h_{\mu\nu}$:
\bea
\frac{1}{2} \int d^4 x \sqrt{\bg}\,h^{\mu\nu} H_{G, \mu\nu\a\b} h^{\a\b},
\label{gravitonh}
\eea
where
\bea
{\bf H_G} = {\bf K}(-\bnabla^2) + {\bf U},
\eea
and
where
\bea
K_{\mu\nu\a\b}&=&\frac{1}{4}(\bar g_{\mu\a} \bar g_{\nu\b}+\bar g_{\mu\b}\bar g_{\nu\a}-\bar g_{\mu\nu}\bar g_{\a\b}),
\\
U_{\mu\nu\a\b}&=& \br K_{\mu\nu\a\b} - \bg_{(\mu(\a}\br_{\b)\nu)} -\br_\mu{}_{(\a}{}_\nu{}_{\b)}
+\frac{1}{2}(\bg_{\mu\nu}\br_{\a\b}+\bg_{\a\b}\br_{\mu\nu}).~~
\label{gravitonh2}
\eea
Here the bracket on indices means symmetrization within the pairs of indices $(\mu,\nu)$ and $(\a,\b)$ in the second
term and within the pair $(\a,\b)$ in the third term in (\ref{gravitonh2}) (with the factor $\frac12$ included).

The diffeomorphism ghost action corresponding to the gauge fixing in (\ref{gfdif}) is
\bea
S_{gh}[h,\bar C,C,\bg] = \int d^d x\sqrt{\bg}\bar C_\mu(-\bar\nabla^2\d_\nu^\mu -\bar R_\nu^\mu ) C^\nu.
\label{gravityghost}
\eea
Hence we can calculate the ghost contribution to the flow equation with the vector-like Hessian 
$(\Delta_{gh})_{\mu\nu}=-\bar g_{\mu\nu}\bnabla^2-\bar R_{\mu\nu}$.

\subsection{Gauge fields}
\label{sec3.2}

The action for the gauge field is
\bea
S_V[V_\mu] = \int d^4 x \sqrt{g}\frac{1}{4} g^{\mu\a} g^{\nu\b} F_{\mu\nu} F_{\a\b},
\label{actionv}
\eea
where the $U(1)$ (abelian) gauge field strength is given by
\bea
F_{\mu\nu} =  \bnabla_\mu V_\nu -\bnabla_\nu V_\mu\,.
\eea
We note that the Christoffel symbols drop out from the antisymmetrized field strength,
but it is convenient to recover the classical part when the derivatives acts on the gauge fields.
The quantum fluctuations are defined as
\bea
V_\mu = \ba_\mu +A_\mu,
\eea
and then the resulting perturbation of the field strength is
\bea
F_{\mu\nu} = \bF_{\mu\nu} + \bnabla_\mu A_\nu -\bnabla_\nu A_\mu\,.
\label{fs}
\eea

The part quadratic in the fluctuations $(h_{\mu\nu},A_\mu)$ of the action (\ref{actionv}) under the volume integral
is given by
\bea
&& \frac{1}{4}\sqrt{\bg}\Big[ h^{\mu\nu}\Big\{  \bg_{\mu\a} \bF_{\rho\nu}\bF^\rho{}_\b + \bF_{\mu\a}\bF_{\nu\b}
+ \bg_{\nu\b} \bF_{\mu\rho}\bF_\a{}^\rho -\frac{1}{2} \bg_{\mu\nu} \bF_{\rho\a}\bF^\rho{}_\b
-\frac{1}{2}\bg_{\a\b} \bF_{\mu\rho}\bF_\nu{}^\rho \nn
&& +\frac{1}{8} \bF_{\rho\la}^2 (\bg_{\mu\nu}\bg_{\a\b} -\bg_{\mu\a}\bg_{\nu\b}-\bg_{\mu\b}\bg_{\nu\a})
 \Big\} h^{\a\b}
+ 2 h^{\mu\nu} \left\{ 2 \bF_{(\mu}{}^\rho \bg_{\nu)\a} \bnabla_{\rho} +2 \bF_{\a(\mu}\bnabla_{\nu)}
 + \bg_{\mu\nu}\bF_{\rho\a}\bnabla^{\rho} \right\} A^\a \nn
&& +2A^\mu \left\{ -\bg_{\mu\nu} \bnabla^2 +\bnabla_\mu \bnabla_\nu +\br_{\mu\nu} \right\} A^\nu \Big].
\label{g1}
\eea

We introduce the gauge fixing term
\bea
\frac{1}{2\a} (\bnabla_\mu A^\mu)^2,
\eea
which contributes to the Hessian
\bea
- \frac{1}{2\a} A^\mu \bnabla_\mu\bnabla_\nu A^\nu.
\label{g2}
\eea
In what follows, we set $\a=1$.
Collecting \p{g1} and \p{g2}, we find that the gauge and gravitational contributions to the quadratic part of
the action (\ref{actionv}) is
\bea
\frac{1}{2} \int d^4 x\sqrt{\bg} (h^{\mu\nu},A^\mu) {\bf H}_V
\left(\begin{array}{c}
h^{\a\b} \\
A^\a
\end{array}
\right),
\label{g0}
\eea
where
\bea
{\bf H}_V = {\bf K}_V (-\bnabla^2) + 2 {\bf V}_V^\d \bnabla_\d +{\bf U}_V,
\label{g3}
\eea
with
\bea
{\bf K}_V &=& \left(\begin{array}{cc}
0 & 0 \\
0 & \bg_{\mu\a}
\end{array}
\right), \label{kv}\\
\label{g4}
{\bf V}_V^\d &=& \left(\begin{array}{cc}
0 & K_{\mu\nu\la\a} \bF^{\la\d}-\frac{1}{2} \d_{\mu\nu}{}^{\la\d}\bF_{\la\a} \\
- K_{\a\b\la\mu} \bF^{\la\d}+\frac{1}{2} \d_{\a\b}{}^{\la\d}\bF_{\la\mu} & 0
\end{array}
\right), \label{vv}\\
{\bf U}_V &=& \left(\begin{array}{cc}
2 K_{\mu\nu\a\la} \bF^{\rho\la} \bF_{\rho\b} + \frac{1}{2} \bF_{\mu\a} \bF_{\nu\b}
-\frac{1}{4} K_{\mu\nu\a\b}\bF_{\rho\la}^2  &0 \\
- 2 K_{\a\b\la\mu} (\bnabla_\rho \bF^{\la\rho})+ \d_{\a\b}{}^{\la\rho} (\bnabla_\rho \bF_{\la\mu}) & \br_{\mu\a}
\end{array}
\right),
\label{g5}
\eea
where
\bea
\d_{\mu\nu}{}^{\c\d} = \frac{1}{2}(\d_\mu^\c \d_\nu^\d + \d_\mu^\d \d_\nu^\c),
\eea
is the unit matrix on the symmetric tensor with two indices.
In various components of ${\bf H}_V$ in Eqs. (\ref{kv})-(\ref{g5}), symmetrization $\mu \lra \nu$, $\a \lra \b$
and $(\mu,\nu) \lra (\a,\b)$ should be understood. Note that ${\bf U}_V$ is not symmetric (as a matrix).
Naively one would expect that $\bf U$ should be symmetric, but it is then not self-adjoint.
Here we have chosen such that the Hessian is self-adjoint (as the operator acting between fluctuations and with
a proper Hermitian complex scalar product in the field space).
See appendix~\ref{appa} for more details.

The gauge ghost contribution to the Hessian consistent with the gauge fixing~\p{g2} is given by
$\Delta_{gh}^G=-\bnabla^2$. This is a scalar operator. 

%We find that its contributions to traces are given by
%\bea
%\tr[{\bm b}_0(\Delta_{gh}^G)] \hs{-2}&=&\hs{-2} \tr({\bm 1}) =1, \\
%\tr[{\bm b}_2(\Delta_{gh}^G)] \hs{-2}&=&\hs{-2} \tr\left({\bm 1}\frac{R}{6} \right)
%=\frac{R}{6}, \\
%\tr[{\bm b}_4(\Delta_{gh}^G)] \hs{-2}&=&\hs{-2} \frac{1}{60} R_{\mu\nu}^2 + \frac{1}{120} R^2.
%\label{vghost}
%\eea

\subsection{Dirac fields}
\label{sec3.3}

The action we consider for Dirac fermions is
\bea
S_f = \int d^4 x \sqrt{g}\, \frac{1}{2} \left(\bar\psi \c^\mu D_\mu \psi- D_\mu \bar\psi \c^\mu \psi
+2m_F \bar \psi \psi\right),
\label{diraca}
\eea
where the covariant derivatives acting on spinors are
\bea
D_\mu \psi&=& (\pa_\mu - i e_F A_\mu)\psi +\frac{1}{2} \omega_{\mu ab}J^{ab} \psi,\\
D_\mu \bar\psi&=& (\pa_\mu + i e_F A_\mu)\bar\psi -\frac{1}{2} \omega_{\mu ab}\bar \psi J^{ab},
\label{dop}
\eea
with $\omega_{\mu ab}$ being spin-connection coefficients and
$
J^{ab}= \frac{1}{4}[\c^a,\c^b]
$
the $O(4)$ generators. From the action (\ref{diraca}) and the definitions of covariant derivatives (\ref{dop}),
one sees that the massive Dirac fermion (with mass $m_F$) is coupled minimally both to the gravitational
and EM fields (via the fermionic electric charge $e_F$).
We consider $N_F$ fermions, but we can calculate the contribution for a single Dirac field
and multiply it by $N_F$ in the end.

We first show that we can express the fluctuation of the vierbein $e^a_\mu$ in terms of that of the metric
$g_{\mu\nu}$ in \p{qf}.
From the soldering relation $g_{\mu\nu}=e_\mu^a e_\nu^b \eta_{ab}$, we find that we can choose~\cite{ZZVP2009}
\bea
e_\mu^a = \bar e_\mu^a +\frac{1}{2} h_\mu^a - \frac{1}{8} h_{\mu\rho} h^{\rho a} + \ldots,
\eea
where we denote
\bea
h_\mu^a = h_\mu^\nu \bar e_\nu^a, \mbox{ etc.}
\eea
The inverse vierbein $e^\mu_a$ and its expansion is given by
\bea
e^\mu_a = \bar e^\mu_a -\frac{1}{2} h^\mu_a + \frac{3}{8} h^\mu_\nu h^\nu_a + \ldots.
\eea
By the tetrad postulate
\bea
\omega_\mu{}^a{}_b=e^a_\rho \G^\rho_{\mu\s}e^\s _b +e^a_\rho \pa_\mu e^\rho_b,
\label{tetradp}
\eea
we can calculate the expansion of the spin-connection:
\bea
\omega_\mu{}^{ab} = \bar \omega_\mu{}^{ab} + \omega_\mu{}^{ab\,(1)} + \omega_\mu{}^{ab\,(2)},
\eea
where
\bea
\omega_\mu{}^{ab\,(1)} &=& \frac{1}{2} \left(\bnabla^\b h^\a_\mu - \bnabla^\a h_\mu^\b\right)
\bar e_\a^a \bar e_\b^b , \\
\omega_\mu{}^{ab\, (2)} &=& \frac{1}{8} \left( 4 h^{\a\rho}\bnabla_\rho h_\mu^\b
- 4h^{\a\rho}\bnabla^\b h_{\mu\rho} - h^{\a\rho}\bnabla_\mu h^\b_\rho
+ h^{\b\rho}\bnabla_\mu h^\a_\rho \right) \bar e_\a^a \bar e_\b^b .
\eea
The Dirac field is decomposed as $\psi \Rightarrow \psi +\chi$. (Here exclusively by bars over the spinorial quantities,
we denote the Dirac conjugate of the spinors, not the background spinors). In this way, we can express the expansion
of the Dirac action (\ref{diraca}) to the quadratic order in fluctuations by only the variations of the covariant
metric tensor $h_{\mu\nu}$, gauge potential $A_\mu$ and the Dirac field $\chi$.
We thus see that while we need to use vierbeins and spin-connections
% (expressed through the former by the tetrad postulate (\ref{tetradp}) and the condition on Levi-Civita
% connection $\Gamma^\mu_{\nu\rho}$ of the metric $g_{\mu\nu}$)
for the explicit construction of the action (\ref{diraca}),
we do not have to use the variation of $e_\mu^a$ nor $\omega_\mu^{ab}$ in the expansion since all fluctuations there are
given in terms of the metric variations $h_{\mu\nu}$.

We then find that the Dirac action (\ref{diraca}) to the quadratic order in fluctuations ($h_{\mu\nu},A_\mu,\chi$)
is given under spacetime integral as
\bea
\frac{\sqrt{\bg}}{2} \hs{-2}&&\hs{-2} \Big[ \bar\chi \c^\mu \bd_\mu \chi -\bd_\mu \bar\chi \c^\mu \chi
+2m_F \bar\chi\chi - 2ie_F A_\mu(\bar\psi \c^\mu \chi+\bar \chi \c^\mu \psi)
+\frac{1}{8} h^{\a\rho}\bnabla^\b h^\mu_\rho \bar\psi \c_{\a\b\mu}\psi
\nn &&
+\frac{1}{2}(h \bg^{\mu\nu} - h^{\mu\nu}) ( \bar\psi \c_\mu \bd_\nu \chi + \bar\chi \c_\mu \bd_\nu \psi
- \bd_\nu \bar\psi \c_\mu  \chi - \bd_\nu \bar\chi \c_\mu  \psi
-2 ie_F A_\mu \bar\psi \c_\nu \psi )
\nn &&
+\frac14 h^{\mu\nu}\bnabla^\rho h_\mu^\b \bpsi \c_{\rho\b\nu}\psi + m_F h(\bar\psi\chi+\bar\chi\psi)
+\frac{1}{8} (3 h^\mu_\rho h_\nu^\rho -2 h h^\mu_\nu)(\bar\psi \c^\nu \bd_\mu \psi - \bd_\mu \bar\psi \c^\nu \psi)
\nn &&
+\frac{1}{8}(h^2-2 h_{\s\la}^2)\,
\left(\bar\psi \c^\mu \bd_\mu \psi - \bd_\mu \bar\psi \c^\mu \psi+2m_F\bar\psi\psi\right) \Big],
\label{hessiandirac}
\eea
where $\bd_\mu$ is the covariant derivative defined in \p{dop} and restricted to the gravitational background
$\bar g_{\mu\nu}$ and EM fields $\bar A_\mu$, and $\c_{\a\b\mu}$ is the antisymmetric product of Dirac gamma matrices.
When we deal with neutral objects (not charged under EM group), we use only spacetime covariant background
derivatives $\bnabla_\mu$.

In the Hessian (\ref{hessiandirac}), we have terms with fermion-fermion ($\bar\chi$-$\chi$), boson-boson ($h$-$h$)
and mixed ($\chi$-$h$ with background $\psi$) fluctuations.
Here we drop the mixed terms since they are intractable by the present methods of computation and moreover
lead to highly nonlocal terms in the effective action.
Let us then first discuss the fermion-fermion terms, and then come back to the boson-boson terms.
The integration over Dirac field $\chi$ in this sector yields the contribution to the effective action
\bea
\G^{Dirac} = - \Tr\log\left[ \bar\Dsl +m_F \right]\!.
\label{gdirac}
\eea

Noting that $\Tr\log\left[ \bar\Dsl +m_F \right]=\Tr\log\left[ \bar\Dsl -m_F \right]$, the formula (\ref{gdirac})
can be rewritten as
\bea
\G^{Dirac} = - \frac12 \Tr\log\left[ -\bar\Dsl^2 +m_F^2 \right].
\label{dirac}
\eea
Now consider
\bea
\bar\Dsl^2 \psi \hs{-2}
&=&\hs{-2} \gamma^a \gamma^b e^\mu_a e^\nu_b \bd_\mu \bd_\nu \psi = \left(\frac{1}{2}\{\c^a,\c^b\}
 + \frac{1}{2}[\c^a,\c^b] \right)
e^\mu_a e^\nu_b \bd_\mu \bd_\nu \psi \nn
&=&\hs{-2} \left( \bd_\mu^2+\frac{\c^{ab}}{2} e^\mu_a e^\nu_b [\bd_\mu,\bd_\nu] \right) \psi \nn
&=&\hs{-2} \left( \bd_\mu^2+\frac{\c^{ab}}{2} e^\mu_a e^\nu_b \left( -ie_F \bar F_{\mu\nu}
 +\frac{1}{4} \br_{\mu\nu}{}^{cd} \c_c \c_d\right) \right) \psi.
\eea
Using the identity
\bea
\c^\mu \c^\nu \c^\rho = \c^{\mu\nu\rho}+\bg^{\mu\nu}\c^\rho+\bg^{\nu\rho}\c^\mu-\bg^{\mu\rho}\c^\nu,
\eea
and Bianchi identity for the Riemann tensor, we find
\bea
\bar\Dsl^2 \psi \hs{-2}&=&\hs{-2} \left( \bd_\mu^2 -ie_F \frac{\c^{\mu\nu}}{2}\bar F_{\mu\nu}
 - \frac{1}{4} \br \right) \psi.
\eea
In this way, we have derived the so-called Lichnerowicz formula relating the square of the gauge-covariant
Dirac operator acting on spinors to standard Bochner Laplacian acting on spinors treated as spacetime scalars
shifted by various curvatures (in the internal and external spaces).

Hence the contribution~\p{dirac} from fermion-fermion sector to the effective action turns out to be
\bea
\G^{Dirac} = - \frac12 \Tr\log\left[ -\bd_\mu^2 +\frac{ie_F}{2}\c^{\mu\nu}\bar F_{\mu\nu} +\frac14 \br +m_F^2 \right].
\eea
We define
\bea
\bm\Delta &\equiv& -\bd_\mu^2 +\frac{ie_F}{2}\c^{\mu\nu} \bar F_{\mu\nu} +\frac14 \br +m_F^2,\\
{\bf U}&=& \frac{ie_F}{2}\c^{\mu\nu}\bar F_{\mu\nu}+ \frac{1}{4}\bar R +m_F^2,\label{deltaU}
\eea
and use the regulator (\ref{litimcf}):
\bea
R_k(\bm\Delta) = (-\bm\Delta+k^2) \t(k^2-\bm\Delta).
\eea
Then we have the contribution from Dirac fields to the FRG flow of the effective action
\bea
\pa_t \G^{Dirac} = - \frac12 \Tr \frac{\pa_t R_k(\bm\Delta)}{\bm\Delta+R_k(\bm\Delta)}.
\label{diracfrge}
\eea
We can now use the technique of nonlocal heat kernel as described before.
Noting that
$\bm\Omega_{\mu\nu} = [\bd_\mu,\bd_\nu] = -ie_F \bar F_{\mu\nu} +\frac{1}{4} \br_{\mu\nu}{}^{cd} \c_c \c_d$,
\p{diracfrge} is cast into
\bea
\hspace{-0.3cm}
\pa_t \G^{Dirac} &=& -\frac12 \int_0^\infty \frac{ds}{(4\pi s)^{2}} \tilde h_k(s)
\int d^4 x\sqrt{g} \Big[4-\left(\frac13\, \br +4 m_F^2 \right) s
+ s^2 \Big\{ 4\br_{\mu\nu}\, f_{Ric}(sz) \br^{\mu\nu} \nn
&&  + 4\br\, f_R(sz) \br +\br f_{RU}(sz)\tr({\bf U}) +\tr({\bf U}f_U(sz) {\bf U})
+\tr(\bm\Omega_{\mu\nu}f_\Omega(sz)\bm\Omega^{\mu\nu})
 \Big\} \Big],~~~~~
\label{diracfrge2}
\eea
with $z=-\bd_\mu^2$. 
Internal traces that we will need are
\begin{equation}
\tr (\mathbf{U}) = \tr{\bf 1}\left(\frac{1}{4}\bar R +m_F^2\right)=\bar R +4m_F^2,
\end{equation}
\begin{equation}
\tr (\mathbf{U}^2) = \tr\!\left(\frac{1}{4}\bar R+m_F^{2}+\frac{1}{2}ie_F\gamma^{\mu\nu} \bar F_{\mu\nu}\right)^{2}
=\frac{1}{4}\bar R^{2}+4m_F^{4}+2\bar Rm_F^{2}+2e_F^{2}\bar F_{\mu\nu}^{2},
\end{equation}
\begin{equation}
\tr\! \left(\bm\Omega_{\mu\nu}\bm\Omega^{\mu\nu}\right)
=\tr\!\left[\left(\frac{1}{4}\bar R_{\mu\nu\rho\sigma}\gamma^{\rho\sigma}
-ie_F \bar F_{\mu\nu}\right)\left(\frac{1}{4}\bar R_{\mu\nu\kappa\lambda}\gamma^{\kappa\lambda}
-ie_F \bar F_{\mu\nu}\right)\right]
=-\frac{1}{2}\bar R_{\mu\nu\rho\sigma}^{2}-4e_F^{2}\bar F_{\mu\nu}^{2}.
\end{equation}

We then find for (\ref{diracfrge2})
\bea
\pa_t \G^{Dirac} &=& -\frac12 \int \frac{d^4 x}{(4\pi)^{2}} \sqrt{g}
\Big[ 4 Q_{2}[h_{k}] - \left( \frac13 \br +4m_F^2 \right) Q_{1}[h_{k}]
+2m_F^2(2g_{RU}+g_{U}) \br
\nn && \hs{-5}
+ 4 \br_{\mu\nu}\, g_{Ric}\, \br^{\mu\nu}
+4m_F^{2}g_{U}m_F^{2}+ \br\left(4g_{R}+g_{RU}+\frac{1}{4}g_{U}\right) \br
- \frac12 \br_{\mu\nu\rho\la}\, g_\Omega \br^{\mu\nu\rho\la}
\nn && \hs{-5}
+ 2 e_F^2 \bar F_{\mu\nu} (g_{U}-2g_\Omega) \bar F^{\mu\nu} \Big],~~~
\label{dcont}
\eea
where $g_{Ric}, g_{R}, g_{RU}, g_{U}$ and $g_\Omega$ are given in \p{g's}.
We note that the term $4m_F^{2}g_{U}m_F^{2}$ gives a constant under the volume integral in (\ref{dcont}),
but we keep it for completeness in our expression.

It should be noticed that by choosing the form of the kinetic operator $\bm\Delta$ and $\bf U$ as in (\ref{deltaU}),
we treated the mass term as an interaction. In this way we can parallel the treatment with
the other massless gauge fields in the derivation of the effective action.
This choice of the leading operator as $z=-\bd_\mu^2$ changes slightly the shape of the cutoff function $R_k(z)$
and the actual form of the flow at intermediate energy scales $k$.
It corresponds to the choice of the cutoff of type I in gravitational theories~\cite{Percacci:2017,CPR},
%From the point of view of a perturbation calculus,
% this choice of treating the fermionic mass term  $m_F \bar\psi\psi$ in (\ref{diraca}) as a perturbative two-leg vertex 
and amounts to taking the massless propagator for fermions.
This is a well-developed technique in renormalization theory and it is used, in particular, in QCD as mass-independent
renormalization. In this case, quark mass is scheme-dependent quantity and it is then the most appropriate scheme.
An alternative choice would be to take $z=-\bd_\mu^2+m_F^2$ (realizing the cutoff of type II).
When the mass interaction is fully resummed, the two schemes do agree. However, we find it better suited to use
the massless scheme for fermions. For the effective action, the differences between two schemes are immaterial.
The difference only shows up in the local terms. The reason for this is that nonlocal logarithmic universal terms
in the effective action $\Gamma_0$ are related to the expressions for the perturbative one-loop beta functions,
and these beta functions in the UV limit are completely insensitive to mass terms.
The fermionic contribution to the gravity is cosnsitent with \cite{DonaP}.

We then come to the boson-boson terms in Eq.~(\ref{hessiandirac}):
\bea
\Delta S &=& \int d^4 x
\sqrt{\bg} \frac{1}{2} \Big[ \frac{1}{8} h^{\a\rho}\bnabla^\b h^\mu_\rho \bar\psi \c_{\a\b\mu}\psi
-i e (h \bg^{\mu\nu} - h^{\mu\nu}) A_\mu \bar\psi \c_\nu \psi
\nn &&
+\frac{1}{4}h^{\mu\nu} \bnabla^\rho h_\mu^\b \bar\psi \c_{\rho\b\nu}\psi
+\frac{1}{8} (3 h^\mu_\rho h_\nu^\rho -2 h h^\mu_\nu)(\bar\psi \c^\nu \bd_\mu \psi - \bd_\mu \bar\psi \c^\nu \psi)
\nn &&
+\frac{1}{8}(h^2-2 h_{\s\la}^2)\,
\left(\bar\psi \c^\mu \bd_\mu \psi - \bd_\mu \bar\psi \c^\mu \psi+2m_F\bar\psi\psi\right) \Big],
\label{dboson}
\eea
which should be written in a self-adjoint form. For this purpose, we check how terms behave under exchange
of left and right fields and integration by parts under volume integral (cf. with appendix \ref{appa}).
The first term in \p{dboson} is transformed under the integration by parts as
\bea
\frac{1}{8}h^{\alpha\rho}\nabla^{\beta}h^{\mu}{}_{\rho}\bar{\psi}\c_{\alpha\beta\mu}\psi
&=& \frac{1}{16}h^{\alpha\rho}\nabla^{\beta}h^{\mu}{}_{\rho}\bar{\psi}\c_{\alpha\beta\mu}\psi
-\frac{1}{16}h^{\mu}{}_{\rho}\nabla^{\beta}\left(h^{\alpha\rho}\bar{\psi}\c_{\alpha\beta\mu}\psi\right) \nn
&=& \frac{1}{16}h^{\alpha\rho}\nabla^{\beta}h^{\mu}{}_{\rho}\bar{\psi}\c_{\alpha\beta\mu}\psi
-\frac{1}{16}h^{\mu}{}_{\rho}\nabla^{\beta}h^{\a\rho}\bar{\psi}\c_{\alpha\beta\mu}\psi,
\label{sad}
\eea
due to complete antisymmetry of the product $\c_{\alpha\beta\mu}$.
No surface term is generated under the partial integration, and then it is clear that
\p{sad} is self-adjoint by itself under the exchange of fields and doing integration by parts.

We then put the quadratized action in~\p{dboson} in the  form:
\bea
\Delta S = \int d^4 x
\sqrt{\bg} \frac{1}{2} (h^{\mu\nu},A^\mu) {\bf H}_D
\left(\begin{array}{c}
h^{\a\b}\\
A^\a
\end{array}\right),
\eea
where the Hessian in the matrix form is
\bea
{\bf H}_D = 2 {\bf V}_D^\d \bd_\d +{\bf U}_D.
\label{HDirac}
\eea
These terms should be combined with other terms from gravity and gauge theory,
and integrated over the fluctuation fields $h_{\mu\nu}$ and gauge fields $A_\mu$.
We remark that in the bosonic terms in (\ref{dboson}) fermionic fields appear only as backgrounds.

Explicitly the components of objects in (\ref{HDirac}) are given by
\bea
{\bf V}_D^\d &=& \left(\begin{array}{cc}
-\frac{1}{16} \bg_{\mu\a}\bpsi \c_{\nu\b}{}^{\d}\psi & 0 \\
0 & 0
\end{array}\right), \label{dirach}\\
{\bf U}_D &=& \left(\begin{array}{cc}
U^{hh}_{\mu\nu\a\b} & U^{hA}_{\mu\nu,\a}\\
U^{Ah}_{\mu,\a\b} & 0%U^{AA}_{\mu\a}
\label{dirach2}
\end{array}\right),
\eea
where
\bea
U^{hh}_{\mu\nu\a\b} &=& \frac{3}{8} \bg_{\a\mu}\left(\bar{\psi}\c_{(\nu}\bd_{\b)}\psi-\bd_{(\nu}\bpsi\cdot\c_{\b)}
\psi\right)-\frac{1}{8}\bar g_{\mu\nu}\left(\bar{\psi}\c_{\a}\bd_{\b}\psi
-\bd_{\a}\bpsi\cdot\c_{\b}\psi\right) \nn
&& 
-\frac{1}{8}\bg_{\a\b}\left(\bar{\psi}\c_{\mu}\bd_{\nu}\psi-\bd_{\mu}\bar{\psi}\cdot\c_{\nu}\psi\right)
-\frac{1}{2} K_{\mu\nu\a\b} \left(\bpsi\c^{\rho}\bd_{\rho}\psi-\bd_{\rho}\bpsi\cdot\c^{\rho}\psi+2m_F\bpsi\psi\right),
\nn
U^{hA}_{\mu\nu,\a} &=& -\frac{1}{2}ie_F\left(\bg_{\mu\nu}\bpsi\c_{\a}\psi
-\bg_{\a(\mu}\bpsi\c_{\nu)}\psi\right),
\nn
U^{Ah}_{\mu,\a\b} &=& -\frac{1}{2}ie_F\left(\bg_{\a\b}\bpsi\c_{\mu}\psi-\bg_{\mu(\a}\bpsi\c_{\b)}\psi\right).%\nn
%U^{AA}_{\mu\a} &=& 0.
\label{dirach3}
\eea
In the above, the symmetrization within each pair of indices $(\mu,\nu)$
and $(\a,\b)$ is understood but not marked.
Here we have also imposed  self-adjointness, which is discussed in appendix~\ref{appa}.

\subsection{Charged scalar}
\label{sec3.4}

We consider charged scalar whose action is
\bea
S_S = \int d^4 x \sqrt{g} \left[g^{\mu\nu} (D_\mu^S \phi^*) (D_\nu^S \phi) +V(|\phi|^2) \right],
\eea
where the covariant derivative on the scalar is
\bea
D_\mu \phi =(\nabla_\mu -ie_S A_\mu)\phi, \qquad
D_\mu \phi^*=(\nabla_\mu +ie_S A_\mu)\phi^* 
\label{covder}
\eea
and where the invariant complex square of the field we write as $|\phi|^2=\phi^*\phi$.
 (By star in superscript, we denote complex conjugation of field.)
It should be understood that when the covariant derivative in (\ref{covder}) acts on fields without charge, it is simply
the gravitational covariant derivative $\nabla_\mu$. When the derivative acts on a scalar quantity as here,
it is simply a partial derivative (without spacetime connection coefficients).

Let us write the field as $\phi=\bp+\vp$, the former being the background and the latter being a fluctuation.
Then the quadratic action is
\bea
S_S \hs{-2}&=&\hs{-2} \int d^4 x \sqrt{g} \left[ \frac{h^2 -2(h_{\rho\la})^2}{8}
\left\{(\bd_\mu\bp^*)(\bd^{ \mu} \bp) +V(|\bp|^2) \right\}
+\Big(h^\mu_\rho h^{\rho\nu}-\frac12 h h^{\mu\nu}\Big)(\bd_\mu\bp^*)(\bd_\nu\bp) \right. \nn
&& + \frac12 h\left\{(\bd_\mu\bp^*) (\bd^{\mu} \vp) +(\bd_\mu\bp)(\bd^{\mu} \vp^*)
-ie_S [ \bp (\bd_\mu\bp^*) -\bp^*(\bd_\mu\bp)] A^\mu \right\} \nn
&& - h^{\mu\nu} \left\{(\bd_\mu\bp^*) (\bd_\nu\vp) +(\bd_\mu\bp)(\bd_\nu \vp^*)
-ie_S [\bp (\bd_\mu\bp^*) -\bp^*(\bd_\mu\bp)] A_\nu \right\} \nn
&& + (\bd_\mu\vp^*)(\bd^{\mu}\vp) +ie_S A_\mu \bp^* (\bd^{\mu}\vp)
-ie_SA^\mu \bp(\bd_\mu\vp^*) +e_S^2|\bp|^2 A_\mu A^\mu \nn
&& -ie_S (\bd_\mu\bp^*) A^\mu \vp + ie_S (\bd_\mu\bp)A^\mu \vp^* +V'(|\bp|^2) \vp\vp^* \nn
&& \left.
+\frac12 V''(|\bp|^2)(\bp^{*2}\vp^2+2|\bp|^2\vp\vp^*+\bp^2 \vp^{*2})
+\frac12 h V'(|\bp|^2)(\bp^{*}\vp+\bp \vp^{*})
\right],
\label{scalarhessian}
\eea
where bars on the covariant derivatives mean that they are made of background fields
and the prime is the derivative with respect to $|\phi|^2$ (not with respect to just $\phi$).
This is put in the form
\bea
\frac12 \int d^4 x \sqrt{g} (h^{\mu\nu}, A^\mu, \vp^*,\vp)
{\bf H}_S
\left(\begin{array}{c}
h^{\a\b} \\
A^\a \\
\vp \\
\vp^*
\end{array}
\right),
\label{sh}
\eea
where the Hessian ${\bf H}_S$ has the following expansion in order of derivatives
\bea
{\bf H}_S = {\bf K}_S(-\bd^2) +2 {\bf V}_S^\d \bd_\d +{\bf U}_S.
\label{scalarhessian_matrix}
\eea
Here we use the ``complex'' basis discussed in the appendix~\ref{appb}.
We find
\bea
{\bf K}_S &=& \left(
\begin{array}{cccc}
0 & 0 & 0 & 0 \\
0 & 0 & 0 & 0 \\
0 & 0 & 1 & 0 \\
0 & 0 & 0 & 1
\end{array}
\right),
\label{kscalar}\\
{\bf V}_S^\d &=& \left(
\begin{array}{cccc}
0 & 0 & -K_{\mu\nu}{}^{\la\d}(\bd_\la \bp^*) &  -K_{\mu\nu}{}^{\la\d}(\bd_\la \bp) \\
0 & 0 & \frac{ie_S}{2}\bp^*\bg_\mu{}^\d & -\frac{ie_S}{2}\bp \bg_\mu{}^\d \\
K_{\a\b}{}^{\la\d}(\bd_\la \bp) & \frac{ie_S}{2}\bp\bg_\a{}^\d & 0 & 0 \\
K_{\a\b}{}^{\la\d}(\bd_\la \bp^*) & -\frac{ie_S}{2}\bp^*\bg_\a{}^\d & 0 & 0
\end{array}
\right), \\
\mbox{and}\nn
{\bf U}_S &=& \left(
\begin{array}{cccc}
U^{hh}_{\mu\nu,\a\b} & U^{hA}_{\mu\nu,\a} & U^{h\vp}_{\mu\nu} & U^{h\vp^*}_{\mu\nu} \\
U^{Ah}_{\mu,\a\b} & U^{AA}_{\mu\a} & U^{A\vp}_{\mu} & U^{A\vp^*}_{\mu} \\
U^{\vp^* h}_{\a\b} & U^{\vp^* A}_{\a} & U^{\vp^* \vp} & U^{\vp^* \vp^*} \\
U^{\vp h}_{\a\b} & U^{\vp A}_{\a} & U^{\vp\vp} & U^{\vp\vp^*}
\end{array}
\right)
\eea
where the various components of the ${\bf U}_S$ tensor are given below
\bea
U^{hh}_{\mu\nu,\a\b} &=& -K_{\mu\nu\a\b} \left[ (\bd_\rho\bp^*)(\bd^{\rho}\bp)+V(|\bp|^2) \right]
+ 2 K_{\mu\nu\la\a}(\bd^{\la} \bp^*) (\bd_\b \bp) + 2 K_{\a\b\la\mu}(\bd^{\la} \bp^*) (\bd_\nu \bp),
\nn
U^{hA}_{\mu\nu,\a} &=& 2ie_S K_{\mu\nu\a}{}^\la \left[ \bp(\bd_\la \bp^*)-\bp^*(\bd_\la\bp)\right],
\nn
U^{Ah}_{\mu,\a\b} &=& 2ie_S K_{\a\b\mu}{}^\la \left[ \bp(\bd_\la \bp^*)-\bp^*(\bd_\la\bp)\right],
\nn
U^{h\vp}_{\mu\nu} &=& \frac12 \bg_{\mu\nu} V' \bp^*,
\qquad \qquad
U^{\vp h}_{\a\b} = 2K_{\a\b}{}^{\rho\la}(\bd_\rho \bd_\la \bp^*) + \frac12 \bg_{\a\b} V' \bp^*,
\nn
U^{h\vp^*}_{\mu\nu} &=& \frac12 \bg_{\mu\nu} V' \bp,
\qquad \qquad
U^{\vp^* h}_{\a\b} = 2K_{\a\b}{}^{\rho\la}(\bd_\rho \bd_\la \bp) + \frac12 \bg_{\a\b} V' \bp,
\nn
U^{AA}_{\mu\a} &=& 2 e_S^2 |\bp|^2 \bg_{\mu\a},
\qquad \qquad
U^{A\vp}_{\mu} = -ie_S (\bd_\mu \bp^*),
\qquad \qquad
U^{\vp A}_{\a} = -2ie_S (\bd_\a \bp^*),
\nn
U^{A\vp^*}_{\mu} &=& ie_S (\bd_\mu \bp),
\qquad \qquad
U^{\vp^* A}_{\a} = 2ie_S (\bd_\a \bp),
\nn
U^{\vp^* \vp} &=& U^{\vp \vp^*} = V' + V'' |\bp|^2,
\quad
U^{\vp \vp} = V'' \bp^{*2},
\quad
U^{\vp^* \vp^*} = V'' \bp^{2}.
\label{scalaru}
\eea

We note that similarly to the case of Dirac fermion, here we treat the mass of the scalar field
(included in the scalar potential $V=V(|\phi|^2)$) as an interaction. In this way we realize cutoff of type I.
For the scalar fluctuations, we are able to deal with all terms including mixed terms in
(\ref{scalarhessian_matrix})), hence our nonlocal contribution to the effective action is complete.
In the next section, we show how to deal with all fluctuations in the scalar sector
within nonlocal heat kernel technique.

\subsection{Total Hessian}
\label{sec3.5}

Collecting all the results for the Hessian (written in (\ref{gravitonh})-(\ref{gravitonh2}), (\ref{g0})-(\ref{g5}),
(\ref{HDirac})-(\ref{dirach3}) and (\ref{sh})-(\ref{scalaru})), we get the bosonic part of the quadratic action
of the total system.
We can write it as
\bea
\frac12 \int d^4 x \sqrt{g} (h^{\mu\nu}, A^\mu, \vp^*,\vp)
{\bf H}_T
\left(\begin{array}{c}
h^{\a\b} \\
A^\a \\
\vp \\
\vp^*
\end{array}
\right),
\eea
where we have expressed the bosonic fluctuations of quantum fields in a multiplet
$(h^{\mu\nu}, A^\mu, \vp^*,\vp)$, and
\bea
{\bf H}_T = {\bf K}_T(-\bd^2) +2 {\bf V}_T^\d \bd_\d +{\bf U}_T.
\label{hessiant}
\eea
The components of the Hessian ${\bf H}_T$ in derivative expansion are given as
\bea
{\bf K}_T = \left(
\begin{array}{cccc}
K_{\mu\nu\a\b} & 0 & 0 & 0 \\
0 & \bg_{\mu\a} & 0 & 0 \\
0 & 0 & 1 & 0 \\
0 & 0 & 0 & 1
\end{array}
\right),
\label{kfactor}
\eea
\bea
{\bf V}_T^\d \hs{-2}&=&\hs{-2} \left(
\begin{array}{cccc}
-\frac{1}{32}\left(\bg_{\mu\a}\bar\psi\c_{\nu\b}{}^\d \psi+\bg_{\nu\b}\bar\psi\c_{\mu\a}{}^\d \psi \right)
 & K_{\mu\nu\la\a} \bF^{\la\d}-\frac{1}{2} \d_{\mu\nu}{}^{\la\d}\bF_{\la\a} & -K_{\mu\nu}{}^{\la\d}(\bd_\la \bp^*)
 &  -K_{\mu\nu}{}^{\la\d}(\bd_\la \bp) \\
- K_{\a\b\la\mu} \bF^{\la\d}+\frac{1}{2} \d_{\a\b}{}^{\la\d}\bF_{\la\mu}  & 0 & \frac{ie_S}{2}\bp^*\bg_\mu{}^\d
 & -\frac{ie_S}{2}\bp \bg_\mu{}^\d \\
K_{\a\b}{}^{\la\d}(\bd_\la \bp) & \frac{ie_S}{2}\bp\bg_\a{}^\d & 0 & 0 \\
K_{\a\b}{}^{\la\d}(\bd_\la \bp^*) & -\frac{ie_S}{2}\bp^*\bg_\a{}^\d & 0 & 0
\end{array}
\right), \nn
\mbox{and}\\
{\bf U}_T \hs{-2}&=&\hs{-2} \left(
\begin{array}{cccc}
U^{hh}_{\mu\nu,\a\b} & U^{hA}_{\mu\nu,\a} & U^{h\vp}_{\mu\nu} & U^{h\vp^*}_{\mu\nu} \\
U^{Ah}_{\mu,\a\b} & U^{AA}_{\mu\a} & U^{A\vp}_{\mu} & U^{A\vp^*}_{\mu} \\
U^{\vp^* h}_{\a\b} & U^{\vp^* A}_{\a} & U^{\vp^* \vp} & U^{\vp^* \vp^*} \\
U^{\vp h}_{\a\b} & U^{\vp A}_{\a} & U^{\vp\vp} & U^{\vp\vp^*}
\end{array}
\right),
\eea
where the components of the ${\bf U}_T$ tensor are
\bea
U^{hh}_{\mu\nu,\a\b} &=& \left[ \br -\frac{1}{4} \bF_{\rho\la}^2 - (\bd_\rho\bp^*)(\bd^{\rho}\bp) - V(|\bp|^2)
- \frac{1}{2} (\bpsi\c^\rho\bd_\rho\psi - \bd_\rho\bpsi\cdot\c^\rho\psi+2m_F\bpsi\psi) \right]K_{\mu\nu,}{}_{\a\b}
\nn &&
- \bar g_{(\mu(\a} \br_{\b)\nu)} -\br_\mu{}_{(\a}{}_\nu{}_{\b)} +\frac{1}{2}(\bg_{\mu\nu}\br_{\a\b}+\bg_{\a\b}\br_{\mu\nu})
+ K_{\mu\nu\la}{}_{(\a} \bF_{\rho}{}_{\b)}\bF^{\rho\la}
+ K_{\a\b}^{\la}{}_{(\mu} \bF^{\rho}{}_{\nu)}\bF_{\rho\la}
\nn &&
+ \frac{1}{2} \bF_{\mu}{}_\a \bF_{\nu}{}_\b
+\frac{3}{16}\bg_{(\mu(\a}(\bar\psi\c_{\b)} \bd_{\nu)}\psi-\bd_{\nu)}\bar\psi\cdot\c_{\b)} \psi)
+\frac{3}{16}\bg_{(\a(\mu}(\bar\psi\c_{\nu)} \bd_{\b)}\psi-\bd_{\b)}\bar\psi\cdot\c_{\nu)} \psi)
\nn &&
-\frac{1}{8}\bg_{\mu\nu}(\bar\psi\c_{(\a} \bd_{\b)}\psi -\bd_{(\b} \bar\psi\cdot\c_{\a)} \psi)
-\frac{1}{8}\bg_{\a\b}(\bar\psi\c_\mu \bd_\nu\psi -\bd_\nu \bar\psi\cdot\c_\mu \psi)
\nn &&
+ 2 K_{\mu\nu}{}^{\la}{}_{(\a}(\bd_\la\bp^*) (\bd_{\b)} \bp) +  2K_{\a\b}{}_{\la\mu}(\bd^{\la} \bp^*) (\bd_\nu \bp),
\nn
U^{hA}_{\mu\nu,\a} &=& 2ie_S K_{\mu\nu\a}{}^\la \left[ \bp(\bd_\la \bp^*)-\bp^*(\bd_\la\bp)\right]
-\frac{ie_F}{2}(\bg_{\mu\nu}\bpsi\c_\a\psi-\bg_{\mu\a}\bpsi\c_\nu\psi),
\nn
U^{Ah}_{\mu,\a\b} &=& - 2 K_{\a\b\la\mu} (\bnabla_\rho \bF^{\la\rho})+ \d_{\a\b}{}^{\la\rho} (\bnabla_\rho \bF_{\la\mu})
+2ie_S K_{\a\b\mu}{}^\la \left[ \bp(\bd_\la \bp^*)-\bp^*(\bd_\la\bp)\right]
\nn &&
-\frac{ie_F}{2}(\bg_{\a\b}\bpsi\c_\mu\psi-\bg_{\a\mu}\bpsi\c_\b\psi),
\nn
U^{h\vp}_{\mu\nu} &=& \frac12 \bg_{\mu\nu} V' \bp^*,
\qquad
U^{\vp h}_{\a\b} = 2K_{\a\b}{}^{\rho\la}(\bd_\rho \bd_\la \bp^*)
+ \frac12 \bg_{\a\b} V' \bp^*,
\nn
U^{h\vp^*}_{\mu\nu} &=& \frac12 \bg_{\mu\nu} V' \bp,
\qquad
U^{\vp^* h}_{\a\b} = 2K_{\a\b}{}^{\rho\la}(\bd_\rho \bd_\la \bp)
 + \frac12 \bg_{\a\b} V' \bp,
\nn
U^{AA}_{\mu\a} &=& \br_{\mu\a}+ 2 e_S^2 |\bp|^2 \bg_{\mu\a},
\qquad
U^{A\vp}_{\mu} = -ie_S (\bd_\mu \bp^*),
\qquad
U^{\vp A}_{\a} = -2ie_S (\bd_\a \bp^*),
\nn
U^{A\vp^*}_{\mu} &=& ie_S (\bd_\mu \bp),
\qquad
U^{\vp^* A}_{\a} = 2ie_S (\bd_\a \bp),
\nn
U^{\vp^* \vp} &=& U^{\vp \vp^*} = V' + V'' |\bp|^2,
\quad
U^{\vp \vp} = V'' \bp^{*2}, \quad
U^{\vp^* \vp^*} = V'' \bp^{2}.~~~
\eea
Here symmetrization $\mu \lra \nu$, $\a \lra \b$ and $(\mu,\nu) \lra (\a,\b)$,
if appropriate, should be understood.

It is convenient to extract an overall factor of ${\bf K}_T$ (\ref{kfactor}) from the full Hessian ${\bf H}_T$
and write it as
\bea
{\bf H}_T = {\bf K}_T \bm {\Delta}
\equiv {\bf K}_T( -\bd^2 {\bf 1}+2 {\bf Y}^\d \bd_\d +{\bf W}).
\label{toth}
\eea
Using
\bea
K^{-1}_{\mu\nu}{}^{\a\b} = 2\d_{\mu\nu}{}^{\a\b}-\bg_{\mu\nu} \bg^{\a\b},
\eea
which is the generalization of the gravitational DeWitt metric in field space, we find
\bea
{\bf Y}^\d \hs{-2}&=&\hs{-2} \left(
\begin{array}{cccc}
-\frac{1}{8}\bg_{(\mu}^{(\a}\bpsi\c_{\nu)}{}^{\b)\d}\psi &
2 K_{\mu\nu\la}{}^{\a} \bF^{\la\d}-\d_{\mu\nu}{}^{\la\d}\bF_{\la}{}^{\a}
& -\d_{\mu\nu}{}^{\la\d}(\bd_\la \bp^*)
& -\d_{\mu\nu}{}^{\la\d}(\bd_\la \bp) \\
- K^{\a\b}{}_{\la\mu} \bF^{\la\d}+\frac{1}{2} \d^{\a\b,\la\d}\bF_{\la\mu}  & 0 & \frac{ie_S}{2}\bp^*\bg_\mu{}^\d
 & -\frac{ie_S}{2}\bp \bg_\mu{}^\d \\
K^{\a\b,\la\d}(\bd_\la \bp) & \frac{ie_S}{2}\bp\bg^\a{}^\d & 0 & 0 \\
K^{\a\b,\la\d}(\bd_\la \bp^*) & -\frac{ie_S}{2}\bp^*\bg^\a{}^\d & 0 & 0
\end{array}
\right), \nn \mbox{and}\\
{\bf W} \hs{-2}&=&\hs{-2} \left(
\begin{array}{cccc}
W^{hh}_{\mu\nu,\a\b} & W^{hA}_{\mu\nu,\a} & W^{h\vp}_{\mu\nu} & W^{h\vp^*}_{\mu\nu} \\
W^{Ah}_{\mu,\a\b} & W^{AA}_{\mu\a} & W^{A\vp}_{\mu} & W^{A\vp^*}_{\mu} \\
W^{\vp^* h}_{\a\b} & W^{\vp^* A}_{\a} & W^{\vp^* \vp} & W^{\vp^* \vp^*} \\
W^{\vp h}_{\a\b} & W^{\vp A}_{\a} & W^{\vp\vp} & W^{\vp\vp^*}
\end{array}
\right).
\label{wtensor}
\eea
In principle, in the tensors $\bf Y$ and $\bf W$ the indices are in natural matrix position
(one index down -- one index up) letting to perform traces (hence contractions) very naturally without usage
of any metric tensor. In particular, all indices $\a$ and $\b$ should be in superscripts.
However, for notational simplicity, from displaying the matrix of $\bf W$ tensor in (\ref{wtensor})
we do not write them in upper position, hoping that this will not lead to any confusion.

The explicit components of the $\bf W$ tensor read
\bea
W^{hh}_{\mu\nu,\a\b} &=& 2 U^{hh}_{\mu\nu,\a\b} - \frac12 \bg_{\mu\nu} \left[
\bF_{\a\rho}\bF_\b{}^\rho-\left(\frac14 \bF_{\rho\la}^2-V-m_F\bpsi \psi\right)\bg_{\a\b}
-\frac14(\bpsi\c_{(\a} \bd_{\b)}\psi -\bd_{(\a} \bpsi\c_{\b)} \psi) \right. \nn
&& \left. \hs{20}
+\frac14(\bpsi\c_{\rho} \bd^\rho \psi -\bd_\rho \c^\rho \psi)\bg_{\a\b} \right], \nn
W^{hA}_{\mu\nu,\a} &=& 2 U^{hA}_{\mu\nu,\a}+\frac{i}{2} \bg_{\mu\nu}[
2 e_S (\bp\bd_\a \bp^*-\bp^*\bd_\a\bp)+3 e_F \bpsi \c_\a\psi], \nn
W^{h\vp^*}_{\mu\nu} &=& - \bg_{\mu\nu} V' \bp,
\nn
W^{h\vp}_{\mu\nu} &=& - \bg_{\mu\nu} V' \bp^*,
\eea
and other components of the $\bf W$ tensor are the same as the corresponding components in ${\bf U}_T$.

We next eliminate the first order term ${\bf Y}^\d \bd_\d$ in (\ref{toth}) by writing
$\tilde {{\bf D}}_\mu={\bf 1}\bd_\mu -{\bf Y}_\mu$ following \cite{barvinsky,CPRT2015}.
Then $\bm\Delta$ in \p{toth} is rewritten as
\bea
\bm\Delta &=& -\tilde {\bf D}_\mu^2 +\tilde {\bf W},
\label{deltashifted}
\eea
where
\bea
\tilde {\bf W} &=& {\bf W}-\bd_\d {\bf Y}^\d +{\bf Y}_\d {\bf Y}^\d.\label{tildeW}
\eea
Here
\bea
\bd_\d {\bf Y}^\d &=&\left(\begin{array}{cccc}
-\frac{1}{8}\bg_{\alpha\mu}\bd^{\delta}\left(\bar{\psi}\c_{\nu\beta\delta}\psi\right)
& 2K_{\mu\nu\lambda\alpha}\bnabla^{\delta}\bF^{\lambda}{}_{\delta}-\bnabla_{\mu}\bF_{\nu\alpha}
& -\bd_{\mu}\bd_{\nu}\bp^{*} & -\bd_{\mu}\bd_{\nu}\bp\\
-K_{\alpha\beta\lambda\mu}\bnabla^{\delta}\bF^{\lambda}{}_{\delta}+\frac{1}{2}\bnabla_{\alpha}\bF_{\beta\mu} & 0 
& \frac{ie_S}{2}\bd_{\mu}\bp^{*} & -\frac{ie_S}{2}\bd_{\mu}\bp\\
K_{\alpha\beta}{}^{\la\d}\bd_\d\bd_{\la}\bp & \frac{ie_S}{2}\bd_{\alpha}\bp & 0 & 0\\
K_{\alpha\beta}{}^{\la\d}\bd_\d\bd_{\lambda}\bp^{*} & -\frac{ie_S}{2}\bd_{\alpha}\bp^{*} & 0 & 0
\end{array}\right),
\nn
{\rm and}&&\\
{\bf Y}_\d {\bf Y}^\d &=& \left(\begin{array}{cccc}
Y_{\mu\nu,\alpha\beta}^{2,hh} & Y_{\mu\nu,\alpha}^{2,hA} & Y_{\mu\nu}^{2,h\vp} & Y_{\mu\nu}^{2,h\vp^{*}}\\
Y_{\mu,\a\b}^{2,Ah} & Y_{\mu,\alpha}^{2,AA} & Y_{\mu}^{2,A\vp} & Y_{\mu}^{2,A\vp^{*}}\\
Y_{\a\b}^{2,\vp^* h} & Y_{\a}^{2,\vp^* A} & Y^{2,\vp^* \vp} & Y^{2,\vp^* \vp^*} \\
Y_{\a\b}^{2,\vp h} & Y_{\a}^{2,\vp A} & Y^{2,\vp\vp} & Y^{2,\vp\vp^*}
\end{array}\right).
\label{yy}
\eea
The explicit forms of the above components of the ${\bf Y}_\d {\bf Y}^\d $ are given in appendix~\ref{appc.1}.

We note that the gauge- and spacetime-covariant shifted derivative $\tilde {\bf D}_\mu$ is matrix-valued
and constructed with backgrounds of the gauge and gravitational fields. By shifting the covariant derivative
$\bar D_\mu$ to the new one $\tilde{\bf D}_\mu$, we reduce the differential operator in (\ref{toth})
to the minimal form with leading symbol with two derivatives. The leading term with two derivatives
is a square of the new covariant derivative $\tilde{\bf D}_\mu$, the term with one derivative less is absent
and finally all the other non-derivative operators (endomorphisms of the internal vector bundle)
are collected in the operator $\tilde {\bf W}$. Now the kinetic operator in (\ref{deltashifted}) has precisely
the form elucidated in (\ref{deltaform}), so the standard method of heat kernel technique can be
applied here to take its functional trace.

For the components of the $\tilde{\bf W}$ tensor, using (\ref{tildeW}) and (\ref{yy}), we find
\bea
\tilde{W}_{\mu\nu,\a\b}^{h h}&=& \Big(2\br-\frac{1}{2}\bF_{\rho\la}^{2}
-2\bd_{\rho}\bp^*\bd^{\rho}\bp\Big)K_{\mu\nu\a\b}
- \bg_{\mu\a}\bg_{\nu\b}\Big[V+\frac{1}{2}\Big(\bar{\psi}\c_{\rho}\bd^{\rho}\psi-\bd_{\rho}\bar{\psi}\c^{\rho}\psi
\nn &&
+2m_F\bar{\psi}\psi\Big)\Big]
+\frac{1}{8}\bg_{\mu\nu}\bg_{\a\b}\left(\bar{\psi}\c_{\rho}\bd^{\rho}\psi-\bd^{\rho}\bar{\psi}\c_{\rho}\psi\right)
-2\bg_{\a\mu}\br_{\b\nu}-2\br_{\mu\a\nu\b}
\nn &&
+\bg_{\mu\nu}\br_{\a\b}+\bg_{\a\b}\br_{\mu\nu}
+2K_{\mu\nu\la\a}\bF^{\rho\la}\bF_{\rho\b}
+\frac{3}{8}\bg_{\a\mu}\left(\bar{\psi}\c_{\b}\bd_{\nu}\psi-\bd_{\nu}\bar{\psi}\c_{\b}\psi\right)
\nn &&
+\frac{3}{8}\bg_{\a\mu}\left(\bar{\psi}\c_{\nu}\bd_{\b}\psi-\bd_{\b}\bar{\psi}\c_{\nu}\psi\right)
-\frac{1}{8}\bg_{\mu\nu}\left(\bar{\psi}\c_{\a}\bd_{\b}\psi-\bd_{\b}\bar{\psi}\c_{\a}\psi\right)
\nn &&
-\frac{1}{4}\bg_{\a\b}\left(\bar{\psi}\c_{\mu}\bd_{\nu}\psi-\bd_{\nu}\bar{\psi}\c_{\mu}\psi\right)
+\frac{1}{8}\bg_{\a\mu}\bd^{\delta}\left(\bar{\psi}\c_{\nu\b\delta}\psi\right)
-\frac{1}{128}\bg_{\mu\a}\bar{\psi}\c_{\nu\la\delta}\psi\bar{\psi}\c_{\b}{}^{\la\delta} \psi
\nn &&
+\frac{1}{128}\bar{\psi}\c_{\mu\a\delta}\psi\bar{\psi}\c_{\nu\b}{}^\delta\psi
+\frac{3}{2}\bg_{\a\mu}\bd_{\nu}\bp^*\bd_{\b}\bp+\frac{3}{2}\bg_{\a\mu}\bd_{\b}\bp^*\bd_{\nu}\bp
-\bg_{\mu\nu}\bd_{\a}\bp^*\bd_{\b}\bp
\nn &&
-\frac{1}{2}\bg_{\a\b}\bd_{\mu}\bp^*\bd_{\nu}\bp ,
\nn
\tilde{W}_{\mu\nu,\a}^{h A}&=&\frac{3}{2}ie_S\bg_{\mu\a}\left(\bp\bd_{\nu}\bp^*-\bp^*\bd_{\nu}\bp\right)
+\frac{ie_S}{2}\bg_{\mu\nu}\bar{\psi}\c_{\a}\psi+ie_S\bg_{\mu\a}\bar{\psi}\c_{\nu}\psi
-2K_{\mu\nu\la\a}\bnabla_{\delta}\bF^{\la\delta}
\nn &&
+\bnabla_{\mu}\bF_{\nu\a}-\frac{1}{16}\bar{\psi}\c_{\mu\a\delta}\psi \bF_{\nu}{}^\delta
-\frac{1}{16}\bg_{\mu\a}\bar{\psi}\c_{\nu\la\delta}\psi \bF^{\la\delta},
\nn
\tilde{W}_{\mu\nu}^{h \varphi}&=&-\bg_{\mu\nu}V'\bp^*+\bd_{\mu}\bd_{\nu}\bp^*,
\nn
\tilde{W}_{\mu\nu}^{h \varphi^*}&=&-\bg_{\mu\nu}V'\bp+\bd_{\mu}\bd_{\nu}\bp,
\nn
\tilde{W}_{\mu,\a\b}^{A h} &=&-K_{\a\b\la\mu}\bnabla_{\rho}\bF^{\la\rho}+\frac{1}{2}\bnabla_{\a}\bF_{\b\mu}
+\frac{3}{2}ie_SK_{\a\b\mu\la}\left(\bd^{\la}\bp^*\bp-\bp^*\bd^{\la}\bp\right)
\nn &&
-\frac{ie_S}{2}\left(\bg_{\a\b}\bar{\psi}\c_{\mu}\psi-\bg_{\a\mu}\bar{\psi}\c_{\b}\psi\right)+\frac{1}{32}\bF_{\a\delta}
\bar{\psi}\c_{\mu\b}{}^\d\psi-\frac{1}{32}\bg_{\a\mu}\bF^{\la\delta}\bar{\psi}\c_{\b\la\delta}\psi,
\nn
\tilde{W}_{\mu\a}^{A A}&=&\br_{\mu\a}+\frac{3}{2}e_S^{2}\bg_{\mu\a}|\bp|^{2}-\frac{1}{2}\bF_{\mu\la}\bF_{\a}{}^{\la}
-\frac{1}{4}\bg_{\mu\a}\bF_{\rho\la}^{2},
\nn
\tilde{W}_{\mu}^{A \varphi}&=&-\frac{3}{2}ie_S\bd_{\mu}\bp^*-\frac{3}{4}\bF_{\la\mu}\bd^{\la}\bp^*,
\nn
\tilde{W}_{\mu}^{A \varphi^*}&=&\frac{3}{2}ie_S\bd_{\mu}\bp-\frac{3}{4}\bF_{\la\mu}\bd^{\la}\bp,
\nn
\tilde{W}_{\a\b}^{\varphi^* h}&=&K_{\a\b\rho\la}\bd^{\rho}\bd^{\la}\bp+\frac{1}{2}\bg_{\a\b}V'\bp,
\nn
\tilde{W}_{\a}^{\varphi^* A}&=&\frac{3}{2}ie_S\bd_{\a}\bp-\frac{3}{4}\bd^{\la}\bp \bF_{\la\a},
\nn
\tilde{W}^{\varphi^* \varphi}&=&V'+\left(V''-e_S^{2}\right)|\bp|^{2}-\bd_{\la}\bp^*\bd^{\la}\bp,
\nn
\tilde{W}^{\varphi^* \varphi^*}&=&\left(V''+e_S^{2}\right)\bp^{2}-\bd_{\la}\bp \bd^{\la}\bp,
\nn
\tilde{W}_{\a\b}^{\varphi h}&=&K_{\a\b\rho\la}\bd^{\rho}\bd^{\la}\bp^*+\frac{1}{2}\bg_{\a\b}V'\bp^*,
\nn
\tilde{W}_{\a}^{\varphi A}&=&-\frac{3}{2}ie_S\bd_{\a}\bp^*-\frac{3}{4}\bd^{\la}\bp^*\bF_{\la\a},
\nn
\tilde{W}^{\varphi \varphi}&=&\left(V''+e_S^{2}\right)\bp^{*2}-\bd_{\la}\bp^*\bd^{\la}\bp^*,
\nn
\tilde{W}^{\varphi \varphi^*}&=&V'+\left(V''-e_S^{2}\right)|\bp|^{2}-\bd_{\la}\bp^*\bd^{\la}\bp .
\eea

Now we wish to calculate the commutator of the shifted covariant derivatives $\tilde {\bf D}_\mu$.
For this purpose the commutator of gauge-covariant derivatives $\bar D_\mu$ has to be found first.
We give it in a matrix form using the general vector of bosonic fluctuations $\Psi=(h_{\a\b},A_\a,\varphi,\varphi^*)^T$:
\bea
\Psi^\dagger \bm\Omega_{\rho\s}\Psi=\left(\begin{array}{cccc}
h_{\mu\nu} & A_{\mu} & \vp^* & \vp\end{array}\right)\bm\Omega_{\rho\s}\left(\begin{array}{c}
h_{\a\b}\\
A_{\a}\\
\vp\\
\vp^*
\end{array}\right)=\Psi^\dagger\left[\bd_{\rho},\bd_{\s}\right]\Psi.
\label{omegapsi}
\eea
Using the commutation relations on particular types of fluctuations
\bea
\left[\bnabla_{\rho},\bnabla_{\s}\right]h_{\mu\nu} &=& 2\br_{\rho\s(\mu}{}^{\kappa}h_{\nu)\kappa}
=2\bg_{(\mu}{}^{\alpha}\br_{\rho\s\nu)}{}^{\b}h_{\alpha\b},\label{f385}\\
\left[\bnabla_{\rho},\bnabla_{\s}\right]A_{\mu} &=& \br_{\rho\s\mu}{}^{\a}A_{\a},\label{f386}\\
\left[\bd_{\rho},\bd_{\s}\right]\bp&=&2\bd_{[\rho}\bd_{\s]}\bp=-2ie_S \bnabla_{[\rho}A_{\s]}\bp=-ie_S \bF_{\rho\s}\bp,
\label{f387}
\eea
we find
\bea
\bm\Omega_{\rho\s}=\left[\bd_{\rho},\bd_{\s}\right]=\left(\begin{array}{cccc}
2\bg_{\mu\a}\br_{\rho\s\nu\b} & 0 & 0 & 0\\
0 & \br_{\rho\s\mu\a} & 0 & 0\\
0 & 0 & -ie_S \bF_{\rho\s} & 0\\
0 & 0 & 0 & ie_S \bF_{\rho\s}
\end{array}\right).
\eea
We have the general formula for the commutator $\tilde{\bm\Omega}_{\rho\s}$ of shifted covariant derivatives
$\tilde {\bf D}_\mu$:
\bea
&&
\tilde{\bm\Omega}_{\rho\s}=\left[\tilde{\bf D}_{\rho},\tilde{\bf D}_{\s}\right]=\bm\Omega_{\rho\s}-2\bd_{[\rho}{\bf Y}_{\s]}
+2{\bf Y}_{[\rho}{\bf Y}_{\s]}
\eea
and in particular for various tensors appearing above
\bea
{\bf Y}_{\s}=\left(\begin{array}{cccc}
-\frac{1}{8}\bg_{\a\mu}\bar{\psi}\c_{\nu\b\s}\psi & 2K_{\mu\nu\la\a}\bF^{\la}{}_\s
-\d_{\mu\nu,\la\s}\bF^\la{}_\a
 & -\d_{\mu\nu,\la\s}\bd^\la\bp^*
 & -\d_{\mu\nu,\la\s}\bd^{\la}\bp\\
-K_{\a\b\lambda\mu}\bF^\la{}_\s+\frac{1}{2}\d_{\a\b,\la\s}\bF^\la{}_\mu
 & 0 
& \frac{ie_S }{2}\bg_{\mu\s}\bp^* & -\frac{ie_S }{2}\bg_{\mu\s}\bp\\
K_{\a\b\la\s}\bd^{\la}\bp & \frac{ie_S }{2}\bg_{\a\s}\bp & 0 & 0\\
K_{\a\b\la\s}\bd^{\la}\bp^* & -\frac{ie_S }{2}\bg_{\a\s}\bp^* & 0 & 0
\end{array}\right),\nn
\eea
and
\bea
&& \hs{-5}
\bd_{[\rho}{\bf Y}_{\s]}=\left(
{\small
\begin{array}{cc}
-\frac{1}{8}\bg_{\a\mu}\bd_{[\rho}\left(\bar{\psi}\c_{\nu\b\s]}\psi\right)
& \frac{1}{2}\bg_{\nu\a}\bd_{\mu}\bF_{\rho\s}-\frac{1}{4}\bg_{\mu\nu}\bd_{\a}\bF_{\rho\s}+\bg_{\mu[\rho}\bd_{\s]}\bF_{\nu\a}
\nn
-\frac{1}{4}\bg_{\a\mu}\bd_{\b}\bF_{\rho\s}+\frac{1}{8}\bg_{\a\b}\bd_{\mu}\bF_{\rho\s}
-\frac{1}{2}\bg_{\a[\rho}\bd_{\s]}\bF_{\b\mu} 
& 0 
\nn
-\frac{1}{2}\bg_{\b[\rho}\bd_{\s]}\bd_{\a}\bp+\frac{ie_S }{8}\bg_{\a\b}\bF_{\rho\s}\bp
& -\frac{ie_S }{2}\bg_{\a[\rho}\bd_{\s]}\bp 
\nn
-\frac{1}{2}\bg_{\b[\rho}\bd_{\s]}\bd_{\a}\bp^*-\frac{ie_S }{8}\bg_{\a\b}\bF_{\rho\s}\bp^* 
& \frac{ie_S }{2}\bg_{\a[\rho}\bd_{\s]}\bp^* 
\end{array}
} \right.
\hs{10}
\\ &&
\left. \hs{50}
{\small
\begin{array}{cc}
\bg_{\mu[\rho}\bd_{\s]}\bd_{\nu}\bp^*
& \bg_{\mu[\rho}\bd_{\s]}\bd_{\nu}\bp \\
-\frac{ie_S }{2}\bg_{\mu[\rho}\bd_{\s]}\bp^* 
& \frac{ie_S }{2}\bg_{\mu[\rho}\bd_{\s]}\bp \\
0 
& 0\\
0 
& 0
\end{array}}
\right),
\eea
where we also used the Bianchi identity
\bea
2\bnabla_{[\rho}\bF_{\b\s]}=\bnabla_{\rho}\bF_{\b\s}-\bnabla_{\s}\bF_{\b\rho}=\bnabla_{\b}\bF_{\rho\s}.
\eea
Finally, we write the components of the $\tilde{\bm\Omega}_{\rho\s}$ tensor in a matrix form
\bea
\tilde{\bm\Omega}_{\rho\s}&=&\left(\begin{array}{cccc}
\tilde{\Omega}_{\mu\nu,\a\b,\rho\s}^{h,h} & \tilde{\Omega}_{\mu\nu,\a,\rho\s}^{h,A}
 & \tilde{\Omega}_{\mu\nu,\rho\s}^{h,\vp} & \tilde{\Omega}_{\mu\nu,\rho\s}^{h,\vp^*}\\
\tilde{\Omega}_{\mu,\a\b,\rho\s}^{A,h} & \tilde{\Omega}_{\mu,\a,\rho\s}^{A,A}
 & \tilde{\Omega}_{\mu,\rho\s}^{A,\vp} & \tilde{\Omega}_{\mu,\rho\s}^{A,\vp^*}\\
\tilde{\Omega}_{\a\b,\rho\s}^{\vp^*,h} & \tilde{\Omega}_{\a,\rho\s}^{\vp^*,A}
 & \tilde{\Omega}_{\rho\s}^{\vp^*,\vp} & \tilde{\Omega}_{\rho\s}^{\vp^*,\vp^*}\\
\tilde{\Omega}_{\a\b,\rho\s}^{\vp,h} & \tilde{\Omega}_{\a,\rho\s}^{\vp,A}
 & \tilde{\Omega}_{\rho\s}^{\vp,\vp} & \tilde{\Omega}_{\rho\s}^{\vp,\vp^*}
\end{array}\right),
\label{ot}
\eea
with the explicit formulas for all components given in appendix~\ref{appc.2}.

Since the Hessian ${\bf H}_T$ of the total system has now the minimal form (\ref{deltaform}),
we can use our master formula for FRGE (\ref{master}) with identifications: ${\bf U}=\tilde {\bf W}$ and
${\bm\Omega}_{\rho\sigma}=\tilde {\bm\Omega}_{\rho\sigma}$. For this we need to evaluate several traces
in the internal spaces of fluctuations, while the functional traces were already done in formula (\ref{master}).
Therefore we need the traces of $\tilde {\bf W}$, $\tilde {\bf W}^2$ and
$\tilde {\bm\Omega}^2=\tilde {\bm\Omega}_{\rho\sigma}\tilde {\bm\Omega}^{\rho\sigma}$.
The results for them are summarized  in appendix~\ref{appc.3}.

\section{Effective action}
\label{sec4}

Now we are ready to derive the flow equation based on Eq. (\ref{master}). In this section, we drop bars on the fields
since all quantities considered are built out of background field values and we will not need any field fluctuation.

We have one graviton, one real vector and one complex scalar (cf. (\ref{omegapsi}) and above it), so in $d=4$
\bea
\tr({\bf 1}) &=& 10+4+2=16.
\eea
From our master formula~\p{master}, we get
\bea
\pa_t\G_k
\hs{-2}&=&\hs{-2} \frac12 \frac{1}{(4\pi)^{2}} \int d^4 x \sqrt{g}\, \Big\{
16\, Q_{2}[h_k] + \Big(\frac{8}{3} R - \tr(\tilde {\bf W})\Big) Q_{1}[h_k]
+16 R_{\mu\nu}\, g_{Ric}\, R^{\mu\nu} + 16 R\, g_{R} R \nn
&& +\, R\, g_{_{RU}} \tr(\tilde {\bf W})+ g_{_{U}} \tr(\tilde {\bf W}^2)
+ g_\Omega\, \tr(\tilde{\bm\Omega}^2) + \ldots \Big\}.
\label{master2}
\eea
In addition, we have the contributions from the coordinate reparametrization (gravitational) and gauge ghosts.
The first one, based on (\ref{gravityghost}), adds to (\ref{master2})
%\bea
%&& \frac12 \frac{1}{(4\pi)^{2}} \int d^4 x \sqrt{g}\, \Big\{
%-8Q_2[h_k] -\frac{10}{3} Q_1[h_k] R
%-2 R_{\mu\nu}\, (4 g_{Ric}+g_U)\, R^{\mu\nu}
%\nn && \hs{20}
%- 2 R\, (4 g_{Ric}-g_{RU}) R - 2 R_{\mu\nu\rho\s} g_\Omega R^{\mu\nu\rho\s} \Big\}.
%\label{ghc1}
%\eea
\bea
&& \frac12 \frac{1}{(4\pi)^{2}} \int d^4 x \sqrt{g}\, \Big\{
-8Q_2[h_k] -\frac{10}{3} Q_1[h_k] R
-2 R_{\mu\nu}\, (4 g_{Ric}+g_U-4 g_\Omega)\, R^{\mu\nu}
\nn && \hs{20}
- 2 R\, (4 g_{R}-g_{RU}+ g_\Omega) R\Big\}.
\label{ghc1}
\eea
The gauge ghost contributes
\bea
&& \frac12 \frac{1}{(4\pi)^{2}} \int d^4 x \sqrt{g}\, \Big\{
-2Q_2[h_k] -\frac{1}{3} Q_1[h_k] R
-2 R_{\mu\nu}\, g_{Ric}\, R^{\mu\nu}
- 2 R\, g_{R} R\Big\}.
\label{ghc2}
\eea
Finally we have a contribution from fermions in ~\p{dcont}.
All these contribute to the FRGE flow of the bosonic effective action $\pa_t \Gamma^{1-loop}_k $.

Plugging the results of the traces of $\tilde {\bf W}, \tilde {\bf W}^2$ and $\tilde{\bm\Omega}^2$ in Eqs.~\p{trw},
\p{trw2} and \p{tro2} from appendix \ref{appc.3} into \p{master2} and collecting all these contributions
(\ref{master2})-(\ref{ghc2}) and (\ref{dcont}), we get the flow equation of the effective action
\bea
\pa_t \Gamma^{1-loop}_k = \pa_t (\G^{gem}_k +\G^{scalar}_k + \G^{fermion}_k),
\label{flow1}
\eea
where $\G^{gem}_k, \G^{scalar}_k$ and $\G^{fermion}_k$ are scale-dependent effective actions
involving gravity and EM fields, scalars and fermions, respectively. Explicitly the flow of them is given by
\bea
&&\hspace{-10mm}
\pa_t\G_k^{gem}= \frac{1}{32\pi^{2}} \int d^4 x \sqrt{g}\, \left\{
(6 -4N_F) Q_2[h_k] +\Big[\!(-8 + \frac{N_F}{3})R +4N_Fm_F^2 +\frac{3}{2} F_{\mu\nu}^2 \Big]Q_1[h_k] 
\right.
\nn &&
+ 2m_F^2 g_1 R
+m_F^2 g_{2} m_F^2
+F_{\mu\nu}g_3 F^{\mu\nu}
+F_{\mu\nu}F_{\rho\sigma}g_{4} \left(F^{\mu\rho}F^{\nu\sigma}\right)
+F_{\mu\nu}F^{\mu\rho}g_{5} \left(F^{\nu\s}F_{\rho\s}\right)
\nn &&
+F_{\mu\nu}F_{\rho\sigma}g_{6} \left(F^{\mu\nu}F^{\rho\sigma}\right)
+F_{\mu\nu}^2g_{7} F_{\rho\s}^2
+  R_{\mu\nu} g_{8} R^{\mu\nu} 
+ R g_{9} R 
+ R_{\mu\nu\rho\s} g_{10} R^{\mu\nu\rho\s}
\nn &&\left.
+ F_{\mu\nu}F^\mu{}_{\rho}g_{11} R^{\nu\rho}
+ F_{\mu\nu}^2 g_{12} R
+ \nabla_\mu F^\mu{}_\nu g_{13} \nabla_\rho F^{\nu\rho}
+ \nabla_\mu F_{\nu\rho} g_{14} \nabla^\mu F^{\nu\rho}\vphantom{\frac{N_F}{3}}\right\},
\label{f46}
\eea
where the structure functions $g_i$ (for $i=1,\ldots,14$) read
\bea
&&
g_1= -N_F(2g_{RU}+g_U), \quad
g_{2}=-4 N_F g_U, \quad
g_3= -2e_S^2 g_{\Omega} -2N_F e_F^2(g_U-2g_\Omega), \quad
g_4= 6g_{\Omega},
\nn &&
g_5= \frac{3}{4} g_U+\frac{5}{2} g_\Omega, \quad
g_6= -\frac{9}{2} g_\Omega, \quad
g_7= \frac{3}{8}g_U- g_\Omega,\quad
g_8= (6-4N_F) g_{Ric}-7g_U+8 g_\Omega, 
\nn &&
g_9= \left(5-\frac{1}{4}N_F\right)g_U+(9-N_F)g_{RU} -2g_\Omega+(6-4N_F)g_R,
\nn &&
g_{10}= 3 g_U -(7-\frac{N_F}{2}) g_{\Omega}, \qquad
g_{11}= - g_U +2 g_{\Omega}, \qquad
g_{12}= -\frac{1}{2} g_U-\frac{3}{2} g_{RU} + g_{\Omega},
\nn &&
g_{13}= -\frac{1}{2} g_U- g_\Omega, \qquad
g_{14}= \frac{3}{4} g_U- \frac{9}{2} g_\Omega,
\eea
and for the flow of the action with scalars
\bea
&&\hspace{-10mm}
\pa_t \G^{scalar}_k=\frac{1}{32\pi^{2}} \int d^4 x \sqrt{g}\, \left\{
\Big[ 10V+ 4(D_\mu\phi^*)(D^\mu\phi) -4e_S^2|\phi|^2-2V'-2|\phi|^2 V'' \Big]Q_1[h_k]
\right.\nn &&
+D_\mu D_\nu\phi^*g_{15} D^\mu D^\nu\phi
+ e_S^2 F_{\mu\nu} \phi^* g_{16} \left(\phi F^{\mu\nu}\right)
+ e_S^2 |\phi|^2 g_{17} F_{\mu\nu}^2
+ D_\mu^2 \phi^* g_{18} D_\nu^2 \phi
\nn &&
+ D_\mu\phi^* D_\nu\phi g_{19} \left(D^\mu\phi^* D^\nu\phi\right)
+ D_\mu\phi^* D_\nu\phi g_{20} \left(D^\nu\phi^* D^\mu\phi\right)
+ 2 \left(D_\mu\phi^*\right)^2 g_{U} \left(D_\nu\phi\right)^2
\nn &&
+ D_\mu\phi^* D^\mu\phi g_{21} \left(D_\nu\phi^* D^\nu\phi\right)
- 2 e_S^2 (D_\mu\phi)^2 g_{U} \phi^{*2}
- 2 e_S^2 (D_\mu\phi^*)^2  g_{U} \phi^2
+ 4 e_S^2 D_\mu\phi^* D^\mu\phi g_{U} |\phi|^2
\nn &&
+ e_S^2 \phi^* D_\mu\phi g_{22} \left(\phi^* D^\mu\phi\right)
+ e_S^2  \phi D_\mu\phi^* g_{22} \left(\phi D^\mu\phi^*\right)
+ e_S^2 \phi^* D_\mu\phi g_{23} \left( \phi D^\mu\phi^*\right)
\nn &&
+ e_S^2 \phi D_\mu\phi^* g_{23} \left(\phi^*D^\mu\phi\right)
+ e_S^4 |\phi|^2 g_{24} |\phi|^2
+ 2 e_S^4 \phi^2 g_{U} \phi^{*2}
+ 2 V' g_{U} V'
+ 10 V g_{U} V
\nn &&
+ 2 \phi V' g_{U} D_\mu^2 \phi^*
+ 2 \phi^* V' g_{U} D_\mu^2 \phi
- 8 \phi V' g_{U} \left(\phi^*V'\right)
- 4 e_S^2 |\phi|^2 g_{U} V'
- 4 e_S^2 |\phi|^2 V'' g_{U} |\phi|^2
\nn &&
+ 2 e_S^2 \phi^2 V'' g_{U} \phi^{*2}
+ 2 e_S^2 \phi^2 g_{U} \left(\phi^{*2}V''\right)
+ 2 |\phi|^2 V'' g_{U} \left( |\phi|^2 V''\right)
+ 2 \phi^2 V'' g_{U} \left(\phi^{*2}V''\right)
\nn &&
+ 4 |\phi|^2 V'' g_{U} V'
- 2 (D_\mu\phi)^2  g_{U} \left(\phi^{*2} V'' \right)
- 2 (D_\mu\phi^*)^2 g_{U} \left(\phi^2 V''\right)
- 4 D_\mu\phi^* D^\mu\phi g_{U} \left(|\phi|^2 V''\right)
\nn &&
+ 4 D_\mu\phi^* D^\mu\phi g_{U} V
- 4 D_\mu\phi^* D^\mu\phi g_{U} V'
+ e_S^2 |\phi|^2 g_{25} R
+ V g_{26} R 
+ 2 V'g_{RU} R
\nn &&
+ 2|\phi|^2 V''g_{RU} R
+ D_\mu\phi^* D^\mu\phi g_{27} R
+ 4  D_\mu\phi^* D_\nu\phi g_\Omega R^{\mu\nu}
+ i e_S \phi D_\mu\phi^* g_{17}\nabla_\nu F^{\mu\nu}
\nn &&
- i e_S \phi^*D_\mu\phi g_{17} \nabla_\nu F^{\mu\nu}
+ 4 i e_S D_\mu\phi^* D_\nu\phi g_\Omega F^{\mu\nu}
- i e_S D_\mu\phi F^{\mu\nu} g_{17} D_\nu\phi^*
\nn &&
+ i e_S D_\mu\phi^* F^{\mu\nu} g_{17} D_\nu\phi
+ D_\mu\phi^* D^\mu\phi g_{28} F_{\nu\rho}^2
-3 D_\mu\phi^* F_{\nu\rho} g_\Omega \left(D^\mu\phi F^{\nu\rho}\right)
\nn &&\left.
+ 5 D_\mu\phi^* D_\nu\phi g_{U} \left(F^{\mu}{}_\rho F^{\nu\rho}\right)
+ 6 D_\mu\phi F_{\nu\rho} g_\Omega \left(D^\nu\phi^* F^{\mu\rho}\right)
+ D_\mu\phi F^\mu{}_\nu g_{29} \left(D_\rho\phi^* F^{\nu\rho}\right)\vphantom{\int d^4x}\right\},
\label{f48}
\eea
where the structure functions $g_i$ (for $i=15,\ldots,29$) read
\bea
&&
g_{15} = 2g_U -8 g_\Omega, \qquad
g_{16} = -\frac{1}{2}g_U +g_\Omega, \qquad
g_{17} = -\frac{9}{2} g_U-3g_\Omega, \qquad
g_{18} = - g_U +2 g_\Omega,
\nn &&
g_{19} = \frac{17}{4} g_U -\frac32 g_\Omega, \qquad
g_{20} = \frac{17}{4} g_U-\frac72 g_\Omega, \qquad
g_{21} = \frac12 g_U -g_\Omega, \qquad
g_{22} = - \frac92 g_U + 3 g_\Omega,
\nn &&
g_{23} = 9 g_U-6 g_\Omega, \quad
g_{24} = 11 g_U-6 g_\Omega, \quad
g_{25} = 3 g_U + 2 g_\Omega + 4 g_{RU}, \quad
g_{26} = -12 g_U -10 g_{RU},
\nn &&
g_{27} = - 3 g_U+ 2 g_\Omega - 4 g_{RU}, \qquad
g_{28} = - \frac54 g_U -\frac32 g_\Omega, \qquad
g_{29} = - \frac94 g_U+ \frac32 g_\Omega,
\eea
and the fermionic terms are
\bea
&&\hspace{-10mm}
\partial_t\G^{fermion}_k=\frac{1}{32\pi^{2}} \int d^4 x \sqrt{g}\, \left\{\Big[3\bar\psi \gamma_\mu D^\mu \psi
 -3 D_\mu \bar\psi \gamma^\mu \psi + 10m_F \bar\psi \psi +
 \frac{3}{128} \bar\psi \gamma_{\mu\nu\rho}\psi \bar\psi \gamma^{\mu\nu\rho} \psi \Big]Q_1[h_k] \right.
\nn &&
+ \frac{25}{32} \left(\bar\psi \gamma_\mu D^\mu \psi - D_\mu \bar\psi \gamma^\mu \psi\right) g_{U}
\left(\bar\psi \gamma_\nu D^\nu \psi - D_\nu \bar\psi \gamma^\nu \psi\right)
+  6 m_F\left( \bar\psi \gamma_\mu D^\mu \psi - D_\mu \bar\psi \gamma^\mu \psi\right) g_{U} \left
(\bar\psi \psi\right)\nn &&
+ 10 m_F^2 \left(\bar\psi \psi\right) g_{U} \left(\bar\psi \psi\right)
+ \frac{21}{1024} \left(\bar\psi \gamma_\mu D^\mu \psi - D_\mu \bar\psi \gamma^\mu \psi\right) g_{U}
 \left(\bar\psi \gamma_{\nu\rho\s} \psi \bar\psi \gamma^{\nu\rho\s} \psi\right)
\nn &&
+ \frac{3}{64} m_F \bar\psi \psi g_{U}\left(\bar\psi \gamma_{\mu\nu\rho} \psi\right)^2
-\frac{3}{128} \left(\bar\psi \gamma_\mu D_\nu \psi - D_\nu \bar\psi \gamma_\mu \psi\right) g_{U}
 \left(\bar\psi \gamma^{\mu}{}_{\rho\s} \psi \bar\psi \gamma^{\nu\rho\s} \psi \right)
\nn &&
+\bar\psi \gamma_{\mu\nu\rho} \psi \bar\psi \gamma^{\mu}{}_{\s\kappa} \psi g_{30}\left(\bar\psi
 \gamma^{\nu\s}{}_{\la}
 \psi \bar\psi \gamma^{\rho\kappa\la} \psi\right)
+ \bar\psi \gamma_{\mu\nu\rho} \psi \bar\psi \gamma^{\mu}{}_{\sigma\kappa} \psi g_{31}\left( \bar
\psi \gamma^{\nu\rho}{}_\lambda \psi \bar\psi \gamma^{\sigma\kappa\lambda} \psi\right)
\nn &&
+  \frac{5}{32768}\bar\psi \gamma_{\mu\nu\rho} \psi \bar\psi \gamma^{\mu\nu}{}_{\sigma} \psi g_{U}
\left(\bar\psi \gamma^{\rho}{}_{\kappa\lambda} \psi \bar\psi \gamma^{\sigma\kappa\lambda} \psi\right)
+\frac{1}{65536}\bar\psi \gamma_{\mu\nu\rho} \psi \bar\psi \gamma^{\mu\nu\rho} \psi g_{U}
\left(\bar\psi \gamma_{\s\kappa\la} \psi \right)^2
\nn &&
+\frac{17}{64}\left(\bar\psi \gamma_\mu D_\nu \psi - D_\nu \bar\psi \gamma_\mu \psi\right)
g_{U} \left(\bar\psi \gamma^\mu D^\nu \psi - D^\nu \bar\psi \gamma^\mu \psi
+ \bar\psi \gamma^\nu D^\mu \psi - D^\mu \bar\psi \gamma^\nu \psi\right)
\nn &&
-\frac{7}{16} \left(\bar\psi \gamma_\mu D^\mu \psi - D_\mu \bar\psi \gamma^\mu \psi\right)g_{U} F_{\nu\rho}^2
+ (\bar\psi \gamma_{\mu\nu\rho} \psi)^2 g_{32} F_{\s\kappa}^2
\nn &&
-\frac{1}{32} \bar\psi \gamma_{\mu\rho\sigma} \psi \bar\psi \gamma_{\nu}{}^{\rho\sigma} \psi g_{U}\left(F^
{\mu}{}_{\kappa} F^{\nu\kappa} \right)
+ \frac{7}{4}\left(\bar\psi \gamma_\mu D_\nu \psi - D_\nu \bar\psi \gamma_\mu \psi\right) g_{U}\left(F^{\mu}
{}_{\rho} F^{\nu\rho}\right)
\nn &&
+ \frac{1}{32}\bar\psi \gamma_{\mu\rho\kappa} \psi \bar\psi \gamma_{\nu\sigma}{}^{\kappa} \psi g_{\Omega}\left( F^
{\mu\nu} F^{\rho\sigma}\right)
-\frac{1}{64} \bar\psi \gamma_{\mu\nu\rho} \psi \bar\psi \gamma^\rho{}_{\sigma\kappa} \psi g_{\Omega}\left( F^{\mu
\nu} F^{\sigma\kappa}\right)
\nn &&
-\frac{3}{256} \bar \psi \gamma_{\mu\nu\rho} \psi F_{\sigma\kappa} g_{\Omega} \left(\bar \psi \gamma^{\mu\nu\rho} 
\psi F^{\sigma\kappa} \right)
+ \frac{1}{128}\bar \psi \gamma_{\mu\nu\rho} \psi F_{\sigma\kappa} g_{\Omega} \left( \bar \psi \gamma^{\mu\nu\sigma} 
\psi F^{\rho\kappa} \right)
\nn &&
+ \bar \psi \gamma_{\mu\nu\rho} \psi F^{\rho\kappa} g_{33} \left( \bar \psi \gamma^{\mu\nu\sigma} 
\psi F_{\sigma\kappa} \right)
+\frac{5}{256} \bar \psi \gamma_{\mu\nu\rho} \psi F_{\sigma\kappa} g_{\Omega} \left( \bar \psi \gamma^{\rho\sigma
\kappa} \psi F^{\mu\nu} \right)
\nn &&
+ \bar \psi \gamma_{\mu\nu\rho} \psi F^{\mu\sigma} g_{34} \left( \bar \psi \gamma^{\rho}{}_{\sigma
\kappa} \psi F^{\nu\kappa} \right)
+ \bar \psi \gamma_{\mu\nu\rho} \psi F^{\mu\nu} g_{35} \left( \bar \psi \gamma^{\rho}{}_{\sigma
\kappa} \psi F^{\sigma\kappa} \right)
\nn &&
-\frac{3}{8}i e_F\bar\psi \gamma_{\mu} \psi  g_U\left(\bar\psi \gamma^{\mu}{}_{\nu\rho} \psi  F^{\nu\rho} \right)
+\frac{3}{64}\bar\psi \gamma_{\mu\nu\rho} \psi \bar\psi \gamma^{\mu\nu}{}_{\s} \psi g_{U} R^{\rho\s}
+ \left(\bar\psi \gamma_\mu D^\mu \psi - D_\mu \bar\psi \gamma^\mu \psi\right)g_{36} R
\nn &&
+ m_F (\bar\psi \psi) g_{37} R
+ (\bar\psi \gamma_{\mu\nu\rho} \psi)^2 g_{38} R
+ \bar\psi \gamma_{\mu\nu\rho} \psi \bar\psi \gamma^{\mu}{}_{\s\kappa} \psi g_{39} R^{\nu\rho\s\kappa}
\nn &&
+ 6 \left(\bar\psi \gamma_\mu D^\mu \psi - D_\mu \bar\psi \gamma^\mu \psi\right) g_{U} V 
+ 20 m_F \bar\psi \psi g_{U} V
+\frac{3}{64}(\bar\psi \gamma_{\mu\nu\rho} \psi)^2 g_{U} V
\nn &&
+ \bar\psi \gamma_{\mu\nu\rho} \psi F^{\mu}{}_{\sigma} g_{40}\nabla^{\s}F^{\nu\rho} 
+\frac{9}{2}e_{F}e_{S}\phi^{*} D_{\mu}\phi g_{U}\left(\bar\psi \gamma^\mu \psi \right)
-\frac{9}{2}e_{F}e_{S} \phi D_{\mu}\phi^{*} g_{U}\left(\bar\psi \gamma^\mu \psi \right)
\nn &&
+ ie_S \phi^{*} D_{\mu}\phi g_{41}\left(\bar\psi \gamma^{\mu\nu\rho} \psi F_{\nu\rho}\right)
- ie_S  \phi D_{\mu}\phi^{*} g_{41}\left(\bar\psi \gamma^{\mu}{}_{\nu\rho} \psi   F^{\nu\rho}\right)
\nn &&
+\frac{1}{8}\left(\bar\psi \gamma_\mu D^\mu \psi - D_\mu \bar\psi \gamma^\mu \psi\right) g_{U}\left(D_{\nu}
\phi^{*}{} D^{\nu}\phi \right)
+ 4m_F \bar\psi \psi g_{U} \left(D_{\mu}\phi^{*}{} D^{\mu}\phi\right)
\nn &&
+ (\bar\psi \gamma_{\mu\nu\rho} \psi)^2 g_{42}\left(D_{\s}\phi^{*}{} D^{\s}\phi\right)
-\frac{1}{64} D_\mu \phi^* \bar\psi \gamma_{\nu\rho\s} \psi g_{\Omega} \left( D^\mu \phi \bar\psi
 \gamma^{\nu\rho\s} \psi\right)
-3ie_{F} \bar\psi \gamma_{\mu} \psi g_{U} \nabla_{\nu} F^{\mu\nu}
\nn &&
+ \frac{19}{8}\left(\bar\psi \gamma_\mu D_\nu \psi - D_\nu \bar\psi \gamma_\mu \psi\right)g_{U}
\left( D^{\mu}\phi^{*}{} D^{\nu}\phi + D^{\mu}\phi D^{\nu}\phi^{*} \right)
+\bar\psi \gamma_{\mu\nu\rho}\psi \bar\psi \gamma^{\mu\nu}{}_\s \psi g_{43}\left(D^{\rho}\phi^{*}
{}D^{\s}\phi \right)
\nn &&
-\frac{1}{64} D_\mu\phi\bar\psi\gamma_{\nu\rho\sigma}\psi g_{\Omega}\left(D^\nu\phi^*\bar\psi\gamma^{\mu\rho
\sigma}\psi\right)
-\frac{1}{16} D_\mu\phi\bar\psi\gamma^{\mu}{}_{\nu\rho}\psi g_{\Omega}\left(D_\sigma\phi^*\bar\psi\gamma^{\nu\rho
\sigma}\psi\right)
\nn &&
+ \bar\psi \gamma_{\mu\nu\rho} \psi F^{\mu\nu} g_{44}\nabla_{\s}F^{\rho\s}
+\frac{3}{8}ie_S F_{\mu\nu}\phi^* g_{\Omega} \left(D_\rho\phi \bar\psi\gamma^{\mu\nu\rho}\psi \right)
-\frac{3}{8}ie_S F_{\mu\nu}\phi\, g_{\Omega} \left(D_\rho\phi^* \bar\psi\gamma^{\mu\nu\rho}\psi \right)
\nn &&
-\frac{3}{4} \nabla_\mu \left(\bar\psi\gamma_{\nu\rho\sigma}\psi\right) g_{\Omega} \left(F^{\mu\nu}F^{\rho\sigma}\right)
-\frac{3}{128} \nabla_{\mu}\left(\bar\psi \gamma^{\mu}{}_{\nu\rho} \psi\right) g_{U}\nabla_{\s}\left(\bar\psi
 \gamma^{\nu\rho\s} \psi\right)
-\frac{3}{2} \nabla_{\mu}\left(\bar\psi \gamma_{\nu\rho\sigma} \psi\right) g_{\Omega} R^{\mu\nu\rho\sigma}
\nn &&
+ \frac{3}{64}\nabla_{\mu}\left(\bar\psi \gamma_{\nu\rho\sigma} \psi\right) g_{\Omega} \nabla^{\nu}\left(\bar\psi 
\gamma^{\mu\rho\sigma} \psi\right) 
-\frac{3}{64} \nabla_{\mu}\left(\bar\psi \gamma_{\nu\rho\sigma} \psi\right) g_{\Omega} \nabla^{\mu}\left(\bar\psi 
\gamma^{\nu\rho\sigma} \psi\right)
\nn &&
+ \bar\psi \gamma_{\mu\nu\rho} \psi F^{\mu}{}_{\sigma} g_{45} \nabla^{\nu}F^{\rho\sigma}
\left.-\frac{3}{256} \bar\psi \gamma_{\mu\nu\rho} \psi \bar\psi \gamma^{\mu}{}_{\sigma\kappa} \psi g_{\Omega} \nabla^\nu 
\left( \bar\psi \gamma^{\rho\sigma\kappa} \psi  \right)\vphantom{\int d^4x}\!\right\},
\label{f410}
\eea
where the structure functions $g_i$ (for $i=30,\ldots,4$5) read
\bea
&&
g_{30}=\frac{1}{32768}g_{U}+\frac{3}{16384}g_{\Omega},\qquad
g_{31}=\frac{1}{32768}g_{U}-\frac{3}{16384}g_{\Omega},\qquad
g_{32}=\frac{1}{128}g_{U}-\frac{1}{256}g_{\Omega},
\nn &&
g_{33}=\frac{1}{512}g_{U}-\frac{13}{256}g_{\Omega},\qquad
g_{34}=\frac{1}{512}g_{U}+\frac{13}{256}g_{\Omega},\qquad
g_{35}=\frac{7}{512}g_{U}+\frac{1}{128}g_{\Omega},
\nn &&
g_{36}=-\frac{15}{4}g_{U}-3g_{RU},\qquad
g_{37}=-12g_{U}-10g_{RU},\qquad
g_{38}=-\frac{5}{128}g_{U}-\frac{3}{128}g_{RU},\qquad
\nn &&
g_{39}=-\frac{3}{128}g_{U}+\frac{3}{64}g_{\Omega},\qquad
g_{40}=\frac{1}{16}g_{U}+\frac{1}{8}g_{\Omega},\qquad
g_{41}=\frac{9}{16}g_{U}+\frac{3}{8}g_{\Omega},
\nn &&
g_{42}=\frac{9}{256}g_{U}-\frac{1}{128}g_{\Omega},\qquad
g_{43}=-\frac{3}{32}g_{U}-\frac{1}{64}g_{\Omega},\qquad
g_{44}=\frac{5}{16}g_{U}-\frac{1}{8}g_{\Omega},
\nn &&
g_{45}=-\frac{1}{16}g_{U}-\frac{1}{8}g_{\Omega}
\,.
\eea
In deriving these results, we have to be careful in making partial integration because
nonlocal operators are inserted between the factors in each of the terms above.
To know precise positions, where to insert nonlocal operators in the form of functions $g_i$,
one must distinguish between left and right tensors, that is one needs to perform traces:
${\rm tr}\left(\tilde{\bf W}_L\tilde{\bf W}_R\right)$ and
${\rm tr}\left(\tilde{\bm\Omega}_{L\rho\sigma}\tilde{\bm\Omega}_R^{\rho\sigma}\right)$.
Subsequently, one inserts $g_i$ factors acting on a bracket collecting product of all right fields.
The distinction between left and right fields here is symmetric and their roles can be reversed.
This is reflected by performing integration by parts under spacetime volume integral and
the fact that form-factors $g_i$ are functions of the integration by parts-invariant gauge-covariant operator
$z=-D_\mu^2$. We have also used Bianchi identities for the gauge field strengths and Riemann curvatures
to simplify some terms.

Now our task is to integrate the flow equation~\p{flow1} from the UV scale $\Lambda$ down to zero.
We first note that the $g_a$ functions have the general form
\bea
g_a = A_a +\left(-A_a +\frac{B_a}{\tz} +\frac{C_a}{\tz^2}\right) \sqrt{1-\frac{4}{\tz}}\, \t(\tz-4),~~~
(a=U, RU,\Omega,1,2,\ldots, 45),
\label{coef}
\eea
and the coefficients are given in Table~\ref{t1}. As we will see in a moment, to obtain the effective action $\Gamma_0$,
it is enough to focus only on $A_a$ coefficients. We can drop the term with the structure function $g_2$
since it gives a constant after spacetime integration.

\begin{table}[tb]
\begin{minipage}[cbt]{.41\linewidth}
\hspace*{-10mm}
\begin{tabular}{|l|c|c|r|}
\hline
$a$ &  $A_a$ & $B_a$ & $C_a$ \\
\hline
$U$ &  $1$ & $0$ & $0$ \\
\hline
$RU$ &  $-\frac{1}{3}$ & $\frac{2}{3}$ & $0$ \\
\hline
$\Omega$ &  $\frac{1}{6}$ & $\frac{2}{3}$ & $0$ \\
\hline
$1$ &  $-\frac{N_F}{3}$ & $-\frac{4}{3}N_F$ & $0$ \\
\hline
$2$ & $-4N_F$ & $0$ & $0$ \\
\hline
$3$ &  $-\frac{e_S^2+4e_F^2 N_F}{3}$ & $\frac{-4 e_S^2+8e_F^2 N_F}{3}$ & $0$ \\
\hline
$4$ &  $1$ & $4$ & $0$ \\
\hline
$5$ &  $\frac{7}{6}$ & $\frac{5}{3}$ & $0$ \\
\hline
$6$ & $-\frac{3}{4}$ & $-3$ & $0$ \\
\hline
$7$ &  $\frac{5}{24}$ & $-\frac{2}{3}$ & $0$ \\
\hline
$8$ &  $-\frac{82+2N_F}{15}$ & $\frac{104-16N_F}{15} $ & $\frac{-48+32N_F}{15}$ \\
\hline
$9$ &  $\frac{106+N_F}{60}$ & $\frac{52+2N_F}{15} $ & $\frac{6-4N_F}{15}$ \\
\hline
$10$ & $\frac{22+N_F}{12}$ & $\frac{-14+N_F}{3} $ & $0$ \\
\hline
$11$ & $-\frac{2}{3}$ & $\frac{4}{3}$ & $0$ \\
\hline
$12$ & $\frac{1}{6}$ & $-\frac{1}{3}$ & $0$ \\
\hline
$13$ &  $-\frac{2}{3}$ & $-\frac{2}{3}$ & $0$ \\
\hline
\end{tabular}
\end{minipage}
\begin{minipage}[cbt]{.28\linewidth}
\begin{center}
\hs{3}
\begin{tabular}{|l|r|r|r|}
\hline
$a$ &  $A_a$ & $B_a$ & $C_a$ \\
\hline
$14$ &  $0$ & $-3$ & $0$ \\
\hline
$15$ &  $\frac{2}{3} $ & $-\frac{16}{3} $ & $0$ \\
\hline
$16$ &  $-\frac{1}{3}$ & $\frac{2}{3}$ & $0$ \\
\hline
$17$ &  $-5$ & $-2$ & $0$ \\
\hline
$18$ &  $-\frac{2}{3}$ & $\frac{4}{3}$ & $0$ \\
\hline
$19$ &  $4$ & $-1$ & $0$ \\
\hline
$20$ &  $\frac{11}{3} $ & $-\frac{7}{3}$ & $0$ \\
\hline
$21$ &  $\frac{1}{3}$ & $-\frac{2}{3}$ & $0$ \\
\hline
$22$ &  $-4$ & $2$ & $0$ \\
\hline
$23$ &  $8$ & $-4$ & $0$ \\
\hline
$24$ &  $10$ & $-4$ & $0$ \\
\hline
$25$ &  $2$ & $4$ & $0$ \\
\hline
$26$ &  $-\frac{26}{3}$ & $-\frac{20}{3}$ & $0$ \\
\hline
$27$ &  $-\frac{4}{3}$ & $-\frac{4}{3}$ &0$ $ \\
\hline
$28$ &  $-\frac{3}{2}$ & $-1$ & $0$ \\
\hline
$29$ &  $-2$ & $1$ & $0$ \\
\hline
\end{tabular}
\end{center}
\end{minipage}
\begin{minipage}[cbt]{.28\linewidth}
\begin{center}
\begin{tabular}{|l|r|r|r|}
\hline
$a$ &  $A_a$ & $B_a$ & $C_a$ \\
\hline
$30$ &  $\frac{1}{16384}$ & $\frac{1}{8192}$ & $0$ \\
\hline
$31$ &  $0$ & $-\frac{1}{8192}$ & $0$ \\
\hline
$32$ &  $\frac{11}{1536}$ & $-\frac{1}{384}$ & $0$ \\
\hline
$33$ &  $-\frac{5}{768}$ & $-\frac{13}{384}$ & $0$ \\
\hline
$34$ &  $\frac{1}{96}$ & $\frac{13}{384}$ & $0$ \\
\hline
$35$ &   $\frac{23}{1536}$ & $\frac{1}{192}$ & $0$ \\
\hline
$36$ &   $-\frac{11}{4}$ & $-2$ & $0$ \\
\hline
$37$ &  $-\frac{26}{3}$ & $-\frac{20}{3}$ & $0$ \\
\hline
$38$ &   $-\frac{1}{32}$ & $-\frac{1}{64}$ & $0$ \\
\hline
$39$ &   $-\frac{1}{64}$ & $\frac{1}{32}$ & $0$\\
\hline
$40$ &  $\frac{1}{12}$ & $\frac{1}{12}$ & $0$ \\
\hline
$41$ &  $\frac{5}{8}$ & $\frac{1}{4}$ & $0$ \\
\hline
$42$ &  $\frac{13}{384}$ & $-\frac{1}{192}$ & $0$ \\
\hline
$43$ &  $-\frac{37}{384}$ & $-\frac{1}{96}$ & $0$ \\
\hline
$44$ &  $\frac{7}{24}$ & $-\frac{1}{12}$ & $0$ \\
\hline
$45$ &  $-\frac{1}{12}$ & $-\frac{1}{12}$ & $0$ \\
\hline
\end{tabular}
\end{center}
\end{minipage}
\caption{Coefficients $A_a$, $B_a$ and $C_a$ in $g_a$ in ~\p{coef} for $a=U, RU,\Omega,1,2,\ldots, 45$.}
\label{t1}
\end{table}

We can integrate the flow equation for $g_a$ over $k$ from $k=0$ to $k=\Lambda$ to obtain
\bea
\int_0^\Lambda g_a(\tz) \frac{dk}{k} \equiv G_{a,\Lambda}(z) -G_{a,0}(z),
\label{nonl1}
\eea
where
\bea
G_{a,\Lambda}(z) -G_{a,0}(z)
= \frac{A_a}{2} \log \frac{\Lambda^2}{z} + A_a 
+ \frac{B_a}{12} 
+ \frac{C_a}{120}.
\label{nonl2}
\eea

We find for large $\Lambda$, for gravity and electromagnetism (cf. (\ref{f46}))
\bea
\G_\Lambda^{gem} &&\hspace{-5mm}=
 \frac{1}{32\pi^{2}} \int d^4 x \sqrt{g}\, \left\{
(3 -2N_F) \frac{\Lambda^4}{2} +\Big[(-8 + \frac{N_F}{3})R +4N_Fm_F^2 +\frac{3}{2} F_{\mu\nu}^2 \Big]\Lambda^2\right.
\nn &&
+ \frac{1}{2} \Big[ \frac{2N_F}{3} m_F^2 R
+ 4 N_F m_F^4
- \frac{e_S^2+4e_F^2 N_F}{3} F_{\mu\nu}^2
+ \frac{13}{6} F_{\mu\nu}F^{\mu\rho} F^{\nu\s}F_{\rho\s}
- \frac{13}{24} (F_{\mu\nu}^2)^2
\nn &&
- \frac{82+2N_F}{15} R_{\mu\nu}^2 
+ \frac{106+N_F}{60} R^2
+ \frac{22+N_F}{12} R_{\mu\nu\rho\s}^2
- \frac{2}{3} F_{\mu\nu}F^\mu{}_{\rho} R^{\nu\rho}
\nn &&\left.
+ \frac{1}{6} F_{\mu\nu}^2 R
- \frac{2}{3} \nabla_\mu F^\mu{}_\nu \nabla_\rho F^{\nu\rho} \Big] \log\left(\frac{\Lambda^2}{\mu^2}\right)
\right\},
\label{e1}
\eea
for scalar sector (cf. (\ref{f48}))
\bea
\G^{scalar}_\Lambda &&\hspace{-6mm} =\frac{1}{32\pi^{2}} \int d^4 x \sqrt{g}\, \left\{
\Big[ 10V+ 4(D_\mu\phi^*)(D^\mu\phi) -4e_S^2|\phi|^2-2V'-2|\phi|^2 V'' \Big] \Lambda^2
\right.\nn &&
+ \frac{1}{2}\Big[ \frac{2}{3} D_\mu D_\nu\phi^* D^\mu D^\nu\phi
- \frac{16}{3} e_S^2 |\phi|^2 F_{\mu\nu}^2
- \frac{2}{3} D_\mu^2 \phi^* D_\nu^2 \phi
+ 6 (D_\mu\phi^*)^2 (D_\nu\phi)^2
\nn &&
+ 4 D_\mu\phi^* D^\mu\phi \left(D_\nu\phi^* D^\nu\phi\right)
- 6 e_S^2 \phi^{*2} (D_\mu\phi)^2
- 6 e_S^2 \phi^2 (D_\mu\phi^*)^2
+ 12 e_S^2 |\phi|^2 D_\mu\phi^* D^\mu\phi
\nn &&
+ 8 e_S^2 D_\mu\phi^* D^\mu\phi
+ 12 e_S^4 |\phi|^4
+ 2 (V')^2
+ 10 V^2
+ 2 \phi V' D_\mu^2 \phi^*
+ 2 \phi^* V' D_\mu^2 \phi
\nn &&
- 8 |\phi|^2 (V')^2
- 4 e_S^2 |\phi|^2 V'
+ 4 |\phi|^4 (V'')^2
+ 4 |\phi|^2 V'' V'
- 2 \phi^{*2} (D_\mu\phi)^2 V''
\nn &&
- 2 \phi^2 (D_\mu\phi^*)^2 V''
- 4 |\phi|^2 D_\mu\phi^* D^\mu\phi V''
+ 4 D_\mu\phi^* D^\mu\phi (V- V')
+ 2 e_S^2 |\phi|^2 R
\nn &&
- \frac{26}{3} V R 
- \frac{2}{3} V' R
-\frac{2}{3} |\phi|^2 V'' R
-\frac{4}{3} D_\mu\phi^* D^\mu\phi R
+ \frac{2}{3} D_\mu\phi^* D_\nu\phi R^{\mu\nu}
\nn &&
- 5 i e_S \phi D_\mu\phi^* \nabla_\nu F^{\mu\nu}
+ 5 i e_S \phi^*D_\mu\phi \nabla_\nu F^{\mu\nu}
- \frac{28}{3} i e_S D_\mu\phi^* D_\nu\phi F^{\mu\nu}
\nn &&\left.
- 2 D_\mu\phi^* D^\mu\phi F_{\nu\rho}^2
+ 8 D_\mu\phi^* D_\nu\phi \left(F^{\mu}{}_\rho F^{\nu\rho}\right) \Big] \log\left(\frac{\Lambda^2}{\mu^2}\right)
\right\},
\label{e2}
\eea
and for fermionic sector (cf. (\ref{f410}))
\bea
&&\hspace{-10mm}
\G^{fermion}_\Lambda=\frac{1}{32\pi^{2}} \int d^4 x \sqrt{g}\, \left\{\Big[3\bar\psi \gamma_\mu D^\mu \psi
 -3 D_\mu \bar\psi \gamma^\mu \psi + 10m_F \bar\psi \psi +
 \frac{3}{128} \bar\psi \gamma_{\mu\nu\rho}\psi \bar\psi \gamma^{\mu\nu\rho} \psi \Big] \Lambda^2 \right.
\nn &&
+ \frac{1}{2}\Big[ \frac{25}{32} \left(\bar\psi \gamma_\mu D^\mu \psi - D_\mu \bar\psi \gamma^\mu \psi\right)^2
+  6 m_F\left( \bar\psi \gamma_\mu D^\mu \psi - D_\mu \bar\psi \gamma^\mu \psi\right)\left(\bar\psi \psi\right)
+ 10 m_F^2 \left(\bar\psi \psi\right)^2
\nn &&
+ \frac{21}{1024} \left(\bar\psi \gamma_\mu D^\mu \psi - D_\mu \bar\psi \gamma^\mu \psi\right)
 \left(\bar\psi \gamma_{\nu\rho\s} \psi \bar\psi \gamma^{\nu\rho\s} \psi\right)
+ \frac{3}{64} m_F \bar\psi \psi \left(\bar\psi \gamma_{\mu\nu\rho} \psi \right)^2
\nn &&
-\frac{3}{128} \left(\bar\psi \gamma_\mu D_\nu \psi - D_\nu \bar\psi \gamma_\mu \psi\right)
 \left(\bar\psi \gamma^{\mu}{}_{\rho\s} \psi \bar\psi \gamma^{\nu\rho\s} \psi \right)
+\frac{1}{16384} \bar\psi \gamma_{\mu\nu\rho} \psi \bar\psi \gamma^{\mu}{}_{\s\kappa}\psi
\left(\bar\psi \gamma^{\nu\s}{}_{\la} \psi \bar\psi \gamma^{\rho\kappa\la} \psi\right)
\nn &&
+  \frac{5}{32768}\bar\psi \gamma_{\mu\nu\rho} \psi \bar\psi \gamma^{\mu\nu}{}_{\sigma} \psi
\left(\bar\psi \gamma^{\rho}{}_{\kappa\lambda} \psi \bar\psi \gamma^{\sigma\kappa\lambda} \psi\right)
+\frac{1}{65536}(\bar\psi \gamma_{\mu\nu\rho} \psi)^2
\left(\bar\psi \gamma_{\s\kappa\la} \psi \right)^2
\nn &&
+\frac{17}{64}\left(\bar\psi \gamma_\mu D_\nu \psi - D_\nu \bar\psi \gamma_\mu \psi\right)^2
+\frac{17}{64}\left(\bar\psi \gamma_\mu D_\nu \psi - D_\nu \bar\psi \gamma_\mu \psi\right)
\left(\bar\psi \gamma^\nu D^\mu \psi - D^\mu \bar\psi \gamma^\nu \psi\right)
\nn &&
-\frac{7}{16} \left(\bar\psi \gamma_\mu D^\mu \psi - D_\mu \bar\psi \gamma^\mu \psi\right) F_{\nu\rho}^2
+ \frac{1}{192} \bar\psi \gamma_{\mu\nu\rho} \psi \bar\psi \gamma^{\mu\nu\rho} \psi F_{\s\kappa}^2
-\frac{7}{192} \bar\psi \gamma_{\mu\rho\sigma} \psi \bar\psi \gamma_{\nu}{}^{\rho\sigma} \psi
\left(F^{\mu}{}_{\kappa} F^{\nu\kappa} \right)
\nn &&
+ \frac{7}{4}\left(\bar\psi \gamma_\mu D_\nu \psi - D_\nu \bar\psi \gamma_\mu \psi\right)
\left(F^{\mu}{}_{\rho} F^{\nu\rho}\right)
+ \frac{1}{64}\bar\psi \gamma_{\mu\rho\kappa} \psi \bar\psi \gamma_{\nu\sigma}{}^{\kappa} \psi
\left( F^{\mu\nu} F^{\rho\sigma}\right)
\nn &&
+\frac{1}{64} \bar\psi \gamma_{\mu\nu\rho} \psi \bar\psi \gamma^\mu{}_{\sigma\kappa} \psi
\left( F^{\nu\rho} F^{\sigma\kappa}\right)
-\frac{3}{8}i e_F\bar\psi \gamma_{\mu} \psi \left(\bar\psi \gamma^{\mu}{}_{\nu\rho} \psi  F^{\nu\rho} \right)
+\frac{3}{64}\bar\psi \gamma_{\mu\nu\rho} \psi \bar\psi \gamma^{\mu\nu}{}_{\s} \psi R^{\rho\s}
\nn &&
-\frac{11}{4} \left(\bar\psi \gamma_\mu D^\mu \psi - D_\mu \bar\psi \gamma^\mu \psi\right) R
-\frac{26}{3} m_F \bar\psi \psi R
- \frac{1}{32} \bar\psi \gamma_{\mu\nu\rho} \psi \bar\psi \gamma^{\mu\nu\rho} \psi R
\nn &&
- \frac{1}{64} \bar\psi \gamma_{\mu\nu\rho} \psi \bar\psi \gamma^{\mu}{}_{\s\kappa} \psi R^{\nu\rho\s\kappa}
+ 6 \left(\bar\psi \gamma_\mu D^\mu \psi - D_\mu \bar\psi \gamma^\mu \psi\right) V 
+ 20 m_F \bar\psi \psi V
\nn &&
+\frac{3}{64} (\bar\psi \gamma_{\mu\nu\rho} \psi)^2 V
+ \frac{1}{12} \bar\psi \gamma_{\mu\nu\rho} \psi F^{\mu}{}_{\sigma}\nabla^{\s}F^{\nu\rho} 
+\frac{9}{2}e_{F}e_{S}\phi^{*} D_{\mu}\phi \left(\bar\psi \gamma^\mu \psi \right)
\nn &&
-\frac{9}{2}e_{F}e_{S} \phi D_{\mu}\phi^{*} \left(\bar\psi \gamma^\mu \psi \right)
+ \frac{11}{16}ie_S \phi^{*} D_{\mu}\phi \left(\bar\psi \gamma^{\mu\nu\rho} \psi F_{\nu\rho}\right)
- \frac{11}{16}ie_S \phi D_{\mu}\phi^{*} \left(\bar\psi \gamma^{\mu\nu\rho} \psi F_{\nu\rho}\right)
\nn &&
+\frac{1}{8}\left(\bar\psi \gamma_\mu D^\mu \psi - D_\mu \bar\psi \gamma^\mu \psi\right) \left(D_{\nu}
\phi^{*}{} D^{\nu}\phi \right)
+ 4m_F \bar\psi \psi \left(D_{\mu}\phi^{*}{} D^{\mu}\phi\right)
\nn &&
+ \frac{1}{32} \bar\psi \gamma_{\mu\nu\rho} \psi \bar\psi \gamma^{\mu\nu\rho} \psi\left(D_{\s}\phi^{*}D^{\s}\phi\right)
-3ie_{F} \bar\psi \gamma_{\mu} \psi \nabla_{\nu} F^{\mu\nu}
\nn &&
+ \frac{19}{8}\left(\bar\psi \gamma_\mu D_\nu \psi - D_\nu \bar\psi \gamma_\mu \psi\right)
\left( D^{\mu}\phi^{*}{} D^{\nu}\phi \right)
+ \frac{19}{8}\left(\bar\psi \gamma_\mu D_\nu \psi - D_\nu \bar\psi \gamma_\mu \psi\right)
\left( D^{\nu}\phi^{*}{} D^{\mu}\phi \right)
\nn &&
-\frac{7}{64} \bar\psi \gamma_{\mu\nu\rho}\psi \bar\psi \gamma^{\mu\nu}{}_\s \psi \left(D^{\rho}\phi^{*}
{}D^{\s}\phi \right)
+ \frac{7}{24} \bar\psi \gamma_{\mu\nu\rho} \psi F^{\mu\nu} \nabla_{\s}F^{\rho\s}
-\frac{1}{8} \nabla_\mu \left(\bar\psi\gamma_{\nu\rho\sigma}\psi\right) \left(F^{\mu\nu}F^{\rho\sigma}\right)
\nn &&
-\frac{3}{128} \nabla_{\mu}\left(\bar\psi \gamma^{\mu}{}_{\nu\rho} \psi\right)\nabla_{\s}\left(\bar\psi
 \gamma^{\nu\rho\s} \psi\right)
-\frac{1}{4} \nabla_{\mu}\left(\bar\psi \gamma_{\nu\rho\sigma} \psi\right) R^{\mu\nu\rho\sigma}
\nn &&
+ \frac{1}{128}\nabla_{\mu}\left(\bar\psi \gamma_{\nu\rho\sigma} \psi\right) \nabla^{\nu}\left(\bar\psi 
\gamma^{\mu\rho\sigma} \psi\right) 
-\frac{1}{128} \nabla_{\mu}\left(\bar\psi \gamma_{\nu\rho\sigma} \psi\right) \nabla^{\mu}\left(\bar\psi 
\gamma^{\nu\rho\sigma} \psi\right)
\nn &&
-\frac{1}{12} \bar\psi \gamma_{\mu\nu\rho} \psi F^{\mu}{}_{\sigma} \nabla^{\nu}F^{\rho\sigma}
\left.-\frac{1}{512} \bar\psi \gamma_{\mu\nu\rho} \psi \bar\psi \gamma^{\mu}{}_{\sigma\kappa} \psi \nabla^\nu 
\left( \bar\psi \gamma^{\rho\sigma\kappa} \psi  \right) \Big] \log\left(\frac{\Lambda^2}{\mu^2}\right) \right\},
\label{e3}
\eea
where we have introduced an arbitrary renormalization scale $\mu$.
These UV-divergences (when $\Lambda\to\infty$) can be subtracted by choosing $G_{a,\Lambda}$ in (\ref{nonl1})
to eliminate the $\Lambda$-dependence.
We set
\bea
G_{a,\Lambda}=\frac{A_a}{2} \log\left(\frac{\Lambda^2}{\mu^2}\right)+\c_a,
\label{f418}
\eea
where $\c_a$ are arbitrary constants. Then the form-factors in \p{nonl2} are
\bea
G_{a,0}=\frac{A_a}{2} \log\left(\frac{z}{\mu^2}\right)+\c_a- \left(A_a+\frac{B_a}{12}+\frac{C_a}{120}\right).
\label{f419}
\eea
Thus the effective action has the local parts~\p{e1}, \p{e2} and \p{e3} with the divergent coefficients replaced by
arbitrary constants, and the following nonlocal parts with {\it definite} and universal coefficients:
\bea
\G_0=\G_0^{gem}+\G_0^{scalar}+\G_0^{fermion}\,.
\label{g0sum}
\eea

We find the following three pieces of the effective action:
\bea
&&\hspace{-10mm}
\G_0^{gem}= \frac{1}{32\pi^{2}} \int d^4 x \sqrt{g}\, \left\{
%\frac{N_F}{3}m_F^2 \log\left(\frac{-\nabla^2}{\mu^2}\right) R
-\frac{e_S^2+4e_F^2 N_F}{6} F_{\mu\nu} \log\left(\frac{-\nabla^2}{\mu^2}\right) F^{\mu\nu}
\right. \nn &&
+ \frac{1}{2} F_{\mu\nu}F_{\rho\sigma} \log\left(\frac{-\nabla^2}{\mu^2}\right) \left(F^{\mu\rho}F^{\nu\sigma}\right)
+ \frac{7}{12} F_{\mu\nu}F^{\mu\rho} \log\left(\frac{-\nabla^2}{\mu^2}\right) \left(F^{\nu\s}F_{\rho\s}\right)
\nn &&
- \frac{3}{8} F_{\mu\nu}F_{\rho\sigma} \log\left(\frac{-\nabla^2}{\mu^2}\right) \left(F^{\mu\nu}F^{\rho\sigma}\right)
+ \frac{5}{48}F_{\mu\nu}^2 \log\left(\frac{-\nabla^2}{\mu^2}\right) F_{\mu\nu}^2
\nn &&
- \frac{41+N_F}{15} R_{\mu\nu} \log\left(\frac{-\nabla^2}{\mu^2}\right) R^{\mu\nu} 
+ \frac{106+N_F}{120} R \log\left(\frac{-\nabla^2}{\mu^2}\right) R
% version with Riem log Riem eliminated
%+\frac{28+3N_F}{30} R_{\mu\nu} \log\left(\frac{-\nabla^2}{\mu^2}\right) R^{\mu\nu} 
%- \frac{3+N_F}{30} R \log\left(\frac{-\nabla^2}{\mu^2}\right) R
\nn &&
+ \frac{22+N_F}{24} R_{\mu\nu\rho\s} \log\left(\frac{-\nabla^2}{\mu^2}\right) R^{\mu\nu\rho\s}
% no above line if Riem log Riem eliminated
- \frac{1}{3} F_{\mu\nu}F^\mu{}_{\rho} \log\left(\frac{-\nabla^2}{\mu^2}\right) R^{\nu\rho}
\nn &&\left.
+ \frac{1}{12} F_{\mu\nu}^2 \log\left(\frac{-\nabla^2}{\mu^2}\right) R
-\frac{1}{3} \nabla_\mu F^\mu{}_\nu \log\left(\frac{-\nabla^2}{\mu^2}\right) \nabla_\rho F^{\nu\rho}
\right\},
\label{f421}
\eea
\bea
&&\hspace{-5mm}
\G^{scalar}_0 = \frac{1}{32\pi^{2}} \int d^4 x \sqrt{g}\, \left\{
\frac{1}{3} D_\mu D_\nu\phi^* \log\left(\frac{-D^2}{\mu^2}\right) D^\mu D^\nu\phi
-\frac{e_S^2}{6} F_{\mu\nu} \phi^* \log\left(\frac{-D^2}{\mu^2}\right) \left(\phi F^{\mu\nu}\right)
\right.\nn &&
- \frac{5e_S^2}{2} |\phi|^2 \log\left(\frac{-D^2}{\mu^2}\right) F_{\mu\nu}^2
-\frac{1}{3} D_\mu^2 \phi^* \log\left(\frac{-D^2}{\mu^2}\right) D_\nu^2 \phi
\nn &&
+ 2 D_\mu\phi^* D_\nu\phi \log\left(\frac{-D^2}{\mu^2}\right) \left(D^\mu\phi^* D^\nu\phi\right)
+ \frac{11}{6} D_\mu\phi^* D_\nu\phi \log\left(\frac{-D^2}{\mu^2}\right) \left(D^\nu\phi^* D^\mu\phi\right)
\nn &&
+ \left(D_\mu\phi^*\right)^2 \log\left(\frac{-D^2}{\mu^2}\right) \left(D_\nu\phi\right)^2
+ \frac{1}{6} D_\mu\phi^* D^\mu\phi \log\left(\frac{-D^2}{\mu^2}\right) \left(D_\nu\phi^* D^\nu\phi\right)
\nn &&
- e_S^2 (D_\mu\phi)^2 \log\left(\frac{-D^2}{\mu^2}\right) \phi^{*2}
- e_S^2 (D_\mu\phi^*)^2 \log\left(\frac{-D^2}{\mu^2}\right) \phi^2
\nn &&
+ 2 e_S^2 D_\mu\phi^* D^\mu\phi \log\left(\frac{-D^2}{\mu^2}\right) |\phi|^2
-2 e_S^2 \phi^* D_\mu\phi \log\left(\frac{-D^2}{\mu^2}\right) \left(\phi^* D^\mu\phi\right)
\nn &&
-2 e_S^2  \phi D_\mu\phi^* \log\left(\frac{-D^2}{\mu^2}\right) \left(\phi D^\mu\phi^*\right)
+4 e_S^2 \phi^* D_\mu\phi \log\left(\frac{-D^2}{\mu^2}\right) \left( \phi D^\mu\phi^*\right)
\nn &&
+4 e_S^2 D_\mu\phi^* \log\left(\frac{-D^2}{\mu^2}\right) D^\mu\phi
+5 e_S^4 |\phi|^2 \log\left(\frac{-D^2}{\mu^2}\right) |\phi|^2
+ e_S^4 \phi^2 \log\left(\frac{-D^2}{\mu^2}\right) \phi^{*2}
\nn &&
+ V' \log\left(\frac{-D^2}{\mu^2}\right) V'
+ 5 V \log\left(\frac{-D^2}{\mu^2}\right) V
+ \phi V' \log\left(\frac{-D^2}{\mu^2}\right) D_\mu^2 \phi^*
\nn &&
+ \phi^* V' \log\left(\frac{-D^2}{\mu^2}\right) D_\mu^2 \phi
- 4 \phi V' \log\left(\frac{-D^2}{\mu^2}\right) \left(\phi^*V'\right)
- 2 e_S^2 |\phi|^2 \log\left(\frac{-D^2}{\mu^2}\right) V'
\nn &&
- 2 e_S^2 |\phi|^2 V'' \log\left(\frac{-D^2}{\mu^2}\right) |\phi|^2
+ e_S^2 \phi^2 V'' \log\left(\frac{-D^2}{\mu^2}\right) \phi^{*2}
+ e_S^2 \phi^2 \log\left(\frac{-D^2}{\mu^2}\right) \left(\phi^{*2}V''\right)
\nn &&
+ |\phi|^2 V'' \log\left(\frac{-D^2}{\mu^2}\right) \left( |\phi|^2 V''\right)
+ \phi^2 V'' \log\left(\frac{-D^2}{\mu^2}\right) \left(\phi^{*2}V''\right)
+ 2 |\phi|^2 V'' \log\left(\frac{-D^2}{\mu^2}\right) V'
\nn &&
- (D_\mu\phi)^2 \log\left(\frac{-D^2}{\mu^2}\right) \left(\phi^{*2} V'' \right)
- (D_\mu\phi^*)^2 \log\left(\frac{-D^2}{\mu^2}\right) \left(\phi^2 V''\right)
\nn &&
- 2 D_\mu\phi^* D^\mu\phi \log\left(\frac{-D^2}{\mu^2}\right) \left(|\phi|^2 V''\right)
+ 2 D_\mu\phi^* D^\mu\phi \log\left(\frac{-D^2}{\mu^2}\right) V
\nn &&
- 2 D_\mu\phi^* D^\mu\phi \log\left(\frac{-D^2}{\mu^2}\right) V'
+ e_S^2 |\phi|^2 \log\left(\frac{-D^2}{\mu^2}\right) R
-\frac{13}{3} V \log\left(\frac{-D^2}{\mu^2}\right) R 
\nn &&
-\frac{1}{3} V' \log\left(\frac{-D^2}{\mu^2}\right) R
-\frac{1}{3} |\phi|^2 V'' \log\left(\frac{-D^2}{\mu^2}\right) R
-\frac{2}{3} D_\mu\phi^* D^\mu\phi \log\left(\frac{-D^2}{\mu^2}\right) R
\nn &&
+ \frac{1}{3} D_\mu\phi^* D_\nu\phi \log\left(\frac{-D^2}{\mu^2}\right) R^{\mu\nu}
-\frac{5i e_S}{2} \phi D_\mu\phi^*\log\left(\frac{-D^2}{\mu^2}\right) \nabla_\nu F^{\mu\nu}
\nn &&
+ \frac{5i e_S}{2} \phi^*D_\mu\phi \log\left(\frac{-D^2}{\mu^2}\right) \nabla_\nu F^{\mu\nu}
+ \frac{1}{3} i e_S D_\mu\phi^* D_\nu\phi \log\left(\frac{-D^2}{\mu^2}\right) F^{\mu\nu}
\nn &&
+\frac{5}{2} i e_S D_\mu\phi F^{\mu\nu} \log\left(\frac{-D^2}{\mu^2}\right) D_\nu\phi^*
-\frac{5i e_S}{2} D_\mu\phi^* F^{\mu\nu} \log\left(\frac{-D^2}{\mu^2}\right) D_\nu\phi
\nn &&
-\frac{3}{4} D_\mu\phi^* D^\mu\phi \log\left(\frac{-D^2}{\mu^2}\right) F_{\nu\rho}^2
-\frac{1}{4} D_\mu\phi^* F_{\nu\rho} \log\left(\frac{-D^2}{\mu^2}\right) \left(D^\mu\phi F^{\nu\rho}\right)
\nn &&
+ \frac{5}{2} D_\mu\phi^* D_\nu\phi \log\left(\frac{-D^2}{\mu^2}\right) \left(F^{\mu}{}_\rho F^{\nu\rho}\right)
+ \frac{1}{2} D_\mu\phi F_{\nu\rho} \log\left(\frac{-D^2}{\mu^2}\right) \left(D^\nu\phi^* F^{\mu\rho}\right)
\nn &&\left.
- D_\mu\phi F^\mu{}_\nu \log\left(\frac{-D^2}{\mu^2}\right) \left(D_\rho\phi^* F^{\nu\rho}\right)
\right\},
\label{f422}
\eea
and
\bea
&&\hspace{-5mm}
\G^{fermion}_0 = \frac{1}{64\pi^{2}} \int d^4 x \sqrt{g}\, \left\{ \frac{25}{32}\left(\bar\psi \gamma_\mu D^\mu \psi
 - D_\mu \bar\psi \gamma^\mu \psi\right) \log\left(\frac{-D^2}{\mu^2}\right)
 \left(\bar\psi \gamma_\mu D^\mu \psi - D_\mu \bar\psi \gamma^\mu \psi\right)
\right.\nn &&
+6 m_F\left( \bar\psi \gamma_\mu D^\mu \psi - D_\mu \bar\psi \gamma^\mu \psi\right) \log\left(\frac{-D^2}{\mu^2}\right) 
\left(\bar\psi \psi\right)
+10 m_F^2 \left(\bar\psi \psi\right) \log\left(\frac{-D^2}{\mu^2}\right) \left(\bar\psi \psi\right)
\nn &&
+\frac{21}{1024} (\bar\psi\c_\mu D^\mu\psi-D_\mu\bar\psi\c^\mu\psi) \log\left(\frac{-D^2}{\mu^2}\right) 
(\bar\psi \gamma_{\mu\nu\rho} \psi)^2
+\frac{3}{64} m_F \bar\psi \psi \log\left(\frac{-D^2}{\mu^2}\right) 
(\bar\psi \gamma_{\mu\nu\rho} \psi)^2
\nn &&
-\frac{3}{128} \left(\bar\psi \gamma_\mu D_\nu \psi - D_\nu \bar\psi \gamma_\mu \psi\right) \log\left(\frac{-D^2}{\mu^2}
\right) \left(\bar\psi \gamma^{\mu}{}_{\rho\s} \psi \bar\psi \gamma^{\nu\rho\s} \psi \right)
\nn &&
+\frac{1}{16384}\bar\psi \gamma_{\mu\nu\rho} \psi \bar\psi \gamma^{\mu}{}_{\s\kappa}\psi \log\left(\frac{-D^2}{\mu^2}\right)
 \left(\bar\psi \gamma^{\nu\s}{}_{\la} \psi \bar\psi \gamma^{\rho\kappa\la} \psi\right)
\nn &&
+\frac{5}{32768}  \bar\psi \gamma_{\mu\nu\rho} \psi \bar\psi \gamma^{\mu\nu}{}_{\sigma} \psi
 \log\left(\frac{-D^2}{\mu^2}\right) \left(\bar\psi \c^{\rho}{}_{\kappa\la} \psi \bar\psi \c^{\s\kappa\la} \psi\right)
\nn &&
+\frac{1}{65536}\bar\psi \gamma_{\mu\nu\rho} \psi \bar\psi \c^{\mu\nu\rho} \psi \log\left(\frac{-D^2}{\mu^2}\right)
(\bar\psi \gamma_{\s\kappa\la} \psi)^2
\nn &&
+\frac{17}{64}\left(\bar\psi \gamma_\mu D_\nu \psi - D_\nu \bar\psi \gamma_\mu \psi\right)
 \log\left(\frac{-D^2}{\mu^2}\right) \left(\bar\psi \gamma_\mu D_\nu \psi - D_\nu \bar\psi \gamma_\mu \psi
+\bar\psi \gamma_\nu D_\mu \psi - D_\mu \bar\psi \gamma_\nu \psi\right)
\nn &&
-\frac{7}{16} \left(\bar\psi \gamma_\mu D^\mu \psi - D_\mu \bar\psi \gamma^\mu \psi\right) \log\left(\frac{-D^2}{\mu^2}
\right) F_{\nu\rho}^2
\nn &&
+\frac{11}{1536} (\bar\psi \gamma_{\mu\nu\rho} \psi)^2 \log\left(\frac{-D^2}{\mu^2}\right)
\left( F_{\s\kappa} F^{\s\kappa} \right)
-\frac{1}{32} \bar\psi \gamma_{\mu\rho\sigma} \psi \bar\psi \gamma_{\nu}{}^{\rho\sigma} \psi \log\left(\frac{-D^2}{\mu^2}
\right) \left(F^{\mu}{}_{\kappa} F^{\nu\kappa} \right)
\nn &&
+\frac{7}{4}  \left(\bar\psi \gamma_\mu D_\nu \psi - D_\nu \bar\psi \gamma_\mu \psi\right) \log\left(\frac{-D^2}{\mu^2}
\right) \left(F^{\mu}{}_{\rho} F^{\nu\rho}\right)
+\frac{1}{192} \bar\psi \gamma_{\mu\rho\kappa} \psi \bar\psi \gamma_{\nu\sigma}{}^{\kappa} \psi \log\left(\frac{-D^2}{\mu^2}
\right) \left( F^{\mu\nu} F^{\rho\sigma}\right)
\nn &&
-\frac{1}{384} \bar\psi \gamma_{\mu\nu\rho} \psi \bar\psi \gamma^\rho{}_{\sigma\kappa} \psi \log\left(\frac{-D^2}{\mu^2}
\right) \left( F^{\mu\nu} F^{\sigma\kappa}\right)
-\frac{1}{512} \bar \psi \gamma_{\mu\nu\rho} \psi F_{\sigma\kappa} \log\left(\frac{-D^2}{\mu^2}\right)
\left(\bar \psi \gamma^{\mu\nu\rho} \psi F^{\sigma\kappa} \right)
\nn &&
+\frac{1}{768} \bar \psi \gamma_{\mu\nu\rho} \psi F_{\sigma\kappa} \log\left(\frac{-D^2}{\mu^2}\right)
\left( \bar \psi \gamma^{\mu\nu\sigma} \psi F^{\rho\kappa} \right)
-\frac{5}{768} \bar \psi \gamma_{\mu\nu\rho} \psi F^{\rho\kappa} \log\left(\frac{-D^2}{\mu^2}\right) 
\left( \bar \psi \gamma^{\mu\nu\sigma} \psi F_{\sigma\kappa} \right)
\nn &&
+\frac{5}{1536} \bar \psi \gamma_{\mu\nu\rho} \psi F_{\sigma\kappa} \log\left(\frac{-D^2}{\mu^2}\right)
\left( \bar \psi \gamma^{\rho\sigma\kappa} \psi F^{\mu\nu} \right)
+\frac{1}{96} \bar \psi \gamma_{\mu\nu\rho} \psi F^{\mu\sigma} \log\left(\frac{-D^2}{\mu^2}\right)
\left( \bar \psi \gamma^{\rho}{}_{\sigma\kappa} \psi F^{\nu\kappa} \right)
\nn &&
+\frac{23}{1536} \bar \psi \gamma_{\mu\nu\rho} \psi F^{\mu\nu} \log\left(\frac{-D^2}{\mu^2}\right)
 \left( \bar \psi \gamma^{\rho}{}_{\sigma\kappa} \psi F^{\sigma\kappa} \right)
-\frac{3}{8}i e_F\bar\psi \gamma_{\mu} \psi \log\left(\frac{-D^2}{\mu^2}\right)
\left(\bar\psi \gamma^{\mu}{}_{\nu\rho} \psi  F^{\nu\rho} \right)
\nn &&
+\frac{3}{64}\bar\psi \gamma_{\mu\nu\rho} \psi \bar\psi \gamma^{\mu\nu}{}_{\s} \psi \log\left(\frac{-D^2}{\mu^2}\right)
 R^{\rho\s}
-\frac{11}{4} \left(\bar\psi \gamma_\mu D^\mu \psi - D_\mu \bar\psi \gamma^\mu \psi\right) \log\left(\frac{-D^2}{\mu^2}
\right) R
\nn &&
-\frac{26}{3} m_F \bar\psi \psi \log\left(\frac{-D^2}{\mu^2}\right) R
-\frac{1}{32} \bar\psi \gamma_{\mu\nu\rho} \psi \bar\psi \gamma^{\mu\nu\rho} \psi \log\left(\frac{-D^2}{\mu^2}\right) R
\nn &&
-\frac{1}{64} \bar\psi \gamma_{\mu\nu\rho} \psi \bar\psi \gamma^{\mu}{}_{\s\kappa} \psi \log\left(\frac{-D^2}{\mu^2}\right)
R^{\nu\rho\s\kappa}
+6 \left(\bar\psi \gamma_\mu D^\mu \psi - D_\mu \bar\psi \gamma^\mu \psi\right) \log\left(\frac{-D^2}{\mu^2}\right) V
\nn &&
+20 m_F \bar\psi \psi \log\left(\frac{-D^2}{\mu^2}\right) V
+\frac{3}{64} (\bar\psi \gamma_{\mu\nu\rho} \psi)^2 \log\left(\frac{-D^2}{\mu^2}\right) V
\nn &&
+\frac{1}{12} \bar\psi \c_{\mu\nu\rho} \psi F^{\mu}{}_{\sigma} \log\left(\frac{-D^2}{\mu^2}\right)\nabla^{\s}F^{\nu\rho} 
+\frac{9}{2}e_{F}e_{S}\phi^{*} D_{\mu}\phi \log\left(\frac{-D^2}{\mu^2}\right) \left(\bar\psi \gamma^\mu \psi \right)
\nn &&
-\frac{9}{2}e_{F}e_{S} \phi D_{\mu}\phi^{*} \log\left(\frac{-D^2}{\mu^2}\right) \left(\bar\psi \gamma^\mu \psi \right)
 +\frac{5}{8}i e_{S} \phi^{*} D_{\mu}\phi \log\left(\frac{-D^2}{\mu^2}\right)
\left(\bar\psi \gamma^{\mu\nu\rho} \psi F_{\nu\rho}\right)
\nn &&
-\frac{5}{8}i e_{S} \phi D_{\mu}\phi^{*} \log\left(\frac{-D^2}{\mu^2}\right)
\left(\bar\psi \gamma^{\mu\nu\rho} \psi   F_{\nu\rho}\right)
+\frac{1}{8}\left(\bar\psi \gamma_\mu D^\mu \psi - D_\mu \bar\psi \gamma^\mu \psi\right) \log\left(\frac{-D^2}{\mu^2}\right)
\left(D_{\nu}\phi^{*}{} D^{\nu}\phi \right)
\nn &&
+4 m_F (\bar\psi \psi) \log\left(\frac{-D^2}{\mu^2}\right) \left(D_{\mu}\phi^{*}{} D^{\mu}\phi\right)
+\frac{13}{384} (\bar\psi \gamma_{\mu\nu\rho} \psi)^2 \log\left(\frac{-D^2}{\mu^2}\right)
\left(D_{\s}\phi^{*}{} D^{\s}\phi\right)
\nn &&
-\frac{1}{384} D_\mu \phi^* \bar\psi \gamma_{\nu\rho\s} \psi \log\left(\frac{-D^2}{\mu^2}\right)
\left( D^\mu \phi \bar\psi \gamma^{\nu\rho\s} \psi\right)
-3ie_{F} \bar\psi \gamma_{\mu} \psi \log\left(\frac{-D^2}{\mu^2}\right) \nabla_{\nu} F^{\mu\nu}
\nn &&
+\frac{19}{8} \left(\bar\psi \gamma_\mu D_\nu \psi - D_\nu \bar\psi \gamma_\mu \psi\right) \log\left(\frac{-D^2}{\mu^2}
\right) \left( D^{\mu}\phi^{*}{} D^{\nu}\phi + D^{\nu}\phi^{*}{} D^{\mu}\phi \right)
\nn &&
-\frac{37}{384}\bar\psi \gamma_{\mu\nu\rho}\psi \bar\psi \gamma^{\mu\nu}{}_\s \psi \log\left(\frac{-D^2}{\mu^2}\right)
\left(D^{\rho}\phi^{*} {}D^{\s}\phi \right)
-\frac{1}{384} D_\mu\phi\bar\psi\gamma_{\nu\rho\sigma}\psi \log\left(\frac{-D^2}{\mu^2}\right)
\left(D^\nu\phi^*\bar\psi\gamma^{\mu\rho\sigma}\psi\right)\nn &&
-\frac{1}{96}D_\mu\phi\bar\psi\gamma^{\mu}{}_{\nu\rho}\psi \log\left(\frac{-D^2}{\mu^2}\right) 
\left(D_\sigma\phi^*\bar\psi\gamma^{\nu\rho\sigma}\psi\right)
+\frac{7}{24} \bar\psi \gamma_{\mu\nu\rho} \psi F^{\mu\nu} \log\left(\frac{-D^2}{\mu^2}\right) \nabla_{\s}F^{\rho\s}
\nn &&
+\frac{ie_S}{16} F_{\mu\nu} \phi^* \log\left(\frac{-D^2}{\mu^2}\right)
\left(D_\rho\phi \bar\psi\gamma^{\mu\nu\rho}\psi \right)
-\frac{ie_S}{16} F_{\mu\nu} \phi \log\left(\frac{-D^2}{\mu^2}\right) 
\left(D_\rho\phi^* \bar\psi\gamma^{\mu\nu\rho}\psi \right)
\nn &&
-\frac{1}{8} \nabla_\mu \left(\bar\psi\gamma_{\nu\rho\sigma}\psi\right) \log\left(\frac{-D^2}{\mu^2}\right) 
\left(F^{\mu\nu}F^{\rho\sigma}\right)
-\frac{3}{128} \nabla_{\mu}\left(\bar\psi \gamma^{\mu}{}_{\nu\rho} \psi\right) \log\left(\frac{-D^2}{\mu^2}\right)
\nabla_{\s}\left(\bar\psi \gamma^{\nu\rho\s} \psi\right)
\nn &&
-\frac{1}{4} \nabla_{\mu}\left(\bar\psi \gamma_{\nu\rho\sigma} \psi\right) \log\left(\frac{-D^2}{\mu^2}\right)
R^{\mu\nu\rho\sigma}
+\frac{1}{128} \nabla_{\mu}\left(\bar\psi \gamma_{\nu\rho\sigma} \psi\right) \log\left(\frac{-D^2}{\mu^2}\right)
\nabla^{\nu}\left(\bar\psi \gamma^{\mu\rho\sigma} \psi\right) 
\nn &&
-\frac{1}{128} \nabla_{\mu}\left(\bar\psi \gamma_{\nu\rho\sigma} \psi\right) \log\left(\frac{-D^2}{\mu^2}\right)
\nabla^{\mu}\left(\bar\psi \gamma^{\nu\rho\sigma} \psi\right)
-\frac{1}{12} \bar\psi \gamma_{\mu\nu\rho} \psi F^{\mu}{}_{\sigma} \log\left(\frac{-D^2}{\mu^2}\right)
\nabla^{\nu}F^{\rho\sigma}
\nn &&\left.
-\frac{1}{512} \bar\psi \gamma_{\mu\nu\rho} \psi \bar\psi \gamma^{\mu}{}_{\sigma\kappa} \psi
\log\left(\frac{-D^2}{\mu^2}\right) \nabla^\nu \left( \bar\psi \gamma^{\rho\sigma\kappa} \psi  \right)\!\right\}.
\label{f423}
\eea

In preparing the above results~(\ref{f421})-(\ref{f423}), we used extensively commutation of covariant derivatives
(\ref{f385})-(\ref{f387}), Bianchi identities for Riemann and field strength tensors and the antisymmetric property
of the product of three Dirac gamma matrices $\gamma_{\mu\nu\rho}$. We also integrated by parts (where it was possible)
to put the terms in the most symmetric form. (The operator $\log\left(\frac{-D^2}{\mu^2}\right)$ is always invariant
under integration by parts.)
Finally, in the scalar and fermionic sectors, we used the square of the gauge-covariant background derivative $D_\mu$
as the argument of the logarithmic form-factors. This is certainly correct. However, in some cases when the operator
acts on the fields in a right bracket, which all together are electromagnetically neutral, the EM background connection
$A_\mu$ completely drops out from $D_\mu^2$ and it is also fine to substitute $\nabla_\mu^2$.
This was already done for all terms in (\ref{f421}) which from construction are uncharged.

The results~\p{f421}-\p{f423} for the quantum effective action are written to the quadratic order
in ``generalized curvatures'' $\cal R$. These generalized curvatures are matrices
$\tilde {\bf{W}}$, $\tilde{ \bm \Omega}_{\mu\nu}$ and the Riemann tensor $R_{\mu\nu\rho\sigma}$ and its contractions.
All terms in all components of $\tilde {\bf{W}}$, $\tilde{\bm \Omega}_{\mu\nu}$ counts as one unit of
generalized curvature. 

Another remark is that according to the discussion in sects.~\ref{sec2} and \ref{sec3.5}, the argument of
the logarithm in Eqs. (\ref{f421})-(\ref{f423}) should be the two-derivative operator $z=-\tilde{\bf D}_\mu^2$,
which is common for all fluctuations, divided by $\mu^2$. The nonlocal part of the effective action~(\ref{f419})
have the same structure as in the local
UV-divergences in \p{e1}, \p{e2} and \p{e3}, with the factors of $\log\left(\frac{\Lambda^2}{\mu^2}\right)$
replaced by $\log\left(\frac{-D^2}{\mu^2}\right)$. We also see that the nonlocal logarithmic pieces of the effective
action depend only on $A_a$ coefficients in  structure functions $g_a$ for each term counted by a different value
of the index $a$. If one looks for a shortcut to get quickly the nonlocal terms in the effective action $\Gamma_0$,
then one can concentrate on $A_a$ coefficients in the structure functions.
One also notes that terms proportional to $Q_2$ and $Q_1$ in (\ref{master2})
do not contribute at all to nonlocal logarithmic terms of the effective action $\Gamma_0$. 
Local operators in the effective action in \p{e1}, \p{e2} and \p{e3} are subject to renormalization
and the coefficients of independent operators are arbitrary since they depend on the renormalization conditions.
In contrast, the coefficients of the nonlocal terms are uniquely determined.

The effective action with the local and nonlocal terms is the main result of this paper.

\section{Discussion}
\label{sec5}

In this paper we have calculated the effective action in the gravity theory coupled to gauge, Dirac and
charged scalar fields. In particular we have obtained its nonlocal part unambiguously.
Most of the work so far on asymptotically safe gravity focused on the existence proof that there are UV FPs.
However, the object directly connected to most important physical quantities is the effective action $\Gamma_0$
(effective average action at $k=0$).
We have used the FRGE and integrate it from high energy scale down to $k=0$ to obtain the effective action.
The nice property of this method is that there are no divergences in contrast to computing the functional integrals.
We have shown that the resulting effective action has nonlocal terms with definite coefficients.
The physical effects corresponding to nonlocal terms in the effective action are non-analytic in momentum,
which are different from those generated by the local and analytic divergent terms.
The local terms are affected by divergences and must be renormalized so that their coefficients depend on
the renormalization conditions.
The result on nonlocal terms can be used to calculate gravitational scattering amplitudes
without ambiguity~\cite{Donoghue1994,BDH,KK} and these predictions can be tested experimentally.

One could also evaluate the effective action perturbatively in the weak field limit
using the covariant nonlocal expansion of the heat kernel in \cite{BV3}.
The effective action in this framework was worked out in \cite{BV4}.
Similar results have been obtained in \cite{DE:2015} using Feynman diagrams.
This result may be used to try to generate cosmological magnetic fields.
This primordial magnetogenesis from anomalies relies on the effective action derived in the
weak field approximation~\cite{El-Menoufi}.
The result is valid in the regime in which the curvature satisfies the condition $R^2 \ll \nabla^2 R$.
During the slow-roll inflation, the geometric curvature satisfies $\nabla^2 R \ll R^2$,
while during matter domination we have $\nabla^2 R \sim R^2$.
So to study primordial magnetogenesis reliably over a long range of cosmological evolution,
it is necessary to go beyond the weak field approximation.
The results in this case have been obtained for the effective action from the Weyl anomaly
in the theory of gauge bosons interacting with Dirac fermion in curved backgrounds~\cite{BBD2017}.
However, there the metric was assumed to be conformally flat. While this restriction has no problem
in the application to Friedmann-Lema\^itre-Robertson-Walker metric, our result is valid for arbitrary backgrounds
and quantum effects from gravity are also taken into account.
The magnetogenesis has been discussed in~\cite{DK2018} with negative conclusion.
Our effective action has nonlocal terms with a similar structure, and could be used to see
if the magnetic fields can be generated.
It would be very interesting to study this problem with the effective action derived here which includes
the quantum gravity effects.

Some comments are in order on the results for $\Gamma_0$ in Eqs. (\ref{g0sum})-(\ref{f423}).
We have found so many terms in the total expression for the effective action, in comparison with quite
a small number of terms present in the original action of the total system (in sect.~\ref{sec3.5}).
We started out with Einsteinian gravitation and renormalizable matter models (of abelian gauge fields,
fermions and scalars). There are many virtual processes that contribute to scattering amplitudes,
which gives all possible terms. The only restriction is the reparametrization and gauge invariance.
If this condition is satisfied, nothing forbids the invariant operators and the corresponding processes appear.
This is true even if one starts with renormalizable or nonrenormalizable theories.
The only difference between these theories is whether we would have divergent coefficients beyond those
operators already present in the bare action.

It is known that  in the perturbative framework (using for example Feynman diagrams as
in~\cite{Burgess,RSW,DE:2015}), one gets the similar expression for the effective action of the model at the one-loop level.
However, the procedure is different. There one isolates the perturbative UV-divergences, finds corresponding beta
functions and renormalizes the theory by introducing an arbitrary energy scale of the renormalization point $\mu$.
Then the couplings of the theory are promoted from constants to scale-dependent one, the scale being represented by $\mu$.
But in the full effective action we have implicit independence on this $\mu$ parameter, because
the quantum effective action is RG transformation invariant. When the logarithmic UV-divergences are taken
into consideration, the RG running of dimensionless couplings is logarithmic with the energy scale $\mu$.
The above-mentioned RG-invariance of the effective action is only achieved if we have very special nonlocal
terms in the effective action built with the logarithm of the momentum.
In covariant language, these nonlocal universal terms take the form with the insertion of nonlocal logarithmic
form-factor of the gauge- and GR-covariant operator $\log\left(\frac{-D^2}{\mu^2}\right)$.
In our case the terms under consideration are of the quadratic order in ``generalized curvatures'' $\cal R$,
and the insertion of the nonlocal operator $\log\left(\frac{-D^2}{\mu^2}\right)$ is unambiguous.
This is in accordance with the fact that we used nonlocal heat kernel expansion to the second order
in curvatures $\cal R$ in Eq. (\ref{master}); in the flow equations (\ref{f46}), (\ref{f48}) and (\ref{f410}),
terms quadratic in $\cal R$ were multiplied by the functional $Q_0[h_k]$, which is a dimensionless numerical
constant, and in the Eqs. (\ref{e1})--(\ref{e3}), the terms quadratic in curvatures $\cal R$
were all proportional to $\log\left(\frac{\Lambda^2}{\mu^2}\right)$. It is consistent that perturbatively
we would consider logarithmic UV-divergences in the Eqs. (\ref{e1})--(\ref{e3}) related to local
terms quadratic in curvatures. The form of the universal quadratic in curvatures $\cal R$ part of
the nonlocal effective action is dictated by RG-invariance and should be read together with
the scale-dependence (RG running) of couplings.
Difference between one-loop effective action obtained by perturbation and that obtained from the FRGE is
also discussed in \cite{LP}.

We emphasize that the main novelty of this work is the computation of nonlocal universal terms in the effective action
$\Gamma_0$ of the total system in sect.~\ref{sec3.5} using the FRGE methods and integration of the RG flow equation.
These results can be also viewed as listing of all one-loop UV-divergences for our total quantum system.
These divergences are encoded in terms, which are at most quadratic in generalized curvatures $\cal R$.

Our results for the effective action $\Gamma_0$ should be understood in certain expansion schemes.
Let us concentrate on nonlocal universal terms which contain logarithmic form-factor of the covariant operator $D^2$.
Besides these terms, there are also some local higher curvature terms with universal (RG-independent) coefficients
and local terms with coefficients dependent on the renormalization conditions.
We performed a ``generalized curvature'' expansion coinciding with the scheme of heat kernel expansion.
Our results are to the second order and we expect that higher orders in $\cal R$ exist too and they could be relevant
for computation of various physical observables. For example, from the quadratic terms in $\cal R$ in $\Gamma_0$,
we can unambiguously derive the quantum finite corrections to the graviton polarization functions,
but to know a precise form of 4-gravitons scattering amplitudes around flat spacetime,
we would need to possess the knowledge of $\Gamma_0$ to order quartic in curvatures $\cal R$.
The curvature expansion is different from derivative expansion which is however a common practice
for the effective field theory approaches. In our terms we have collected all powers of energy/momentum
into covariant logarithmic form-factors of the operator $D^2$.
Within these approximation schemes and limitations, the results in (\ref{g0sum})-(\ref{f423}) are complete
in the sense that there are no other terms except those written and with the precise coefficients that we have determined.
It is also an important problem which kind of terms may be generated at two-loop and beyond.

One can also ask a question about a possible extension of this type of computation for a model of renormalizable gravity
in $d=4$ spacetime dimensions by including higher curvature terms. Such terms may be relevant operators, so
it is important to study the role of these terms in general RG flows.
The first renormalizable model is due to Stelle \cite{Stelle1} and contains generally covariant terms with $R^2$
and $R_{\mu\nu}^2$  (the $R_{\mu\nu\rho\s}^2 $ term can be easily eliminated by using the Gauss-Bonnet identity
valid in $d=4$).
However, due to the higher derivative nature of operators in the gravitational sector, there is an issue with unitarity.
Leaving this problem aside, we may still analyze beta functions of the couplings.
Due to the renormalizability of this model of gravity coupled to matter \cite{matter1,matter2,stw},
the number of UV-divergent terms is smaller.
We note that in gravitational models, in which there are no perturbative UV-divergences (and they are
UV-finite~\cite{fin, univ}) by the above argumentation we will not find any logarithmic term in the effective action
 $\Gamma_0$.

We have not included Yukawa couplings between fermions and scalar in our study.
The reason is that the gauge invariance forbids Yukawa interactions between our scalar field and fermions
because the charges do not match.
However in more general situation, the Yukawa couplings may play significant role in the flow of
the gravitational coupling and a scalar potential $V(\phi)$ \cite{ZZVP2009}.
So it should be significant to extend our work in this direction.
Other interesting subjects include studying quantum effects in other physical processes.
For example, it would be of interest to use the quantum effective action to study if the quantum effects
may tame the singularities in the blackholes and in the early universe, if they can be a rescue for the information
paradox in the Hawking radiation process or whether there is any physical consequence in the observed gravitational waves.

We hope to return to these important subjects in the future.

%Another direction of possible extension 

%%%%%%%%%%%%%%%%%%%%%%%%%
\section*{Acknowledgment}

We thank Roberto Percacci, Aleksandr Pinzul, Ilya Shapiro,  Alberto Tonero, and Omar Zanusso
 for valuable discussions.
This work was supported in part by the Grant-in-Aid for Scientific Research Fund of the JSPS (C) No. 16K05331. L.R. was supported from ESIF, EU Operational Programme Research, Development and Education No. CZ$.02.2.69/0.0/0.0/16_{-}027/0008465$.

\appendix

\section{Self-adjointness of the Hessian}
\label{appa}

\newcommand\cH{{\cal H}}
\newcommand\cG{{\cal G}}

When there are first derivative terms in the Hessian and the coefficients depend on the coordinates, there is an ambiguity
in how to write the Hessian in the matrix form. We remove this ambiguity by imposing the self-adjointness condition.
If we use the metric $\cG$ in field space,
the self-adjointness of the kinetic operator $\Delta$ means that
\bea
\cG(\psi',\Delta\psi)=\cG(\psi,\Delta\psi').
\label{sa}
\eea
In the usual case without first derivative terms, this requirement simply gives the symmetric matrix for $\Delta$,
but this is not true when there are first derivative terms.

We illustrate this by consideration of some simple examples.
In the action~\p{g1}, we have the terms with first derivative:
\bea
h^{\mu\nu} \left\{ \bF_{(\mu}{}^\rho \bg_{\nu)\a} \bnabla_{\rho} + \bF_{\a(\mu}\bnabla_{\nu)}
 + \frac{1}{2}\bg_{\mu\nu}\bF_{\rho\a}\bnabla^{\rho} \right\} A^\a
= h^{\mu\nu}[2 K_{\mu\nu\la\a}\bF^{\la\rho}\bnabla_\rho-\d_{\mu\nu}{}^{\la\rho}\bF_{\la\a}\bnabla_\rho] A^\a.
\label{adj1}
\eea
To clearly formulate this, it is necessary to distinguish the unprimed (left) and primed (right)  fields.
We rewrite this as
\bea
&& h^{\mu\nu}{}'\left[ K_{\mu\nu\la\a}\bF^{\la\rho}\bnabla_\rho-\frac{1}{2}\d_{\mu\nu}{}^{\la\rho}\bF_{\la\a}
\bnabla_\rho\right] A^\a
-A^\a{}'\left[ K_{\mu\nu\la\a}\bF^{\la\rho}\bnabla_\rho-\frac{1}{2}\d_{\mu\nu}{}^{\la\rho}\bF_{\la\a}
\bnabla_\rho\right] h^{\mu\nu} \nn
&& -A^\a{}'\left[ K_{\mu\nu\la\a}(\bnabla_\rho\bF^{\la\rho})-\frac{1}{2}\d_{\mu\nu}{}^{\la\rho}
(\bnabla_\rho\bF_{\la\a})\right] h^{\mu\nu}.
\label{adj2}
\eea
Here, in the second line of (\ref{adj2}), we added the surface terms after the partial integration.
This is the same as \p{adj1} if we identify the fields with and without prime upon partial integration.
We get rid of the derivatives on the unprimed fields by partial integration and then what we get is
\bea
&& - A^\a \left[ K_{\mu\nu\la\a}\bF^{\la\rho}\bnabla_\rho-\frac{1}{2}\d_{\mu\nu}{}^{\la\rho}\bF_{\la\a}
\bnabla_\rho\right] h^{\mu\nu}{}'
-A^\a \left[ K_{\mu\nu\la\a}(\bnabla_\rho\bF^{\la\rho})-\frac{1}{2}\d_{\mu\nu}{}^{\la\rho}
(\bnabla_\rho\bF_{\la\a})\right] h^{\mu\nu}{}' \nn
&& +h^{\mu\nu}\left[ K_{\mu\nu\la\a}\bF^{\la\rho}\bnabla_\rho-\frac{1}{2}\d_{\mu\nu}{}^{\la\rho}\bF_{\la\a}
\bnabla_\rho\right] A^\a{}',
\eea
which is the same as \p{adj2} in which fields with and without primes are interchanged.
Eq.~\p{adj2} is written in the matrix form as in \p{g3} -- \p{g5}, which is therefore self-adjoint
even though it does not look symmetric.
It is important that in the process we use only integration by parts and do not exchange fields unawarely.

Another example is the $h-\vp$ mixing term in the action~\p{scalarhessian}:
\bea
&&\int d^4 x \sqrt{g} \left[ 
\frac12 h\left\{(\bd_\mu\bp^*) (\bd^{\mu} \vp) +(\bd_\mu\bp)(\bd^{\mu} \vp^*) \right\}
- h^{\mu\nu} \left\{(\bd_\mu\bp^*) (\bd_\nu\vp) +(\bd_\mu\bp)(\bd_\nu \vp^*) \right\} \right.\nn
&& \hs{30}\left. +\frac12 h V' (\bp^* \vp+\bp \vp^*) \right] \nn
&& = \int d^4 x \sqrt{g} \left[ 
-2 h^{\mu\nu}K_{\mu\nu\rho\la} \left\{(\bd^\rho\bp^*)\bd^\la \vp + (\bd^\rho\bp)\bd^\la \vp^* \right\}
+\frac12 h V' (\bp^* \vp+\bp \vp^*)
\right].
\label{hshessian}
\eea
%The self-adjoint condition in this case is
%\bea
%\int d^4 x \sqrt{g}
%(h^{\a\b}, \vp^*, \vp) U (h^{\mu\nu}{}', \vp', \vp'{}^*)^T
%= \int d^4 x \sqrt{g}
%(h^{\mu\nu}{}', \vp'{}^*, \vp') U^\dagger (h^{\a\b}, \vp, \vp^*)^T.
%\eea
Let us rewrite \p{hshessian} as
\bea
&& \hs{-5} \int d^4 x \sqrt{g} \left[ 
-h^{\mu\nu}{}' K_{\mu\nu\rho\la} \left\{(\bd^\rho\bp^*)\bd^\la \vp + (\bd^\rho\bp)\bd^\la \vp^* \right\}
+ K_{\a\b\rho\la} \left\{ \vp' (\bd^\rho\bp^*) + \vp'^*(\bd^\rho\bp)\right\}\bd^\la h^{\a\b} \right.
\nn && \hs{-5} \left.
+ K_{\a\b\rho\la} \left\{ \vp'(\bd^\rho \bd^\la\bp^*) + \vp'^*(\bd^\rho\bd^\la\bp)\right\} h^{\a\b}
+ \frac14 h' V'(\bp^* \vp + \bp \vp^*) 
+ \frac14 h V' (\bp^* \vp'+\bp \vp'^*)
\right].
\label{hshessian_self}
\eea
This is the same as \p{hshessian} if we identify the fields with and without prime upon integration by parts.
Upon this, we get
\bea
&& \hs{-5} \int d^4 x \sqrt{g} \left[ 
 K_{\mu\nu\rho\la} \left\{ \vp(\bd^\rho\bp^*) + \vp^* (\bd^\rho\bp) \right\} \bd^\la h^{\mu\nu}{}'
- h^{\a\b} K_{\a\b\rho\la} \left\{ (\bd^\rho\bp^*)\bd^\la\vp' + (\bd^\rho\bp)\bd^\la\vp'^*\right\}
 \right.
\nn && \hs{-5} \left.
+ K_{\mu\nu\rho\la} \left\{ \vp'(\bd^\rho \bd^\la\bp^*) + \vp'^*(\bd^\rho\bd^\la\bp)\right\} h^{\mu\nu}
+ \frac14 h' V'(\bp^* \vp + \bp \vp^*) 
+ \frac14 h V' (\bp^* \vp'+\bp \vp'^*)
\right]\!.\hspace{0.5cm}
\eea
This is the same expression as \p{hshessian_self} in which fields with and without primes are interchanged.
Eq.~\p{hshessian_self} is written in the matrix form as in \p{scalarhessian_matrix}, which is again self-adjoint.
Other terms are similar.

\section{Scalar basis}
\label{appb}

When we have a complex scalar $\phi$, we have two different ways to represent it.
One is to decompose the scalar field into two real fields
\bea
\phi = \frac{\phi_1+i\phi_2}{\sqrt{2}}.
\eea
Suppose we have the action in terms of the real fields:
\bea
S = \int d^4 x (\phi_1,\phi_2) M \left(\begin{array}{c}
\phi_1 \\
\phi_2
\end{array}
\right).
\label{action1}
\eea
The path integrals over the complex scalar field $\phi$ and its conjugate $\phi^*$ produce the factor $
(\det M)^{-1/2}$. Note that this matrix is of the size $2\times 2$.

Now noting that
\bea
\left(\begin{array}{c}
\phi_1 \\
\phi_2
\end{array}
\right)
= \frac{1}{\sqrt{2}} \left(\begin{array}{cc}
1 & 1 \\
-i & i
\end{array}
\right)
\left(\begin{array}{c}
\phi \\
\phi^*
\end{array}
\right)
\equiv
N \left(\begin{array}{c}
\phi \\
\phi^*
\end{array}
\right),
\eea
then \p{action1} can be rewritten as
\bea
S = \int d^4 x\, (\phi^*,\phi)N^\dagger M N \left(\begin{array}{c}
\phi \\
\phi^*
\end{array}
\right)
\equiv \int d^4 x\, \psi^\dagger N^\dagger M N \psi .
\label{action2}
\eea
Though we have a ``complex'' field here, $\psi^\dagger$ and $\psi$ are not independent
($\phi$ and $\phi^*$ are independent).
So the path integral over $\psi$ yields
\bea
(\det N^\dagger M N)^{-1/2} = (\det M)^{-1/2} ,
\eea
the same as the real field basis. We can use either basis whichever convenient.
If the action has only $\phi^*$-$\phi$ terms, we may simply use action of the form $\phi^* M \phi$,
but this is not the case with our action in section \ref{sec3.4}.
In the text, we employ the ``complex'' basis.

\section{Explicit forms of tensors and their traces}
\label{appc}

\subsection{Scalar ${\bf Y}^2$}
\label{appc.1}

Here we collect the explicit form of components of ${\bf Y}^2$ in \p{yy}.
\bea
&&\hs{-5}
Y_{\mu\nu,\a\b}^{2,hh}=-\frac{1}{128}\bg_{\mu\a}\bar{\psi}\c_{\nu\la\d}\psi\bar{\psi}
\c_{\b}{}^{\la\d}\psi+\frac{1}{128}\bar{\psi}\c_{\mu\a\d}\psi\bar{\psi}\c_{\nu\b}{}^\d\psi
-\bF_{\mu\a}\bF_{\nu\b}+\frac{1}{2}\bg_{\mu\nu}\bF_{\a\lambda}\bF_{\b\lambda}
\nn && \hs{5}
+\frac{1}{2}\bg_{\a\b}\bF_{\mu\la}\bF_{\nu}{}^\la-\bg_{\mu\a}\bF_{\nu\la}\bF_{\b}{}^\la
-\frac{1}{8}\bg_{\mu\nu}\bg_{\a\b}\bF_{\rho\lambda}^{2}-K_{\a\b\mu\rho}\bd_{\nu}\bp^*\bd^{\rho}\bp
-K_{\a\b\mu\rho}\bd_{\nu}\bp \bd^{\rho}\bp^*,
\nn &&\hs{-5}
Y_{\mu\nu,\a}^{2,hA}=-\frac{1}{16}\bar{\psi}\c_{\mu\a\d}\psi \bF_{\nu}{}^\d
-\frac{1}{16}\bg_{\mu\a}\bar{\psi}\c_{\nu\lambda\d}\psi \bF_{\lambda}{}^\d-\frac{ie_S}{2}\bg_{\mu\a}\bp\bd_{\nu}\bp^*
+\frac{ie_S}{2}\bg_{\mu\a}\bp^*\bd_{\nu}\bp,
\nn &&\hs{-5}
Y_{\mu\nu}^{2,h\vp}=\left(-\frac{1}{8}\bg_{\a\mu}\bar{\psi}\c_{\nu \b\d}\psi\right)
\left(-\d^{\a\b,\d\rho}\bd_{\rho}\bp^*\right)+\left(2K_{\mu\nu}{}^{\la \a}\bF_{\la\d}
-\d_{\mu\nu,\lambda\d}\bF^{\la \a}\right)\left(\frac{ie_S}{2}\bg_{\a}{}^\d\bp^*\right)=0,
\nn &&\hs{-5}
Y_{\mu\nu}^{2,h\vp^*}=\left(-\frac{1}{8}\bg_{\a\mu}\bar{\psi}\c_{\nu \b\d}\psi\right)
\left(-\d^{\a\b,\d\rho}\bd_{\rho}\bp\right)+\left(2K_{\mu\nu\la}{}^\a \bF^{\la\d}
-\d_{\mu\nu,\lambda\d}\bF_{\lambda \a}\right)\left(-\frac{ie_S}{2}\bg_{\a\d}\bp\right)=0,
\nn &&\hs{-5}
Y_{\mu,\a}^{2,AA}=-\frac{1}{2}\bF_{\mu\lambda}\bF_{\a}{}^\la-\frac{1}{4}\bg_{\mu\a}\bF_{\rho\la}^{2}
-\frac{e_S^{2}}{2}\bg_{\mu\a}|\bp|^2,
\nn &&\hs{-5}
Y_{\mu,\a\b}^{2,Ah}=\frac{1}{32}\bF_{\a\d}\bar{\psi}\c_{\mu\b}{}^\d\psi
-\frac{1}{32}\bg_{\a\mu}\bF_{\la\d}\bar{\psi}\c_{\b\la}{}^\d\psi
+\frac{ie_S}{2}K_{\a\b\mu}{}^\rho\bp^*\bd_{\rho}\bp-\frac{ie_S}{2}K_{\a\b\mu}{}^\rho\bd_{\rho}\bp^*\bp,
\nn &&\hs{-5}
Y_{\mu}^{2,A\vp}=\left(-K_{\a\b\la\mu}\bF^{\la\d}+\frac{1}{2}\d_{\a\b}{}^{\la\d}\bF_{\la\mu}\right)
\left(-\d^{\a\b\rho}{}_{\d}\bd_{\rho}\bp^*\right)=-\frac{3}{4}\bF^{\la}{}_{\mu}\bd_{\la}\bp^*,
\nn &&\hs{-5}
Y_{\mu}^{2,A\vp^*}=\left(-K_{\a\b\la\mu}\bF^{\la}{}^\d+\frac{1}{2}\d_{\a\b,\la}{}^{\d}\bF^{\la}{}_\mu\right)
\left(-\d^{\a\b}{}_{\rho\d}\bd_{\rho}\bp\right)=-\frac{3}{4}\bF^{\la}{}_\mu\bd_{\la}\bp,
\nn &&\hs{-5}
Y_{\alpha\beta}^{2,\vp^* h}=\left(K^{\rho\s\la\d}\bd_{\la}\bp\right)\left(-\frac{1}{8}\bg_{\a(\rho}\bar{\psi}
\c_{\s)\beta\d}\psi\right)+\left(\frac{ie_S}{2}\bg^{\rho\d} \bp\right)\left(-K_{\a\beta\rho}{}^{\la}\bF_{\la\d}
+\frac{1}{2}\d_{\a\b,\d}{}^\la \bF_{\la \rho}\right)=0,
\nn &&\hs{-5}
Y_{\alpha}^{2,\vp^* A}=\left(K^{\rho\s\la\d}\bd_{\la}\bp\right)\left(2K_{\rho\s\a}{}^\b \bF_{\b\d}
-\d_{\rho\s,\d}{}^{\b}\bF_{\b\a}\right)=-\frac{3}{4}\bd^{\la}\bp \bF_{\la\alpha},
\nn &&\hs{-5}
Y^{2,\vp^*\vp}=\left(K_{\a\b}{}^{\la\d}\bd_{\la}\bp\right)\left(-\d^{\a\b}{}_{\rho\d}\bd^{\rho}\bp^*\right)
+\left(\frac{ie_S}{2}\bg^{\a\d}\bp\right)\left(\frac{ie_S}{2}\bg_{\a\d}\bp^*\right)
=-\bd_{\la}\bp \bd^{\la}\bp^*-e_S^{2} |\bp|^2,
\nn &&\hs{-5}
Y^{2,\vp^*\vp^*}=\left(K_{\rho\s}{}^{\la\d}\bd_{\la}\bp\right)\left(-\d^{\rho\s}{}_{\a\d}\bd^{\a}\bp\right)
+\left(\frac{ie_S}{2}\bg^{\rho\d}\bp\right)\left(-\frac{ie_S}{2}\bg_{\rho\d}\bp\right)
=-\bd_{\la}\bp \bd^{\la}\bp+e_S^{2}\bp^2,
\nn &&\hs{-5}
Y_{\a\b}^{2,\vp h}=\left(K^{\rho\s\la\d}\bd_{\la}\bp^*\right)\left(-\frac{1}{8}\bg_{\a(\rho}
\bar{\psi}\c_{\s) \b\d}\psi\right) + \left(-\frac{ie_S}{2}\bg^{\rho\d}\bp^*\right)
\left(-K_{\a\b\rho \la}\bF^{\la}{}_\d+\frac{1}{2}\d_{\alpha\b,\d\la}\bF^\la{}_{\rho}\right)=0,
\nn &&\hs{-5}
Y_{\alpha}^{2,\vp A}=\left(K^{\rho\s\la\d}\bd_{\la}\bp^*\right)\left(2K_{\rho\s\b\alpha}\bF^\b{}_{\d}
-\d_{\rho\s,\b\d}\bF^\b{}_{\a}\right)=-\frac{3}{4}\bd^{\la}\bp^*\bF_{\la\alpha},
\nn &&\hs{-5}
Y^{2,\vp\vp}=\left(K^{\a\b\la\d}\bd_{\la}\bp^*\right)\left(-\d_{\a\b\rho\d}\bd^{\rho}\bp^*\right)
+\left(-\frac{ie_S}{2}\bg^{\a\d}\bp^*\right)\left(\frac{ie_S}{2}\bg_{\a\d}\bp^*\right)
=-\bd^{\la}\bp^*\bd_{\la}\bp^*+e_S^{2}\bp^{*2},
\nn &&\hs{-5}
Y^{2,\vp\vp^*}=\left(K^{\a\b\la\d}\bd_{\la}\bp^*\right)\left(-\d_{\a\b\rho\d}\bd^\rho\bp\right)
+\left(-\frac{ie_S}{2}\bg^{\a\d}\bp^*\right)\left(-\frac{ie_S}{2}\bg_{\a\d}\bp\right)
=-\bd_{\la}\bp^*\bd^{\la}\bp-e_S^{2}|\bp|^2.\nn
\eea

\subsection{Tensor $\tilde{\Omega}_{\rho\s}$}
\label{appc.2}

The components of $\tilde{\bm\Omega}_{\rho\s}$ in \p{ot} are
%({the antisymmetrization with 1/2 in $\rho,\s$})
\bea
\tilde{\Omega}_{\mu\nu,\a\b,\rho\s}^{h,h}&=&
2\bg_{\mu\a}\br_{\rho\s\nu\b}+\frac{1}{4}\bg_{\a\mu}\bd_{[\rho}
\left(\bar{\psi}\c_{\nu\b\s]}\psi\right)-\frac{1}{64}\bg_{\mu\a}\bar{\psi}\c_{\nu[\rho}{}^\kappa \psi\bar{\psi}\c_{\b\s]
\kappa}\psi
+\bg_{\mu[\rho}\bF_{\nu\a}\bF_{\b\s]}
\nn &&
-\bg_{\mu\a}\bF_{\nu[\rho}\bF_{\b\s]}+\bg_{\a[\rho}\bF_{\mu\b}\bF_{\nu\s]}
-\frac{1}{2}\bg_{\mu\nu}\bg_{\a[\rho}\bF_{\b}{}^\kappa \bF_{\s]\kappa}
+\frac{1}{2}\bg_{\a\b}\bg_{\mu[\rho}\bF_{\nu}{}^\kappa \bF_{\s]\kappa}
\nn &&
-\bg_{\mu[\rho}\bg_{\a\s]}\bF_{\nu\kappa}\bF_{\b}{}^\kappa
-\bg_{\mu[\rho}\bg_{\a\s]}\bd_{\nu}\bp^*\bd_{\b}\bp+\frac{1}{2}\bg_{\mu[\rho}\bg_{\a\b}\bd_{\nu}\bp^*\bd_{\s]}\bp
\nn &&
-\bg_{\mu[\rho}\bg_{\a\s]}\bd_{\b}\bp^*\bd_{\nu}\bp+\frac{1}{2}\bg_{\mu[\rho}\bg_{\a\b}\bd_{\s]}\bp^*\bd_{\nu}\bp,
\nn
\tilde{\Omega}_{\mu\nu,\a,\rho\s}^{h,A}&=&-\frac{1}{8}\bar{\psi}\c_{\mu\a[\rho}\psi \bF_{\nu\s]}
-\frac{1}{8}\bar{\psi}\c_{\mu\rho\s}\psi \bF_{\nu\a}-\frac{1}{8}\bg_{\mu\a}\bar{\psi}\c_{\nu\kappa[\rho}\psi
 \bF^\kappa{}_{\s]}
-\frac{1}{8}\bg_{\mu[\rho}\bar{\psi}\c_{\nu\s]\kappa}\psi \bF_{\a}{}^\kappa
\nn &&
-ie_S \bg_{\mu[\rho}\bg_{\s]\a}(\bp\bd_{\nu}\bp^*-\bp^*\bd_{\nu}\bp)
-2\bg_{\mu\a}\bnabla_{[\s}\bF_{\rho]\nu}+\bg_{\mu\nu}\bnabla_{[\s}\bF_{\rho]\a}
-2\bg_{\mu[\rho}\bnabla_{\s]}\bF_{\nu\a},
\nn
\tilde{\Omega}_{\mu\nu,\rho\s}^{h,\vp}
&=&\frac{1}{8}\bg_{\mu[\rho}\bar{\psi}\c_{\nu\s]\kappa}\psi \bd^{\kappa}\bp^*
-\frac{1}{8}\bar{\psi}\c_{\mu\rho\s}\psi \bd_{\nu}\bp^*-2\bg_{\mu[\rho}\bd_{\nu}\bd_{\s]}\bp^*
+\frac{ie_S }{2}\bg_{\mu\nu}\bp^*\bF_{\rho\s},
\nn
\tilde{\Omega}_{\mu\nu,\rho\s}^{h,\vp^*}
&=&\frac{1}{8}\bg_{\mu[\rho}\bar{\psi}\c_{\nu\s]\kappa}\psi \bd^{\kappa}\bp
-\frac{1}{8}\bar{\psi}\c_{\mu\rho\s}\psi \bd_{\nu}\bp-2\bg_{\mu[\rho}\bd_{\nu}\bd_{\s]}\bp
-\frac{ie_S }{2}\bg_{\mu\nu}\bp \bF_{\rho\s},
\nn
\tilde{\Omega}_{\mu,\a\b,\rho\s}^{A,h}
&=& -\frac{1}{16}\bar{\psi}\c_{\mu\a[\rho}\psi \bF_{\b\s]}+\frac{1}{16}\bar{\psi}\c_{\a\rho\s}\psi \bF_{\b\mu}
-\frac{1}{16}\bg_{\a\mu}\bar{\psi}\c_{\kappa\b[\rho}\psi \bF^\kappa{}_{\s]}
-\frac{1}{16}\bg_{\a[\rho}\bar{\psi}\c^\kappa{}_{\b\s]}\psi \bF_{\kappa\mu}
\nn &&
+\frac{ie_S }{2}\bg_{\mu[\rho}\bg_{\s]\a}(\bp^*\bd_{\b}\bp-\bp\bd_{\b}\bp^*)
-\frac{ie_S }{4}\bg_{\a\b}\bg_{\mu[\rho}(\bp^*\bd_{\s]}\bp- \bp\bd_{\s]}\bp^*)
\nn &&
-\bg_{\a\mu}\bnabla_{[\s}\bF_{\b\rho]}-\frac{1}{2}\bg_{\a\b}\bnabla_{[\s}\bF_{\rho]\mu}
+\bg_{\a[\rho}\bnabla_{\s]}\bF_{\b\mu},
\nn
\tilde{\Omega}_{\mu,\a,\rho\s}^{A,A}
&=& \br_{\rho\s\mu\a}-\bF_{\mu\a}\bF_{\rho\s}-2\bF_{\mu[\rho}\bF_{\s]\a}
-\frac{1}{2}\bg_{\mu[\rho}\bF_{\a}{}^\kappa \bF_{\s]\kappa}+\frac{1}{2}\bg_{\a[\rho}\bF_{\mu}{}^\kappa \bF_{\s]\kappa}
-e_S ^{2}\bg_{\mu[\rho}\bg_{\a\s]}\bp^*\bp,
\nn
\tilde{\Omega}_{\mu,\rho\s}^{A,\vp}
&=&-\frac{1}{2}\bd_{\mu}\bp^*\bF_{\rho\s}-\frac{1}{2}\bg_{[\rho\mu}\bd^{\kappa}\bp^*\bF_{\kappa\s]}
+\bd_{[\rho}\bp^*\bF_{\mu\s]}+ie_S \bg_{\mu[\rho}\bd_{\s]}\bp^*,
\nn
\tilde{\Omega}_{\mu,\rho\s}^{A,\vp^*}
&=& -\frac{1}{2}\bd_{\mu}\bp \bF_{\rho\s}-\frac{1}{2}\bg_{[\rho\mu}\bd^{\kappa}\bp \bF_{\kappa\s]}
+\bd_{[\rho}\bp \bF_{\mu\s]}-ie_S \bg_{\mu[\rho}\bd_{\s]}\bp,
\nn
\tilde{\Omega}_{\a\b,\rho\s}^{\vp^*,h}
&=& \frac{1}{16}\bd_{\a}\bp\bar{\psi}\c_{\b\rho\s}\psi-\frac{1}{16}\bg_{\a[\rho}\bd^{\kappa}\bp\bar{\psi}\c_{\b\s]\kappa}
\psi +\bg_{\a[\rho}\bd_{\b}\bd_{\s]}\bp,
\nn
\tilde{\Omega}_{\a,\rho\s}^{\vp^*,A}
&=& \frac{1}{2}\bg_{\a[\rho}\bd^{\kappa}\bp \bF_{\kappa\s]}+\frac{1}{2}\bd_{\a}\bp \bF_{\rho\s}
+\bd_{[\rho}\bp \bF_{\s]\a}+ie_S \bg_{\a[\rho}\bd_{\s]}\bp,
\nn
\tilde{\Omega}_{\rho\s}^{\vp^*,\vp}
&=&-ie_S \bF_{\rho\s}-\bd_{[\rho}\bp^*\bd_{\s]}\bp,
\nn
\tilde{\Omega}_{\rho\s}^{\vp^*,\vp^*}&=&0,
\nn
\tilde{\Omega}_{\a\b,\rho\s}^{\vp,h}
&=& \frac{1}{16}\bd_{\a}\bp^*\bar{\psi}\c_{\b\rho\s}\psi-\frac{1}{16}\bg_{\a[\rho}\bd^{\kappa}\bp^*\bar{\psi}
\c_{\b\s]\kappa}\psi+\bg_{\a[\rho}\bd_{\b}\bd_{\s]}\bp^*,
\nn
\tilde{\Omega}_{\a,\rho\s}^{\vp,A}
&=& \frac{1}{2}\bg_{\a[\rho}\bd^{\kappa}\bp^*\bF_{\kappa\s]}+\frac{1}{2}\bd_{\a}\bp^*\bF_{\rho\s}
+\bd_{[\rho}\bp^*\bF_{\s]\a}-ie_S \bg_{\a[\rho}\bd_{\s]}\bp^*,
\nn
\tilde{\Omega}_{\rho\s}^{\vp,\vp}&=&0,
\nn
\tilde{\Omega}_{\rho\s}^{\vp,\vp^*}&=&ie_S \bF_{\rho\s}+\bd_{[\rho}\bp^*\bd_{\s]}\bp.
\eea
Here it is understood that the antisymmetrization is made only for the indices $\rho$ and $\s$ and it has the weight $1/2$.
We have also used the relations
\bea
&&
ie_S \bg_{\a[\rho}\bp^*\bF_{\b\s]}+\bg_{\a[\rho}\bd_{\s]}\bd_{\b}\bp^*=\bg_{\a[\rho}\bd_{\b}\bd_{\s]}\bp^*,
\\ &&
\left[\bd_{\rho},\bd_{\s}\right]\bp^*=ie_S \bF_{\rho\s}\bp^*.
\eea

\subsection{Traces of various tensors}
\label{appc.3}

In this part we collect internal traces of various tensors, which we needed in section \ref{sec4}. In particular, they are essential information to compute explicitly results in formula (\ref{master2}).

We find that the trace of $\tilde {\bf W}$ is
\bea
\tr(\tilde {\bf W})=
7 \br -10 V -4 \bd_\mu \bp^*\bd^\mu \bp - \frac32 \bF_{\mu\nu}^2 +4e_S^2 |\bp|^2 +2 V' + 2 |\bp|^2 V'' \nn
-3 \bar\psi \gamma_\mu \bd^\mu \psi +3 \bd_\mu \bar\psi \gamma^\mu \psi - 10m_F \bar\psi \psi - 
 \frac{3}{128} \bar\psi \gamma_{\mu\nu\rho}\psi \bar\psi \gamma^{\mu\nu\rho} \psi.
\label{trw}
\eea
The trace of $\tilde {\bf W}^2$ is
\bea
&& \hs{-7}
\tr(\tilde {\bf W}^2) = 5 \br^2 - 12 \br V + 10 V^2 %- \frac12 \bnabla_\rho \bF_{\mu\nu} \bnabla^\mu \bF^{\nu\rho}
-\frac{1}{2} \bnabla_\mu \bF^{\mu}{}_{\nu} \bnabla_\rho \bF^{\nu\rho}
+\frac{3}{4} \bnabla_\mu \bF_{\nu\rho} \bnabla^\mu \bF^{\nu\rho}
\nn &&
+\frac{9}{2} i e_S \bp^* \bd_\mu \bp \bnabla_\nu \bF^{\mu\nu}
-\frac{9}{2} i e_S\bd_\mu \bp^*\bp\bnabla_\nu \bF^{\mu\nu}
+ \frac{19}{4} (\bd_\mu \bp^*\bd^\mu \bp )^2
+\frac{25}{4}(\bd_\mu\bp^*)^2(\bd_\mu\bp)^2
\nn &&
+2\bd_\mu \bd_\nu\bp^*\bd^\mu \bd^\nu\bp
- \bd_\mu^2 \bp^*\bd_\mu^2 \bp
-9i e_S \bd_\mu \bp^*\bd_\nu\bp \bF^{\mu\nu}
+\frac{29}{4}\bd_\mu\bp^* \bd_\nu\bp  \bF^{\mu}{}_{\rho}\bF^{\nu\rho }
\nn &&
+\frac{3}{4}\bF_{\mu\nu}\bF^{\mu}{}_{\rho}\bF^{\nu}{}_{\sigma}\bF^{\rho\sigma}
+\frac{3}{8}(\bF_{\mu\nu}^2)^2
- \bF_{\mu\nu} \bF^{\mu}{}_{\rho}\br^{\nu\rho}
-5 \br_{\mu\nu}^2
+3 \br_{\mu\nu\rho\sigma}^2
+ 3 e_S^2 |\bp|^2 \br
\nn &&
+ 13 e_S^4 |\bp|^4
-\frac{5}{4}\bd_\mu \bp^* \bd^\mu\bp\bF_{\mu\nu}^2 -5 e_S^2 |\bp|^2\bF_{\mu\nu}^2-\frac{1}{2} \bF_{\mu\nu}^2\br  
+  2\bp^* \bd_\mu^2 \bp V'+2  \bd_\mu^2\bp^*\bp  V'
\nn &&
+2V'^2
-8|\bp|^2 V'^2
+4|\bp|^2 V' V''
+4|\bp|^4 V''^2
-2(\bd_\mu\bp^*)^2\bp^2 V'' - 2\bp^{*2} (\bd_\mu\bp)^2 V''
\nn &&
-\frac{13}{2}e_S^2(\bd_\mu\bp^*)^2\bp^2 - \frac{13}{2}e_S^2\bp^{*2} (\bd_\mu\bp)^2
-4e_S^2 |\bp|^2 V'+ 4 \bd_\mu\bp^*\bd^\mu\bp V -3 \bd_\mu\bp^*\bd^\mu\bp\br
\nn &&
 + 9 e_S^2\bd_\mu\bp^*\bd^\mu\bp+13 e_S^2 |\bp|^2\bd_\mu\bp^*\bd^\mu\bp- 4\bd_\mu\bp^*\bd^\mu\bp V'
-4|\bp|^2 \bd_\mu\bp^*\bd^\mu\bp V''
\nn &&
+ \frac{25}{32} \left(\bar\psi \gamma_\mu \bd^\mu \psi - \bd_\mu \bar\psi \gamma^\mu \psi\right)^2
+ 6m_F \left( \bar\psi \gamma_\mu \bd^\mu \psi - \bd_\mu \bar\psi \gamma^\mu \psi\right)  \bar\psi \psi
+ 10 m_F^2 \left(\bar\psi \psi\right)^2
\nn &&
+ \frac{17}{64} \left(\bar\psi \gamma_\mu \bd_\nu \psi - \bd_\nu \bar\psi \gamma_\mu \psi\right)^2
- \frac{17}{64} \left(\bar\psi \gamma_\mu \bd_\nu \psi - \bd_\nu \bar\psi \gamma_\mu \psi\right)
 \left(\bar\psi \gamma^\nu \bd^\mu \psi - \bd^\mu \bar\psi \gamma^\nu \psi\right) 
\nn &&
+ \frac{21}{1024}\left(\bar\psi \gamma_\mu \bd^\mu \psi - \bd_\mu \bar\psi \gamma^\mu \psi\right)
 \bar\psi \gamma_{\mu\nu\rho} \psi \bar\psi \gamma^{\mu\nu\rho} \psi
+ \frac{3}{64} m_F \bar\psi \psi \bar\psi \gamma_{\mu\nu\rho} \psi \bar\psi \gamma^{\mu\nu\rho} \psi
\nn &&
-  \frac{3}{128} \left(\bar\psi \gamma_\mu \bd_\nu \psi - \bd_\nu \bar\psi \gamma_\mu \psi\right)
 \bar\psi \gamma^{\mu}{}_{\rho\s} \psi \bar\psi \gamma^{\nu\rho\s} \psi 
+ \frac{1}{32768} \bar\psi \gamma_{\mu\nu\rho} \psi \bar\psi \gamma^{\mu}{}_{\s\kappa} \psi \bar\psi
 \gamma^{\nu\s}{}_{\la}
 \psi \bar\psi \gamma^{\rho\kappa\la} \psi
\nn &&
+ \frac{3}{16384} \bar\psi \gamma_{\mu\nu\rho} \psi \bar\psi \gamma^{\mu\nu}{}_{\sigma} \psi \bar\psi
 \gamma^{\rho}{}_{\kappa\lambda}
 \psi \bar\psi \gamma^{\sigma\kappa\lambda} \psi
+ \frac{1}{65536}   \bar\psi \gamma_{\mu\nu\rho} \psi
 \bar\psi \gamma^{\mu\nu\rho} \psi \bar\psi \gamma_{\s\kappa\la} \psi \bar\psi \gamma^{\s\kappa\la}\psi
\nn &&
-  \frac{7}{16} \left(\bar\psi \gamma_\mu \bd^\mu \psi - \bd_\mu \bar\psi \gamma^\mu \psi\right)
 \bF_{\nu\rho} \bF^{\nu\rho}
+ \frac{1}{128} \bar\psi \gamma_{\mu\nu\rho} \psi \bar\psi \gamma^{\mu\nu\rho} \psi \bF_{\s\kappa} \bF^{\s\kappa}
\nn &&
-  \frac{15}{512} \bar\psi \gamma_{\mu\rho\sigma} \psi \bar\psi \gamma_{\nu}{}^{\rho\sigma} \psi
 \bF^{\mu}{}_{\kappa} \bF^{\nu\kappa} 
+ \frac{7}{4} \left(\bar\psi \gamma_\mu \bd_\nu \psi - \bd_\nu \bar\psi \gamma_\mu \psi\right) \bF^{\mu}{}_{\rho}
 \bF^{\nu\rho}
\nn &&
+ \frac{1}{512} \bar\psi \gamma_{\mu\rho\kappa} \psi \bar\psi \gamma_{\nu\sigma}{}^{\kappa} \psi \bF^{\mu\nu}
 \bF^{\rho\sigma}
+ \frac{7}{512} \bar\psi \gamma_{\mu\nu\rho} \psi \bar\psi \gamma^{\mu}{}_{\sigma\kappa} \psi \bF^{\nu\rho}
\bF^{\sigma\kappa}
\nn &&
-  \frac{3}{8}i e_F \bar\psi \gamma_{\mu\nu\rho} \psi \bar\psi \gamma^{\mu} \psi \bF^{\nu\rho} 
+ \frac{3}{64} \bar\psi \gamma_{\mu\nu\rho} \psi \bar\psi \gamma^{\mu\nu}{}_{\s} \psi \br^{\rho\s}
-  \frac{15}{4} \left(\bar\psi \gamma_\mu \bd^\mu \psi - \bd_\mu \bar\psi \gamma^\mu \psi\right) \br
\nn &&
- 12m_F \bar\psi \psi \br
-  \frac{5}{128} \bar\psi \gamma_{\mu\nu\rho} \psi \bar\psi \gamma^{\mu\nu\rho} \psi \br
-  \frac{1}{64} \bar\psi \gamma_{\mu\nu\rho} \psi \bar\psi \gamma^{\mu}{}_{\s\kappa} \psi \br^{\nu\rho\s\kappa}
\nn &&
-  \frac{1}{64} \bar\psi \gamma_{\mu\nu\rho} \psi \bar\psi \gamma^{\mu}{}_{\s\kappa} \psi \br^{\nu\s\rho\kappa}
+ 6 \left(\bar\psi \gamma_\mu \bd^\mu \psi - \bd_\mu \bar\psi \gamma^\mu \psi\right) V + 20m_F \bar\psi \psi V
\nn &&
+ \frac{3}{64} \bar\psi \gamma_{\mu\nu\rho} \psi \bar\psi \gamma^{\mu\nu\rho} \psi V
+  \frac{1}{16} \bar\psi \gamma_{\mu\nu\rho} \psi \bF^{\mu}{}_{\sigma} \bnabla^{\s}\bF^{\nu\rho} 
+ \frac{9}{2} e_F e_S \bar\psi \gamma_\mu \psi \bp^{*}{} \bd^{\mu}\bp
\nn &&
-  \frac{9}{2} e_F e_S \bar\psi \gamma_{\mu} \psi \bd^{\mu}\bp^* \bp
+ \frac{9}{16}i e_S \bar\psi \gamma_{\mu\nu\rho} \psi \bp^{*}{} \bd^{\mu}\bp  \bF^{\nu\rho}
-  \frac{9}{16}i e_S \bar\psi \gamma_{\mu\nu\rho} \psi \bd^{\mu}\bp^{*} \bp {} \bF^{\nu\rho}
\nn &&
+ 4m \bar\psi \psi \bd_{\mu}\bp^{*}{} \bd^{\mu}\bp
+ \frac{9}{256} \bar\psi \gamma_{\mu\nu\rho} \psi \bar\psi \gamma^{\mu\nu\rho} \psi \bd_{\s}\bp^{*}{} \bd^{\s}\bp
+ \frac{1}{8} \left(\bar\psi \gamma_\mu \bd^\mu \psi - \bd_\mu \bar\psi \gamma^\mu \psi\right) \bd_{\nu}\bp^{*}{}
 \bd^{\nu}\bp 
\nn &&
- 3i e_F \bar\psi \gamma_{\mu} \psi \bnabla_{\nu}\bF^{\mu\nu}
+ \frac{19}{8} \left(\bar\psi \gamma_\mu \bd_\nu \psi - \bd_\nu \bar\psi \gamma_\mu \psi\right)
 \bd^{\mu}\bp^{*}{} \bd^{\nu}\bp
\nn &&
+ \frac{19}{8} \left(\bar\psi \gamma_\mu \bd_\nu \psi-\bd_\nu \bar\psi \gamma_\mu \psi\right)
\bd^{\nu}\bp^{*}{} \bd^{\mu}\bp 
-  \frac{3}{32} \bar\psi \gamma_{\mu\nu\rho}\psi \bar\psi \gamma^{\mu\nu}{}_\s \psi \bd^{\rho}\bp^{*}{}\bd^{\s}\bp
 \nn &&
+  \frac{5}{16} \bar\psi \gamma_{\mu\nu\rho} \psi \bF^{\mu\nu} \bnabla_{\s}\bF^{\rho\s}
-  \frac{3}{128} \bd_{\mu}\left(\bar\psi \gamma^{\mu}{}_{\nu\rho} \psi\right) \bd_{\s}\left(\bar\psi
 \gamma^{\nu\rho\s} \psi\right)
- \frac{1}{16} \bar\psi \gamma_{\mu\nu\rho} \psi \bF^{\mu}{}_{\sigma} \bnabla^{\nu}\bF^{\rho\sigma}.
\label{trw2}
\eea
The trace of $\tilde {\bf \Omega}^2=\tilde {\bf \Omega}_{\rho\s}\tilde {\bf \Omega}^{\rho\s}$ is
\bea
&& \tr(\tilde{\bf\Omega}^2)=
-\frac{9}{2}\left(\bnabla_\mu\bF_{\nu\rho}\right)^2
%+\bnabla^\rho\bF^{\mu\nu} \bnabla_\nu \bF_{\mu\rho}
+\left(\bnabla_\mu \bF^{\mu}{}_{\nu}\right)^2
+3i e_S \left(\bp^* \bd_\mu\bp -\bd_\mu \bp^*\bp \right) \bnabla_\nu \bF^{\mu\nu} 
\nn &&
-6e_S^2\bd_\mu \bp^*\bd^\mu \bp+2\bd_\mu \bp^*\bd^\mu \bp\br 
-\frac{9}{2}\left(\bd_\mu\bp^*\bd^\mu\bp \right)^2
+3e_S^2 \left(\bd_\mu\bp^*\right)^2\bp^2 
+ 3 e_S^2 \bp^{*2} \left(\bd_\mu \bp\right)^2
\nn &&
-8 \bd_\mu\bd_\nu \bp^*\bd^\mu\bd^\nu \bp
+2 \bd_\mu^2\bp^*\bd_\mu^2\bp
-2ie_S\bd_\mu \bp^*\bd_\nu\bp  \bF^{\mu\nu}
\nn &&
-2e_S^2\left(\bF_{\mu\nu}\right)^2+\left(\bF_{\mu\nu}\right)^2\br-\frac{9}{2}\bd_\mu\bp^*\bd^\mu\bp
\left(\bF_{\mu\nu}\right)^2
+\frac{9}{2}\bd_\mu \bp^*\bd_\nu \bp \bF^{\mu}{}_{\rho}\bF^{\nu\rho}
+ \frac{17}{2} \bF_{\mu\nu}\bF^{\mu}{}_{\rho}\bF^{\nu}{}_{\s}\bF^{\rho\s}
\nn &&
-\frac{11}{2}\left(\bF_{\mu\nu}^2\right)^2
+4 \bd_\mu\bp^*\bd_\nu\bp \br^{\mu\nu}
+2 \bF_{\mu\nu}\bF^\mu{}_{\rho}\br^{\nu\rho}
-7\br_{\mu\nu\rho\s}^2
+ 2e_S^2 |\bp|^2 \br
\nn &&
-6 e_S^2 |\bp|^2 \bd_\mu\bp^*\bd^\mu\bp 
-2 e_S^2|\bp|^2 \bF_{\mu\nu}^2
-6 e_S^4|\bp|^4
-\frac{3}{2}\left(\bd_\mu\bp^*\right)^2\left(\bd_\mu\bp\right)^2 
\nn &&
+ \frac{3}{16384} \bar\psi \gamma_{\mu\nu\rho} \psi \bar\psi \gamma^{\mu}{}_{\sigma\kappa} \psi \bar\psi
 \gamma^{\nu\s}{}_{\lambda}
 \psi \bar\psi \gamma^{\rho\kappa\lambda} \psi
-  \frac{3}{16384} \bar\psi \gamma_{\mu\nu\rho} \psi \bar\psi \gamma^{\mu\nu}{}_{\sigma} \psi \bar\psi
 \gamma^{\rho}{}_{\kappa\lambda}
 \psi \bar\psi \gamma^{\sigma\kappa\lambda} \psi
\nn &&
-  \frac{1}{64} \left(\bar\psi \gamma_{\mu\nu\rho} \psi\right)^2 \left( \bF_{\mu\nu}\right)^2
-  \frac{11}{256} \bar\psi \gamma_{\mu\nu\rho} \psi\bar\psi \gamma^{\mu\nu}{}_{\sigma} \psi
 \bF^{\rho}{}_{\kappa}\bF^{\sigma\kappa}
\nn &&
+ \frac{21}{256} \bar\psi \gamma_{\mu\nu\rho} \psi \bar\psi \gamma^{\mu}{}_{\sigma\kappa} \psi
 \bF^{\nu\sigma} \bF^{\rho\kappa}
+ \frac{3}{256} \bar\psi \gamma_{\mu\nu\rho} \psi \bar\psi \gamma^{\mu}{}_{\sigma\kappa} \psi
 \bF^{\nu\rho} \bF^{\sigma\kappa}
\nn &&
+ \frac{3}{32} \bar\psi \gamma_{\mu\nu\rho} \psi \bar\psi \gamma^{\mu}{}_{\sigma\kappa} \psi \br^{\nu\sigma\rho\kappa}
+  \frac{1}{8} \bar\psi \gamma_{\mu\nu\rho} \psi \bF^{\mu}{}_{\sigma} \bnabla^{\sigma}\bF^{\nu\rho}
- \frac{3}{128} \left(\bar\psi \gamma_{\mu\nu\rho} \psi\right)^2  \bd_{\mu}\bp^{*}{} \bd^{\mu}\bp
\nn &&
+ \frac{3}{8}i e_S \bar\psi \gamma_{\mu\nu\rho} \psi  \bp^{*}{} \bd^{\mu}\bp \bF^{\nu\rho}
-  \frac{3}{8}i e_S \bar\psi \gamma_{\mu\nu\rho} \psi  \bd^{\mu}\bp^{*}{}\bp \bF^{\nu\rho}
- \frac{1}{8} \bar\psi \gamma_{\mu\nu\rho} \psi \bF^{\mu\nu} \bnabla_{\s}\bF^{\rho\s}
\nn &&
-  \frac{3}{32} \bar\psi \gamma_{\mu\nu\rho}\psi\bar\psi\gamma^{\mu\nu}{}_{\s}\psi\bd^{\rho}\bp^{*}{}\bd^{\s}\bp 
+ \frac{3}{4} \bar\psi \gamma_{\mu\nu\rho} \psi\bd^\mu \bd^\nu\bp^{*}\bd^{\rho}\bp
  + \frac{3}{4} \bar\psi \gamma_{\mu\nu\rho} \psi\bd^{\mu}\bp^{*}{}\bd^\nu\bd^\rho\bp  
\nn &&
- \frac{1}{8} \bar\psi \gamma_{\mu\nu\rho} \psi \bF^{\mu}{}_{\s} \bnabla^{\nu}\bF^{\rho\s}
- \frac{3}{4}  \bd_{\mu}\left(\bar\psi \gamma_{\nu\rho\s} \psi\right) \bF^{\mu\nu} \bF^{\rho\s}
- \frac{3}{2}  \bd_{\mu}\left(\bar\psi \gamma_{\nu\rho\s} \psi\right)\br^{\mu\nu\rho\s}
\nn &&
-  \frac{3}{64} \bd_{\mu}\left(\bar\psi \gamma_{\nu\rho\s} \psi\right) \bd^{\mu}\left(\bar\psi
 \gamma^{\nu\rho\s} \psi\right)
+ \frac{3}{64} \bd_{\mu}\left(\bar\psi \gamma_{\nu\rho\s} \psi\right) \bd^{\nu}\left(\bar\psi \gamma^{\mu\rho\s} \psi\right)
\nn &&
- \frac{3}{256} \bar\psi \gamma_{\mu\nu\rho} \psi \bar\psi \gamma^{\mu}{}_{\s\kappa}\psi\bd^{\nu}\left(
\bar\psi\gamma^{\rho\s\kappa} \psi\right).
\label{tro2}
\eea

%%%%%%%%%%%%%%%%%%%%%%%%%%%%%%%%%

\end{document}